\documentclass[1p]{elsarticle}
\journal{Computer Physics Communications}

\usepackage{lineno}

\biboptions{comma,square,sort&compress}
\usepackage{hyperref}
\hypersetup{pdfborder={0 0 0},
            colorlinks=true,
            urlcolor=blue,
            breaklinks=true,
            extension = }
\usepackage{amssymb}
\usepackage{amsmath}
\DeclareMathSymbol{\shortminus}{\mathbin}{AMSa}{"39}
\newcommand{\pvec}[1]{\vec{#1}\mkern2mu\vphantom{#1}} 

\usepackage{float}
\usepackage[caption=false]{subfig}

\usepackage{multirow}

\usepackage{booktabs}

\newcommand{\citea}[1]{{\protect\NoHyper\citeauthor{#1}\protect\endNoHyper}}

\usepackage{acronym}
\newacro{ABI}[ABI]{Application Binary Interface}
\newacro{API}[API]{Application Programming Interface}
\newacro{BIS}[BIS]{Backward Importance Sampling}
\newacro{cdf}[CDF]{Cumulative Distribution Function}
\newacro{csda}[CSDA]{Continuous Slowing Down Approximation}
\newacro{COMET}[COMET]{COherent Muon to Electron Transition}
\newacro{dcs}[DCS]{Differential Cross-Section}
\newacro{ddcs}[DDCS]{Doubly Differential Cross-Section}
\newacro{DEM}[DEM]{Digital Elevation Model}
\newacro{gos}[GOS]{Generalized Oscillator Strength}
\newacro{GRAND}[GRAND]{Giant Radio Array for Neutrino Detection}
\newacro{IS}[IS]{Importance Sampling}
\newacro{LOC}[LOC]{Lines Of Code}
\newacro{LPM}[LPM]{Landau, Pomeranchuk and Migdal~\citep{Landau1953,Migdal1956}}
\newacro{MDF}[MDF]{Materials Description File}
\newacro{pdf}[PDF]{Probability Density Function}
\newacro{pdg}[PDG]{Particle Data Group\,\citep{Zyla2020}}

\usepackage{etoolbox}
\makeatletter
\patchcmd{\ps@pprintTitle}
  {Preprint submitted to}
  {Accepted for publication in}
  {}{}
\patchcmd{\ps@pprintTitle}
  {\@date}
  {May 31, 2022}
  {}{}
\makeatother

\begin{document}
\begin{frontmatter}
\title{The PUMAS library}

\author[lpc]{Valentin~Niess\corref{cor1}}
\ead{niess@in2p3.fr}

\cortext[cor1]{Corresponding author}

\address[lpc]{
Universit\'e Clermont Auvergne, CNRS/IN2P3, LPC, F-63000 Clermont-Ferrand,
France.}

\begin{abstract}
The PUMAS library is a transport engine for muon and tau leptons in matter. It
can operate with a configurable level of details, from a fast deterministic CSDA
mode to a detailed Monte~Carlo simulation.  A peculiarity of PUMAS is that it
is revertible, i.e. it can run in forward or in backward mode. Thus, the PUMAS
library is particularly well suited for muography applications. In the present
document, we provide a detailed description of PUMAS, of its physics and of its
implementation.
\end{abstract}

\end{frontmatter}


{\bf PROGRAM SUMMARY}

\begin{small}
\noindent
{\em Program Title: The PUMAS library} \\
{\em CPC Library link to program files:} (to be added by Technical Editor) \\
{\em Developer's repository link: https://niess.github.io/pumas-pages } \\
{\em Code Ocean capsule:} (to be added by Technical Editor)\\
{\em Licensing provisions: LGPL-3.0 } \\
{\em Programming language: C99} \\
{\em Nature of problem:
    Transport of high energy muon or tau leptons in matter. } \\
{\em Solution method:
    Transport engine with a configurable level of details, from a fast
    deterministic CSDA mode to a detailed Monte~Carlo simulation. The transport
    engine can operate in both forward and backward modes.
} \\
\end{small}

\section{Introduction}

The PUMAS library was initially designed for solving the muography forward
problem using the \ac{csda}, which is a deterministic transport model. The
muography forward problem consists in computing the flux of atmospheric muons
transmitted through dense targets ($\rho \gtrsim 1\,$g$/$cm$^3$, typically
rocks), with dimensions from metres to kilometres. Inverting this flux in order
to determine the target parameters, e.g. its bulk density, is an inverse problem
which is beyond the scope of PUMAS.

A spectacular result of muography is the discovery of a big void in Khufu's
pyramid, reported by \citet{Morishima2017}. Many other applications of
atmospheric muons have been investigated as well. A detailed overview of this
topic is available in the review of \citet{Bonechi2020}.

Then, PUMAS has been enhanced with detailed Monte~Carlo capabilities similar to
other existing muon transport engines, e.g. MUM\,\citep{Sokalski2001},
MUSIC\,\citep{Kudryavtsev2009} or PROPOSAL\,\citep{Koehne2013,Dunsch2019}. A
particular care was taken in accurately modelling the physics, not only at high
energies $E \gtrsim 10\ $GeV, but also at lower ones down to $\sim$$1\ $MeV
kinetic energy.  While only high energy atmospheric muons are expected to be
transmitted through typical muography targets, low energy ones scattering on the
target surface contribute to the background of muon images, as discussed
by~\citet{Gomez2017}. Thus, accurate muography predictions require to transport
muons precisely over a large range of energies covering the spectrum of
atmospheric muons, e.g. up to $\sim$PeV energies for thick targets.

Though the primary scope of PUMAS is muography it is not limited to that.  The
transport engine has been extended to tau leptons in addition to muons.  PUMAS
is part of various experiments software, not only muography ones.  For
example, it is used by the \ac{GRAND} experiment~\citep{Alvarez-Muniz2020} for
its sensitivity study to neutrinos of cosmic origin.  PUMAS is also used by the
\ac{COMET} experiment~\citep{Abramishvili2020} in order to estimate the rate of
background events induced by atmospheric muons.

From its initial design, PUMAS retains the capacity to operate at various levels
of details. The accuracy of the transport is configurable on the fly, i.e.
during the particle propagation. While PUMAS is able to perform detailed
Monte~Carlo simulations, it can be more relevant to resort to \ac{csda},
depending on the use case. As matter of fact, \ac{csda} is surprisingly accurate
for the muography of small targets, with an extent of less than a few hundred
metres, as illustrated in section~\ref{sec:validation-transmission}.

Being revertible is another peculiarity of the PUMAS transport engine.  PUMAS
can run in forward or in backward mode. The latter is implemented using a
Jacobian weighting procedure described in detail in \citet{Niess2018}. Combined
with an accurate modelling of the physics at low energies, PUMAS is particularly
efficient for simulating the background induced by scattered muons, as
illustrated in section~\ref{sec:validation-showa-shinzan}.

\subsection{Scope of the document}

This article provides a physics oriented description of the PUMAS library. It
does not describe the library \ac{API}. It does not discuss practical examples
of usage either.  Those are available from the PUMAS
website\,\citep{PUMAS:GitHub}. The backward specific implementation details have
been previously addressed by \citet{Niess2018}, and are only briefly reported
herein.  For a complete overview the reader can refer to the original article
on backward Monte~Carlo.

The article is divided in four main sections.  Section~\ref{sec:physics}
discusses the interactions of muon and tau leptons with matter, as implemented
in PUMAS. Section~\ref{sec:condensed} provides a detailed overview of the
transport algorithms available in PUMAS.  Section~\ref{sec:miscellaneous}
contains complementary information for more specific usages of PUMAS, e.g.
composite materials or non uniform densities. Section~\ref{sec:validation}
discusses a selection of validation tests of PUMAS.

This article concerns version~$1.2$ of PUMAS, the latest at the time of this
writing. However, most physics aspects discussed herein are not bound to a
specific version of PUMAS.

\subsection{Common definitions}

Before getting into the details, let us introduce some notations and definitions
used throughout the discussion. A summary of those can be found in
table~\ref{tab:variables}. It should be noted that we use a natural system of
units where $c = 1$ and $4 \pi \epsilon_0 = 1$.

\begin{table}
    \caption{Definition of physical constants and variables used in the text.  A
    natural system of units is used where $c = 1$ and $4 \pi \epsilon_0 =
    1$.  The values of physical constants are also indicated using PUMAS system
    of units, i.e. GeV and m.
    \label{tab:variables}}
\center
\begin{tabular}{lll}
    \toprule
    Symbol & Description & Value (GeV, m) \\
    \midrule
    $\alpha$ & Fine structure constant, $\alpha = e^2 / \hbar$. &
        $7.29735\cdot 10^{-3}$ \\
    $\hbar$  & Reduced plank constant. & $1.97327\cdot 10^{-16}$ \\
    $\mathcal{N}_A$ & Avogadro’s number. & $6.02214\cdot 10^{23}$ \\
    \midrule
    $m_e$ & Electron rest mass. & $0.51100\cdot 10^{-3}$ \\
    $r_e$ & Classical electron radius, $r_e = \alpha \hbar / m_e$.
        & $2.81794\cdot 10^{-15}$ \\
    \midrule
    $m$ & Projectile (muon or tau) rest mass. & \\
    $E$ & Projectile total energy, $E = \gamma m$. & \\
    $p$ & Projectile momentum, $p = \beta E$. & \\
    $T$ & Projectile kinetic energy, $T = E - m$. & \\
    $z$ & Projectile charge number, $z = \pm 1$. & \\
    \midrule
    $Z$ & Target atomic charge number. & \\
    $A$ & Target atomic mass number. & \\
    $M$ & Target average molar mass, \eqref{eq:molar-mass}. & \\
    $I$ & Target mean excitation energy,
        \eqref{eq:mean-excitation-energy}. & \\
    $a_S$ & Target electronic shells scaling factor,
        \eqref{eq:resonance-energies}. & \\
    \midrule
    $\sigma$ & Reaction cross-section. & \\
    $\nu$ ($x$) & Projectile (fractional) energy loss,
        \eqref{eq:fractional-loss}. & \\
    $\theta$ ($\mu$) & Projectile deflection angle (parameter),
        \eqref{eq:angular-parameter}. & \\
    $\nu_{\scriptscriptstyle{C}}$ ($\mu_{\scriptscriptstyle{C}}$) & Energy
    (angular) cutoff for hard collisions. & \\
    $\Lambda_h$ ($\Sigma_h$) & Hard interaction length
        ($\Sigma_h = 1 / \Lambda_h$), \eqref{eq:hard-interaction-length}. & \\
    $S$ ($S_s$) & Material (soft) stopping power,
        \eqref{eq:csda-energy-loss}.& \\
    $\lambda_1$ ($\lambda_{1,s}$) & (Restricted) transport mean free path,
        \eqref{eq:transport-path}.& \\
    $\Omega$ ($\Omega_s$) & (Soft) energy straggling,
        \eqref{eq:soft-straggling}.& \\
    \bottomrule
\end{tabular}
\end{table}

Let us first consider a single collision of a muon or tau projectile on a target
atom. Let $m$, $E$ and $p$ denote the rest mass, the total energy and the
momentum of the projectile before the collision. The relativistic factors of
the projectile are $\beta = p / E$ and $\gamma = E / m$. For transport problems,
we find more natural to consider the kinetic energy $T = E - m$ instead of the
total energy $E$ of the projectile. If not explicitly specified, the term
``energy'' stands for the projectile kinetic energy herein.

The parameters of the target atom are its charge number $Z$ and its mass number
$A$. Let $\sigma$ denote the microscopic cross-section of this reaction.  The
collision parameters of interest to us are the projectile energy loss $\nu$ and
its deflection angle $\theta$, defined as $\cos\theta = \vec{p}\cdot\pvec{p}' /
(p p')$, with $p'$ the projectile momentum after the collision. The \ac{dcs} or
\ac{ddcs} w.r.t. to collision parameters are written using Leibniz notation,
e.g. $d\sigma/d\nu$.

In some cases, it is convenient to consider reduced collision parameters $x$ and
$\mu$ instead of $\nu$ and $\theta$. Those are defined as
\begin{align}
    x =& \nu / T \label{eq:fractional-loss}, \\
    \mu =& (1 - \cos\theta) / 2. \label{eq:angular-parameter}
\end{align}

The propagation of a muon or tau leptons through matter can be considered as a
succession of independent collision processes with ``isolated'' atoms
constituting the target material.  An important exception to the isolated
assumption is the collision with bound atomic electrons.  Thus, the target
material reduces to an atomic mixture together with an electronic structure. Let
$f_i$ denote the mass fractions of constituent atomic elements of the material
and $A_i$ their corresponding mass numbers. Let $\sigma_{el}$ be the
cross-section for collisions with bound electrons of the material and
$\sigma_{ij}$ the one for collisions on atom $i$ with physics process $j$.  The
total cross-section $\sigma$ is given by
\begin{equation} \label{eq:total-cross-section}
    \frac{1}{M} \sigma =
        \sum_{i,j}{\frac{f_i}{A_i} \sigma_{ij}} + \frac{1}{M} \sigma_{el},
\end{equation}
where $M$ is the average molar mass of the atomic mixture, defined as
\begin{equation} \label{eq:molar-mass}
    \frac{1}{M} = \sum_i{\frac{f_i}{A_i}} .
\end{equation}

The transport of muon or tau particles through matter implies a large number of
collisions, most of which occur with little energy loss or deflection.  Those
soft events are thus best treated collectively.  Two quantities are of
particular interest when considering collective effects: the stopping power,
denoted $S$, and the transport mean free path, denoted $\lambda_1$. When
considering only soft collisions with energy transfer below some cutoff value
$\nu_{\scriptscriptstyle{C}}$, one gets the soft stopping power
\begin{equation} \label{eq:csda-energy-loss}
    S_s(T, \nu_c) =
        \frac{\mathcal{N}_A}{M}
        \int_{0}^{\nu_{\scriptscriptstyle{C}}}{\frac{d\sigma}{d\nu}(T)\nu d\nu}.
\end{equation}
The total stopping power is $S(T) = S_s(T, T)$. Note that we use the mass
stopping power. In order to get the stopping power per path length, one must
multiply $S$ or $S_s$ by the density of the target material.

The transport mean free path $\lambda_{1,s}$ restricted to soft collisions with
$\mu \leq \mu_{\scriptscriptstyle{C}}$ is
\begin{equation} \label{eq:transport-path}
    \frac{1}{\lambda_{1,s}(T, \mu_{\scriptscriptstyle{C}})} =
        2 \frac{\mathcal{N}_A}{M}
        \int_0^{\mu_{\scriptscriptstyle{C}}}{\frac{d\sigma}{d\mu}(T)\mu d\mu}.
\end{equation}
As for the stopping power, $\lambda_{1,s}$ is defined per mass of the target
material. The total transport mean free path is $\lambda_1(T) = \lambda_{1,s}(T,
1)$.  The transport mean free path is the analogue of the stopping power for
angular deflections. It is directly related to the multiple scattering angle
after a large number of collisions.

A target material of particular importance for muography is ``standard rock''.
This is a fictitious material whose properties approximate an average rock of
the Cayuga Rock Salt Mine near Ithaca, New York. Let us follow the definition of
\citet{Groom2001}. Standard rock is thus made of a single fictitious atom of
``rockium'' (Rk) with $Z = 11$ and $A = 22\ $g/mol. It has the electronic
structure of calcium carbonate, but its density is $2.65\ $g/cm$^3$.  Let us
point out that real rocks are actually composite materials made of several
minerals. However, standard rock has been established as a convenient standard
over years in the astroparticle community.  Hence, it is used as reference
material herein.

\section{Interactions of muons and taus with matter \label{sec:physics}}

In the present section we review the interactions of muon and tau leptons with
matter, as implemented in PUMAS.  The interactions of muons and taus are
described with the same models. These models differ from the one used for
electrons due to the larger masses of muons and taus.  In the following we focus
on the case of muons in order to simplify the discussion. In addition we limit
the discussion to collisions with a single atomic element, except for
electronic\footnote{In this manuscript we use ``electronic'' to refer to
scattering from atomic electrons.} collisions, discussed in
section~\ref{sec:electronic-process}.

\begin{figure}[th]
    \center
    \includegraphics[width=\textwidth]{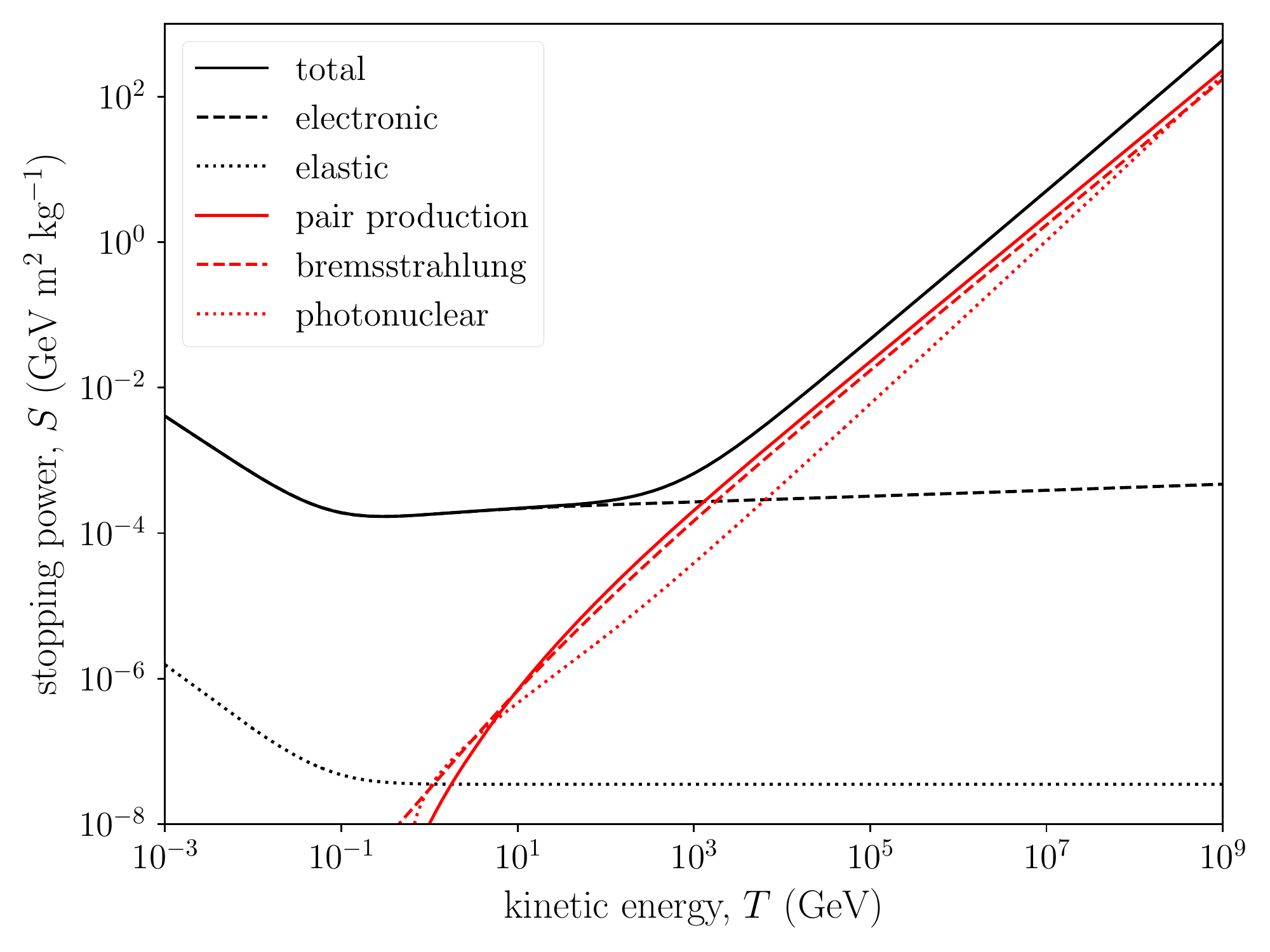}
    \caption{Stopping power, $S$, for a muon in standard rock. The total
    stopping power is indicated as well as the individual contributions of the
    different processes discussed in section~\ref{sec:physics}.
    \label{fig:energy-loss}}
\end{figure}

The interactions of muon and tau leptons with matter consist of five physics
processes, discussed further in this section.  The contributions of these
processes to the total stopping power is shown on figure~\ref{fig:energy-loss},
for a muon in standard rock.

\subsection{Elastic collisions \label{sec:elastic}}

The elastic collision of charged particles with atoms is a scattering process,
$\ell + Z \to \ell + Z$. This collision is described by a single parameter, the
energy loss $\nu$ or alternatively the scattering angle $\theta$. Both are
related by collision kinematics.

Elastic collisions occur with a negligible energy transfer compared to other
processes discussed hereafter (see e.g. figure~\ref{fig:energy-loss}).  They are
however the main contribution to the deflection of muons from their initial
trajectory, as further discussed in section\,\ref{sec:condensed}. Thus, unlike
other processes, elastic collisions are parametrized by an angular coordinate
rather than by the energy loss.

In PUMAS, elastic collisions are modelled following~\citet{Salvat2013}, with
some modifications discussed below.

\subsubsection{Elastic DCS in the centre of mass frame}

For fast projectiles, the elastic process is essentially an electrostatic
interaction between a point-like particle and the target atom field, which is
considered as unaltered over the course of the interaction. The corresponding
radial symmetric interaction potential is
\begin{equation}
    V(r) = \frac{z Z e^2}{r} \omega(r),
\end{equation}
where $\omega(r)$ is a screening function depending on the distance $r$ between
the projectile and the target atom.  For a point-like nucleus, the atomic
screening is well approximated by a sum of exponentials:
\begin{equation} \label{eq:screening}
    \omega(r) = \sum_{i=1}^{n}{a_i e^{-b_i r}} .
\end{equation}
The coefficients $a_i$ and $b_i$ and the number of terms $n$ depend on $Z$. In
PUMAS, we use the results of \citet{Salvat1987} as coefficients values. These
coefficients have been obtained from self consistent Dirac-Hartree-Fock-Slater
(DHFS) calculations.  However, for $Z=1$ we use a single exponential instead, as
explained in \ref{sec:elastic-implementation}.

The target recoil is accounted for by considering the collision in the Centre of
Mass (CM) frame but with an effective projectile mass (see
e.g.~\citet{Boschini2011}). Let $p_0$ denote the projectile momentum in the CM
frame. The projectile and target energies are individually conserved, in the CM
frame, but the momentum direction changes by an angle $\theta_0$.  Within the
eikonal approximation, the elastic \ac{dcs} in the CM frame is given by
\begin{align} \label{eq:eikonal_dcs}
    \frac{d\sigma_0}{d\mu_0} =& 4 \pi \frac{p_0^2}{\hbar^2}
        \left| \mathcal{H}_0\{e^{i \varphi(r)} - 1\}(k_0) \right|^2, \\
        \varphi(r) =& -2 \xi \int_{r}^{+\infty}{\frac{\omega(r')}{\sqrt{r^2 -
        r'^2}}dr'},
\end{align}
where $k_0 = 2 \frac{p_0}{\hbar} \sqrt{\mu_0}$ and $\mu_0 = \frac{1}{2}(1 -
\cos(\theta_0))$. $\mathcal{H}_0$ is the Hankel transform of order zero, defined
as
\begin{equation}
    \mathcal{H}_0\{f\}(k) = \int_0^{+\infty}{J_0(k r) f(r) r dr},
\end{equation}
where $J_0$ is the Bessel function of the first kind and of order zero (see e.g.
\citet{Piessens2000}).

The prefactor $\xi$ appearing in the eikonal phase $\varphi$ is defined as
\begin{equation} \label{eq:expansion-parameter}
    \xi = \frac{\alpha z Z}{\beta} .
\end{equation}
Note that $\xi$ depends on the projectile relative speed $\beta$ in the
laboratory frame, not in the CM one.

Note also that \citet{Salvat2013} applies an extra correction to the eikonal
phase following \citet{Wallace1971}. However, for the present scope, i.e.  muon
or tau projectiles with kinetic energy $T \geq 1\ $MeV, this correction is
negligible, as shown in \ref{sec:wallace}. It is thus not considered in PUMAS.

For exponential screening functions given by equation~\eqref{eq:screening}, the
eikonal phase is written as
\begin{equation}
    \varphi(r) = -2 \xi \sum_{i=1}^{n}{a_i K_0(b_i r)},
\end{equation}
where $K_0$ is the modified Bessel function of the second kind and of order
zero. The Hankel transforms of $K_0$ and of $K_0^2$ are (see e.g. table 8.3 of
\citet{Bateman1953})
\begin{align}
    \mathcal{H}_0\{K_0(b_i r)\}(k) =& \frac{1}{k^2 + b_i^2}, \\
    \mathcal{H}_0\{K_0^2(b_i r)\}(k) =& \frac{1}{k \sqrt{k^2 + 4 b_i^2}}
        \ln\left(\frac{\sqrt{k^2 + 4 b_i^2} + k}{\sqrt{k^2 + 4 b_i^2} - k}
        \right) .
\end{align}
Expanding the exponential in equation~\eqref{eq:eikonal_dcs} at leading
order in $\xi$ yields
\begin{equation} \label{eq:born_dcs}
    \frac{d\sigma_0}{d\mu_0} = \frac{\pi \hbar^2}{p_0^2} \left(\xi
        \sum_{i=1}^n{\frac{a_i}{\mu_i + \mu_0}} +
        \mathcal{O}\left(\xi^2\right) \right)^2,
\end{equation}
where $\mu_i$ is an angular parameter accounting for the nuclear charge
screening, defined as
\begin{equation} \label{eq:screening-parameter}
    \mu_i = \left( \frac{\hbar b_i}{2 p_0} \right)^2 .
\end{equation}

\subsubsection{Coulomb correction}

The elastic \ac{dcs} given by equation~\eqref{eq:born_dcs} corresponds to the
first Born approximation. In order to improve the accuracy of this result, one
could compute higher order terms in $\xi$ or integrate the eikonal amplitude
numerically.  However, this would not be practical within a Monte~Carlo like
PUMAS.  Thus, it is customary to instead use the \ac{dcs} obtained from the
first Born approximation, but with an effective screening parameter
$\tilde{\mu}_i$ set in order to reproduce the multiple scattering distribution
obtained with the eikonal \ac{dcs}. For this purpose, the Coulomb correction
computed by \citet{Kuraev2014} is used in PUMAS. The screening parameters
$\mu_i$ are increased by
\begin{align}
    \label{eq:coulomb_correction}
    \tilde{\mu}_i   &= \mu_i e^{2 f(\xi)}, \\
    f(\xi) &= \operatorname{\mathbb{R}e}\{\psi(1 + i\xi) - \psi(1)\},
\end{align}
where $\psi$ is the digamma function. For $\xi < 1$, equation~(39) of
\citea{Kuraev2014} provides an accurate approximation of $f(\xi)$. For higher
values of $\xi$, the asymptotic expansion of the digamma function is more
efficient. It is given by
\begin{equation}
    f(\xi) = \frac{1501}{2520} + \ln(\rho) -
        \frac{1}{2 \rho^2} -
        \frac{\cos(2 \phi)}{12 \rho^2} -
        \frac{\cos(4 \phi)}{120 \rho^4} -
        \frac{\cos(6 \phi)}{252 \rho^6} + \mathcal{O}(\frac{1}{\rho^8}),
\end{equation}
where $\rho^2 = 1 + \xi^2$ and $\tan(\phi) = \xi$.

Note that the Coulomb correction computed by \citea{Kuraev2014} assumes a single
exponential term for the atomic screening $\omega$. It is shown in
\ref{sec:coulomb_correction} that the same Coulomb correction can be applied to
a weighted sum of exponentials as given by equation~\eqref{eq:screening}.

\subsubsection{Nuclear form factor}

At large scattering angles, i.e. small impact parameters, the finite size of the
nucleus charge needs to be taken into account. This effectively modifies the
interaction potential $V$ by adding an extra term to the screening function:
\begin{align} \label{eq:nuclear-screening}
    \omega(r) =& \sum_{i=1}^{n}{a_i e^{-b_i r}} + \Delta
    \omega_{\scriptscriptstyle{N}} (r), \\
    \Delta \omega_{\scriptscriptstyle{N}}(r) =&
    \int_0^r{\rho_{\scriptscriptstyle{N}}(r') 4 \pi r'^2 dr'} +
        r \int_r^{+\infty}{\rho_{\scriptscriptstyle{N}}(r') 4 \pi r' dr'} - 1,
\end{align}
where $\rho_{\scriptscriptstyle{N}}$ is the nuclear charge density normalized to
one.

It would be convenient to consider an exponential distribution for the nuclear
charge distribution, as previously. The exponential distribution leads to a
rational fraction in $\mu$ for the \ac{dcs}, which can be integrated
analytically.  However, the nuclear density has a close to uniform core.
Therefore, exponential functions are a rough approximation in this case.
Nevertheless, depending on the use case the exponential density might be
accurate enough, as discussed in \ref{sec:elastic-implementation}.

In order to get a more accurate description of the nucleus charge, the
uniform-uniform model of \citet{Helm1956} is used in PUMAS. The nuclear density
is parametrized by a convolution product of two uniform spheres. Let us further
assume that the spheres are identical of radius $R_N$. The corresponding U$^2$
nuclear density is given by
\begin{equation} \label{eq:nuclear_density}
    \rho_{\scriptscriptstyle{U^2}}(r) =
    \begin{cases}
        \frac{3}{64 \pi R_{\scriptscriptstyle{N}}^3} \left(16 - \frac{12
        r}{R_{\scriptscriptstyle{N}}} +
            \frac{r^3}{R_{\scriptscriptstyle{N}}^3} \right) & \text{if } r < 2
            R_{\scriptscriptstyle{N}} \\
        0 & \text{otherwise}
    \end{cases} .
\end{equation}
The model parameter $R_{\scriptscriptstyle{N}}$ can be determined from
experimental values of nuclear charge radii, e.g. as compiled by
\citet{DeVries1987}. The values used in PUMAS for isotopic mixtures are given in
\ref{sec:nuclear-radius}.

Computing the modified \ac{dcs} using the eikonal approximation in
equation~\eqref{eq:eikonal_dcs} would be rather involved. Fortunately, since
atomic and nuclear screenings operate at different length scales, the nuclear
contribution to the screening function can be taken into account by factorising
the elastic \ac{dcs} with a nuclear form factor $|F_N|^2$ (see e.g.
\citet{Butkevich2002a} or \citet{Salvat2005}).  Within the first Born
approximation, the form factor is the Fourier transform of the charge density.
For the U$^2$ distribution one obtains
\begin{equation}
    F_{\scriptscriptstyle{U^2}}(k_0) = \left| \frac{3}{(k_0
    R_{\scriptscriptstyle{N}})^3}\left(\sin(k_0 R_{\scriptscriptstyle{N}}) - k_0
    R_{\scriptscriptstyle{N}}
        \cos(k_0 R_{\scriptscriptstyle{N}}) \right) \right|^2.
\end{equation}
Note that $F_{\scriptscriptstyle{U^2}} = F_{\scriptscriptstyle{U}}^2$, where
$F_{\scriptscriptstyle{U}}$ is the form factor obtained for a uniform sphere or
radius $R_{\scriptscriptstyle{N}}$.

\begin{figure}[th]
    \centering
    \subfloat{\includegraphics[width=0.5\textwidth]{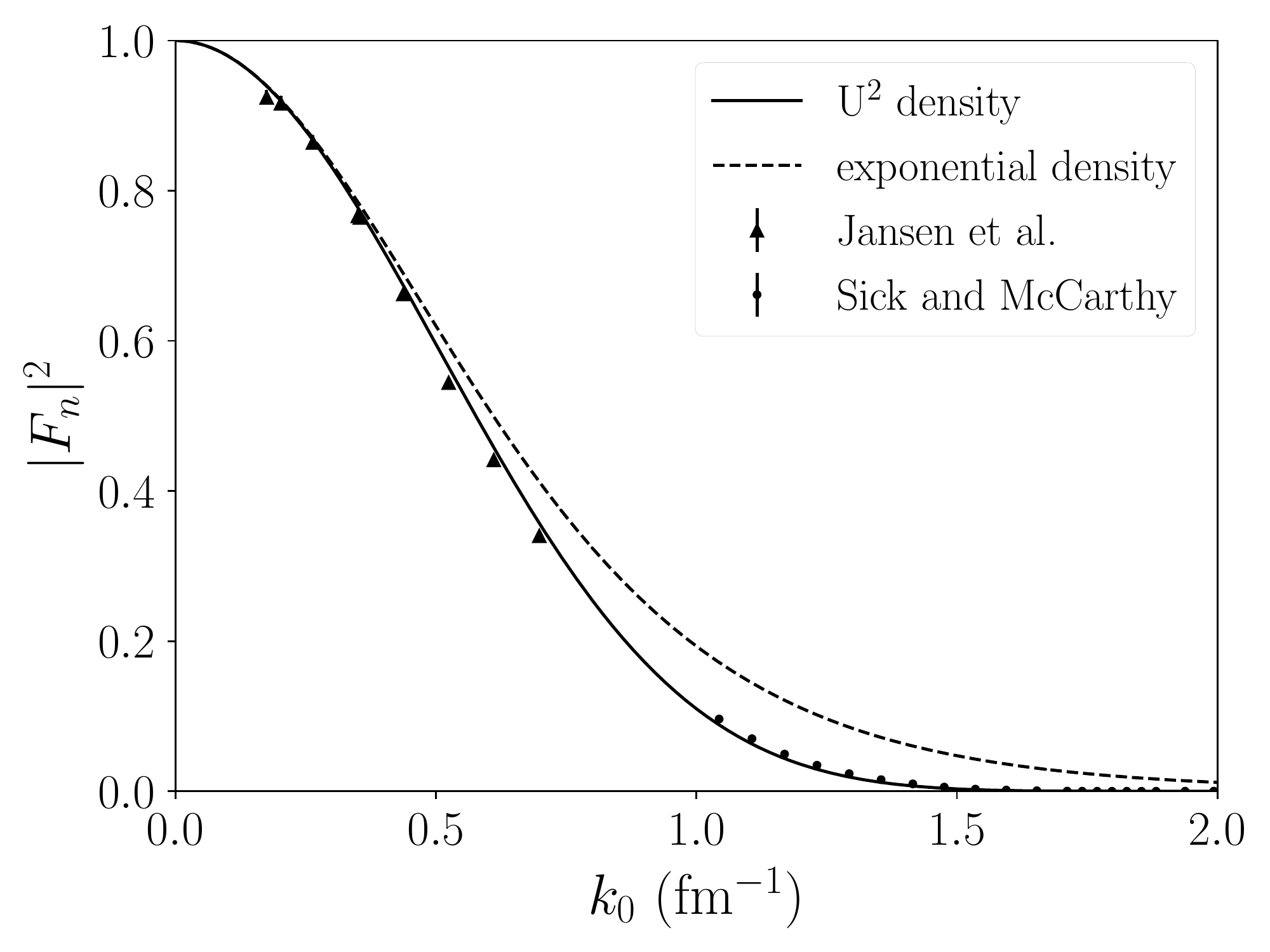}}
    \subfloat{\includegraphics[width=0.5\textwidth]{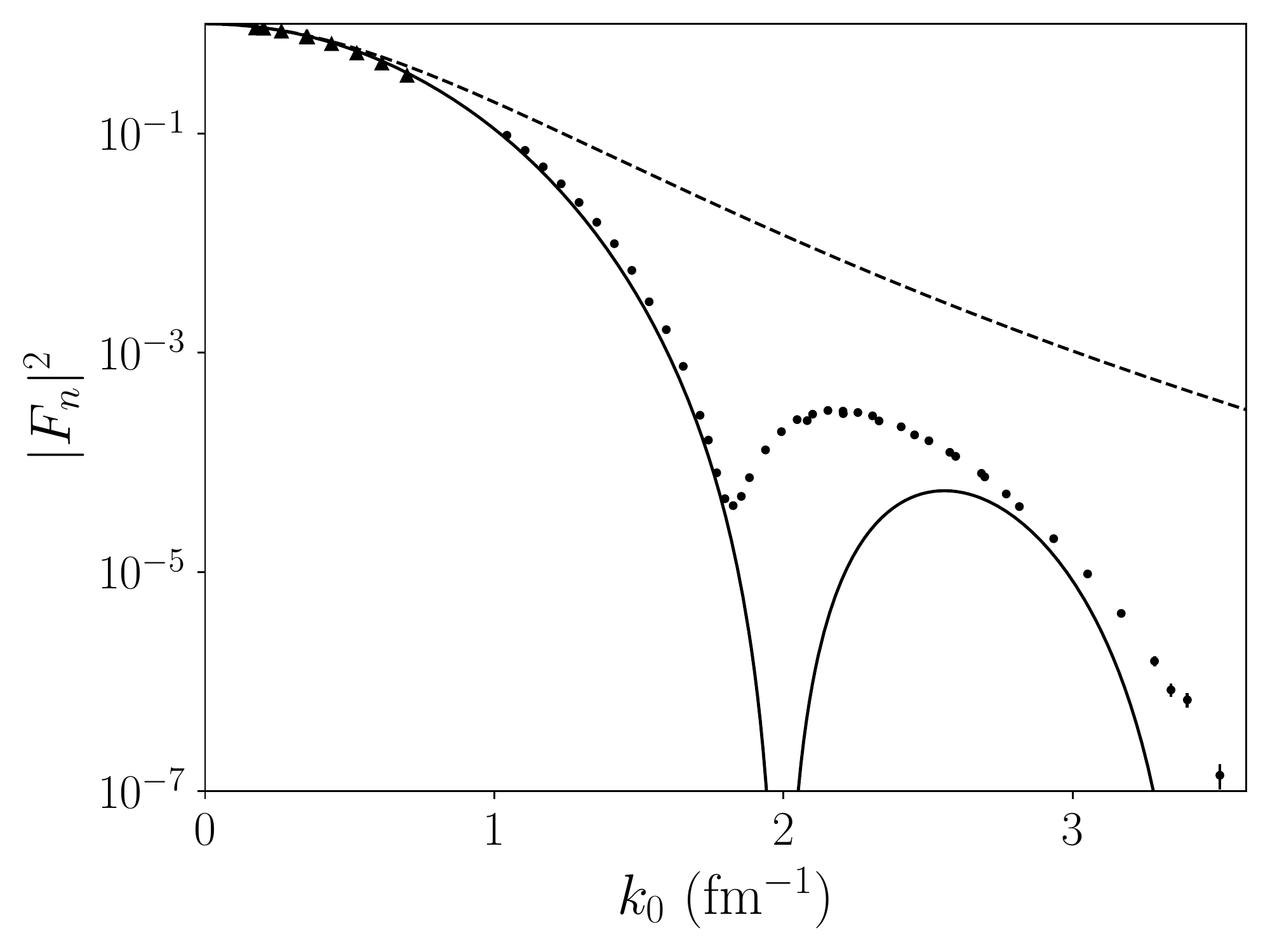}}
    \caption{Nuclear form factor, $|F_N|^2$, for $^{12}$C using a linear (left)
    or logarithmic (right) scale. Markers are experimental results from
    \citet{Jansen1972} and \citet{Sick1970}. The solid and dashed lines
    correspond to parametrizations using a U$^2$ or exponential nuclear charge
    density.
    \label{fig:nuclear_form_factor}}
\end{figure}

Figure~\ref{fig:nuclear_form_factor} shows a comparison of the U$^2$ form factor
to the experimental data of \citet{Jansen1972} and \citet{Sick1970}, obtained
from electron scattering on $^{12}$C. Similar results are obtained for other
atomic elements, e.g. $^{16}$O. The form factor for the exponential density is
also indicated. The U$^2$ distribution reproduces well the main peak of the form
factor using a single parameter $R_N$. But, this simple model fails to reproduce
higher order harmonics. However, this is not considered relevant for the present
purpose. Let us point out that we also investigated the use of two different
radii, $R_1$ and $R_2$, instead of a single one. This was not able to improve
the fit of secondary peaks, while requiring $R_1 \simeq R_2$ in order to fit the
main peak. Note also that using a single sphere instead of a U$^2$ convolution
product results in a too smooth decrease of the form factor.

Let us recall that the nuclear form factors considered so far have been obtained
from the first Born approximation. In the case of an exponential density, it is
shown in \ref{sec:coulomb_correction} that the Coulomb correction from
\citet{Kuraev2014} is still valid. It amounts to rescaling the nuclear radius
parameter, or the momentum transfer $k_0$, by
\begin{equation}
    \tilde{R}_{\scriptscriptstyle{N}} = R_{\scriptscriptstyle{N}} e^{-f(\xi)},
\end{equation}
where $f(\xi)$ was given previously in equation~\eqref{eq:coulomb_correction}.
In the case of the U$^2$ distribution we assume that this rescaling is valid as
well.

\subsubsection{Spin correction}

The elastic \ac{dcs} discussed previously have been derived for spinless
projectiles. In order to account for the muon or tau spin we follow
\citet{Salvat2013} and apply an additional multiplicative factor
$\mathcal{R}_{\scriptscriptstyle{S}}$ defined as the ratio of the spin $1/2$ to
the spin $0$ elastic \ac{dcs} obtained from the first Born approximation. It is
given by
\begin{equation}
    \mathcal{R}_{\scriptscriptstyle{S}} = 1 - \beta^2 \mu_0 .
\end{equation}

\subsubsection{Total elastic DCS}

Collecting all terms, PUMAS uses the following elastic \ac{dcs} in the CM frame:
\begin{equation} \label{eq:elastic-dcs}
    \frac{d\sigma_0}{d\mu_0} = \frac{\pi r_e^2 m_e^2 Z^2}{\beta^2 p_0^2} 
    \left(
        \sum_{i=1}^{n}{\frac{a_i}{\tilde{\mu}_i + \mu_0}} \right)^2
        \left| F_{\scriptscriptstyle{U}}\left(\sqrt{\frac{\mu_0}{\tilde{\mu}_{n+1}}}\right)
        \right|^4 (1 - \beta^2 \mu_0),
\end{equation}
where
\begin{equation}
    F_{\scriptscriptstyle{U}}(x) = \frac{3}{x^3} \left(\sin(x) - x
        \cos(x)\right),
\end{equation}
and
\begin{equation} \label{eq:elastic-nuclear-screening}
    \tilde{\mu}_{n+1} = \left( \frac{\hbar e^{f(\xi)}}{2 p_0 R_N} \right)^2 .
\end{equation}

The elastic \ac{dcs} in the laboratory frame can be computed from the CM one
using Lorentz transform e.g. as detailed by \citet{Salvat2013}. However, in
practice it is seldom required to transform the CM \ac{dcs} in PUMAS. For
example, the scattering angle in an elastic collision is simulated in the CM
frame, and then transformed back to the laboratory frame. Additional technical
details are provided in \ref{sec:elastic-implementation}.

The energy lost in elastic collisions is related to the scattering angle
$\theta$ (see e.g. equation~(49) from \citet{Salvat2013}). For muon and tau
leptons, elastic collisions contribute to less than $0.1\,\%$ to the stopping
power, for atomic elements ranging from hydrogen to uranium.  Therefore, elastic
collisions are approximated as lossless in PUMAS.

\subsection{Electronic collisions \label{sec:electronic-process}}

The electronic energy loss consists of inelastic collisions of the projectile
with the bound electrons of the target resulting in an excited electronic state.
Such collisions may lead to the ionisation of the target with ejection of a fast
electron, a.k.a. delta ray. As for elastic collisions, in PUMAS, electronic
collisions are modelled following \citet{Salvat2013} with some modifications
discussed below.

\subsubsection{Electronic DCS}

The target electronic structure is represented by $n$ shells with binding
energies $E_k$ and occupancy numbers $Z_k$. The total electric charge is $Z =
\sum{Z_k}$. Note that the additivity assumption over atomic constituents is not
valid for the electronic \ac{dcs}. I.e. the target electronic structure must be
considered as a whole in this case.

Following \citet{Salvat2013}, the electronic shells are modelled as independent
$\delta$--oscillators with \ac{gos} $f_k$ given by
\begin{equation}
    f_k(\nu, Q) = \begin{cases}
        \delta(\nu - I_k) & \text{if } Q \leq E_k \\
        \delta(\nu - Q) & \text{otherwise}
    \end{cases},
\end{equation}
where $Q$ is the recoil energy of the collided electron and $I_k > E_k$ the
resonance energy of the $\delta$--oscillator.

The electronic \ac{dcs} can be computed from the \acp{gos} as in
\citet{Fano1963}. It is obtained by summing up the contributions of each shell
and by integrating out the electron recoil, taking into account the medium
dielectric polarization. Details of the computation are provided in the
PENELOPE-2014 manual~\citep{Salvat2015}. The result can be expressed as:
\begin{equation}
    \frac{d\sigma}{d\nu} = \sum_{k=1}^n{Z_k \left(
        \frac{d\sigma_{\scriptscriptstyle{C},k}}{d\nu} +
        \frac{d\sigma_{\scriptscriptstyle{D},k}}{d\nu} \right)},
\end{equation}
where $\sigma_{\scriptscriptstyle{C},k}$ is the cross-section for close
collisions ($Q > E_k$) with the $k^\text{th}$ shell and
$\sigma_{\scriptscriptstyle{D},k}$ the cross-section for distant collisions ($Q
\leq E_k$).

The \ac{dcs} for close collisions is given by
\begin{equation}
    \frac{d\sigma_{\scriptscriptstyle{C},k}}{d \nu} = \begin{cases}
        \frac{2 \pi r_e^2 m_e}{\beta^2} \left[\frac{1}{\nu^2} -
            \frac{\beta^2}{\nu_\text{max}} \frac{1}{\nu} +
            \frac{1}{2 E^2} \right]
        & \text{if } E_k \leq \nu \leq \nu_\text{max} \\
        0 & \text{otherwise}
    \end{cases},
\end{equation}
where
\begin{equation} \label{eq:electronic-close-cutoff}
    \nu_\text{max} = \frac{2 m_e \beta^2 \gamma^2}{1 +
        2 \gamma \frac{m_e}{m} + \left(\frac{m_e}{m}\right)^2} .
\end{equation}

The \ac{dcs} for distant collisions has a discrete structure, contrary to the
close one.  Furthermore, following~\citet{Fano1963}, transverse interactions are
impacted by dielectric couplings  leading to the ``density effect'', initially
studied  by \,\citet{Fermi1940}.  The corresponding \ac{dcs} is given by
\begin{equation} \label{eq:dcs_inelastic_distant}
    \frac{d\sigma_{\scriptscriptstyle{D}, k}}{d \nu} = \frac{2 \pi
        r_e^2 m_e}{\beta^2} \left[
        \ln\left(\frac{2 m_e \beta^2 \gamma^2 E_k}{I_k^2 +
        E_{\scriptscriptstyle{F}}^2} \right) +
        \frac{E_{\scriptscriptstyle{F}}^2}{\gamma^2 E_p^2} - \beta^2 \right]
    \frac{\delta(\nu - I_k)}{I_k} .
\end{equation}
The plasma energy is
\begin{equation}
    E_p = \hbar \omega_p \simeq 28.816 \text{ eV}
        \sqrt{\frac{\rho}{\text{g}\ \text{cm}^{-3}}
        \frac{\text{g}\ \text{mol}^{-1}}{M} Z}
\end{equation}
where $\rho$ is the target density. The parameter $E_{\scriptscriptstyle{F}}$
appearing in equation~\eqref{eq:dcs_inelastic_distant} stems from the
screening of distant interactions due to the medium dielectric properties. It
is obtained by solving
\begin{equation} \label{eq:density_effect_L2}
    \sum_{k=1}^n{\frac{Z_k}{I_k^2 + E_{\scriptscriptstyle{F}}^2}} =
        \frac{Z}{\gamma^2 E_p^2} .
\end{equation}
The parameter $E_{\scriptscriptstyle{F}}$ increases with the projectile energy.
At high energies it reaches an asymptotic value, $E_{\scriptscriptstyle{F}}
\to \gamma E_p$, resulting in the density effect saturation of the energy loss.
Note that the previous equation~\eqref{eq:density_effect_L2} has no positive
solution for $E_{\scriptscriptstyle{F}}^2$ when
\begin{equation}
    \sum_{k=1}^n{\frac{Z_k}{I_k^2}} \leq \frac{Z}{\gamma^2 E_p^2} .
\end{equation}
In this case there is no density effect, as can be seen e.g. from Appendix B of
\citet{Fano1956}. Then, it is correct to set $E_{\scriptscriptstyle{F}}=0$ in
equation~\eqref{eq:dcs_inelastic_distant}.

\subsubsection{Radiative correction \label{sec:inelastic-radiative}}

At high energies, the interaction of the projectile with atomic electrons can
result in knock-on electrons with the creation of a bremsstrahlung photon. This
process was studied in detail by \citet{Kelner1997}. The bremsstrahlung photon
can be emitted by the muon or by the knock-on electron. The former case is
considered later in section~\ref{sec:bremsstrahlung}. The latter case is
accounted for by a radiative correction to the close electronic cross-section:
\begin{equation}
    \frac{d\sigma'_{\scriptscriptstyle{C,k}}}{d\nu} =
        \frac{d\sigma_{\scriptscriptstyle{C,k}}}{d\nu} (1 + \Delta_{e\gamma}) .
\end{equation}
Following \citet{Sokalski2001}, the radiative correction is given by
\begin{equation}
    \Delta_{e\gamma}(\nu) = \frac{\alpha}{2 \pi}
        \ln\left(1 + \frac{2 \nu}{m_e}\right)
        \left[ \ln\left(\frac{4 E (E - \nu)}{m^2}\right) -
        \ln\left(1 + \frac{2 \nu}{m_e}\right) \right] .
\end{equation}
Note that this correction is suppressed for $\nu \ll m_e$. Thus, it does not
impact the total cross-section for close interactions, which is essentially due
to collisions with $\nu \simeq E_k \ll m_e$. However, $\Delta_{e\gamma}$
increases the electronic energy loss at high energies. This is discussed further
below.

\subsubsection{Stopping power}

Let $I_{max}$ denote the maximum value of the union of the sets $\{E_k\}$ and
$\{I_k\}$. Then, collecting previous expressions and assuming $I_{max} \ll
\nu_{\scriptscriptstyle{C}} \leq \nu_{max}$, the soft electronic stopping
power is given by
\begin{multline} \label{eq:electronic-energy-loss}
    S_s(T, \nu_{\scriptscriptstyle{C}}) =
        \frac{2 \pi r_e^2 m_e Z}{\beta^2} \frac{\mathcal{N}_A}{M} \left[
        \ln\left(\frac{2 m_e \beta^2 \gamma^2
        \nu_{\scriptscriptstyle{C}}}{I^2} \right) -
        \beta^2 \left(1 +
        \frac{\nu_{\scriptscriptstyle{C}}}{\nu_{max}}\right) \right. \\ \left.
        - \delta_{\scriptscriptstyle{F}} +
        \frac{\nu^2_{\scriptscriptstyle{C}}}{4 E^2} +
        \delta_{e\gamma}(\nu_{\scriptscriptstyle{C}}) \right],
\end{multline}
where the material mean excitation energy $I$ and the Fermi density effect
correction $\delta_{\scriptscriptstyle{F}}$ are
\begin{align} \label{eq:mean-excitation-energy}
    Z \ln I =& \sum{Z_k \ln I_k} \\
    \delta_{\scriptscriptstyle{F}} =& \sum_{k=1}^n{\frac{Z_k}{Z} \ln\left(
        1 + \frac{E^2_{\scriptscriptstyle{F}}}{I_k^2}\right)} -
        \frac{E^2_{\scriptscriptstyle{F}}}{\gamma^2 E^2_p} .
\end{align}

A remarkable feature of the electronic stopping power is that, apart from the
density effect, the details of the electronic structure are summarised by the
mean excitation energy $I$. The latter can be estimated by various means. In
PUMAS, we use the values compiled by the \ac{pdg} for different materials,
available from their website. When no data are available for a specific
material, then Bragg additivity rule is used by summing up the contributions to
the electronic stopping power of its atomic elements, as in
equation~\eqref{eq:mean-excitation-energy}.

The factor $\delta_{e\gamma}$ in equation~\eqref{eq:electronic-energy-loss}
arises from the radiative correction discussed previously in
section~\ref{sec:inelastic-radiative}. An analytical approximation for
$\delta_{e\gamma}$ is derived in \ref{sec:electronic-radiative-correction}. One
obtains
\begin{equation} \label{eq:electronic-radiative-energy}
    \delta_{e\gamma}(\nu) \simeq \frac{\alpha}{2\pi}
        \ln^2\left(1 + \frac{2\nu}{m_e}\right)
        \left[\ln\left(2\gamma\right) -
        \frac{1}{3} \ln\left(1 + \frac{2\nu}{m_e}\right) \right] .
\end{equation}
The radiative correction to the electronic stopping power is significant only at
high energies. At EeV it increases the electronic stopping power by 20\,\%.
However, as it becomes significant so do also other radiative processes.
Overall, the contribution of $\delta_{e\gamma}$ to the total stopping power is
negligible when considering all radiative processes. Nevertheless, for
consistency with other computations we include this term in the electronic
stopping power.

The total electronic stopping power $S(T)$ is obtained by setting the upper
bound $\nu_{\scriptscriptstyle{C}} = \nu_{max}$ in
equation~\eqref{eq:electronic-energy-loss}. Note that $\nu_{max} \geq 2 m_e \gg
I_{max}$ in practice. Apart from the density effect, the resulting expression is
identical to \citet{Groom2001}. We refer to the latter for a more in depth
discussion of other potential corrections to the electronic stopping power, not
included in PUMAS, since those are not relevant for the transport of muons or
taus.

\subsubsection{Density effect \label{sec:density-effect}}

Computing the density effect correction $\delta_{\scriptscriptstyle{F}}$
requires specifying the electronic structure of the target medium (actually its
dielectric response). However, as was pointed out by \citet{Fano1963}, ``crude
information on the $Z_k$ and $I_k$ suffices for an estimate of
$E_{\scriptscriptstyle{F}}$''. Therefore, let us simply assume that the
electronic shells are the union of the atomic shells of the constituent atoms
of the target, considered as isolated. I.e. the $E_k$ are taken as the
ionisation energies of free atoms, and the $Z_k$ are the corresponding
occupancies weighted by the relative densities of constituent atoms. Let us
further assume that the oscillator's resonance energies are
\begin{equation} \label{eq:resonance-energies}
    I_k = a_S E_K ,
\end{equation}
where $a_S$ is computed from the material mean excitation energy, as
\begin{equation} \label{eq:sternheimer-scaling}
    \ln a_S = \ln I - \frac{1}{Z} \sum_{k=1}^n{Z_k \ln E_k} .
\end{equation}

The latter model is similar to \citet{Sternheimer1952}, but with a simpler
relationship between $I_k$ and $E_K$. Instead of
equation~\eqref{eq:resonance-energies}, \citea{Sternheimer1952} derives the
following relationship
\begin{equation} \label{eq:sternheimer-energy}
    I_k^2 = \begin{cases}
        \frac{Z_k}{Z} E_p^2 & \text{for conduction electrons} \\
        a_S^2 E_k^2 + \frac{2}{3} \frac{Z_k}{Z} E_p^2 & \text{otherwise}
    \end{cases} .
\end{equation}
Note that Sternheimer's result implies assumptions, one of which is $|E_{i} -
E_{j}| \gg E_p$ for all pairs of shells $i, j$. This is not valid in condensed
media, e.g water, where hydrogen and oxygen have similar binding energies w.r.t.
the plasma energy.

\begin{figure}[th]
    \centering
    \subfloat{\includegraphics[width=0.5\textwidth]{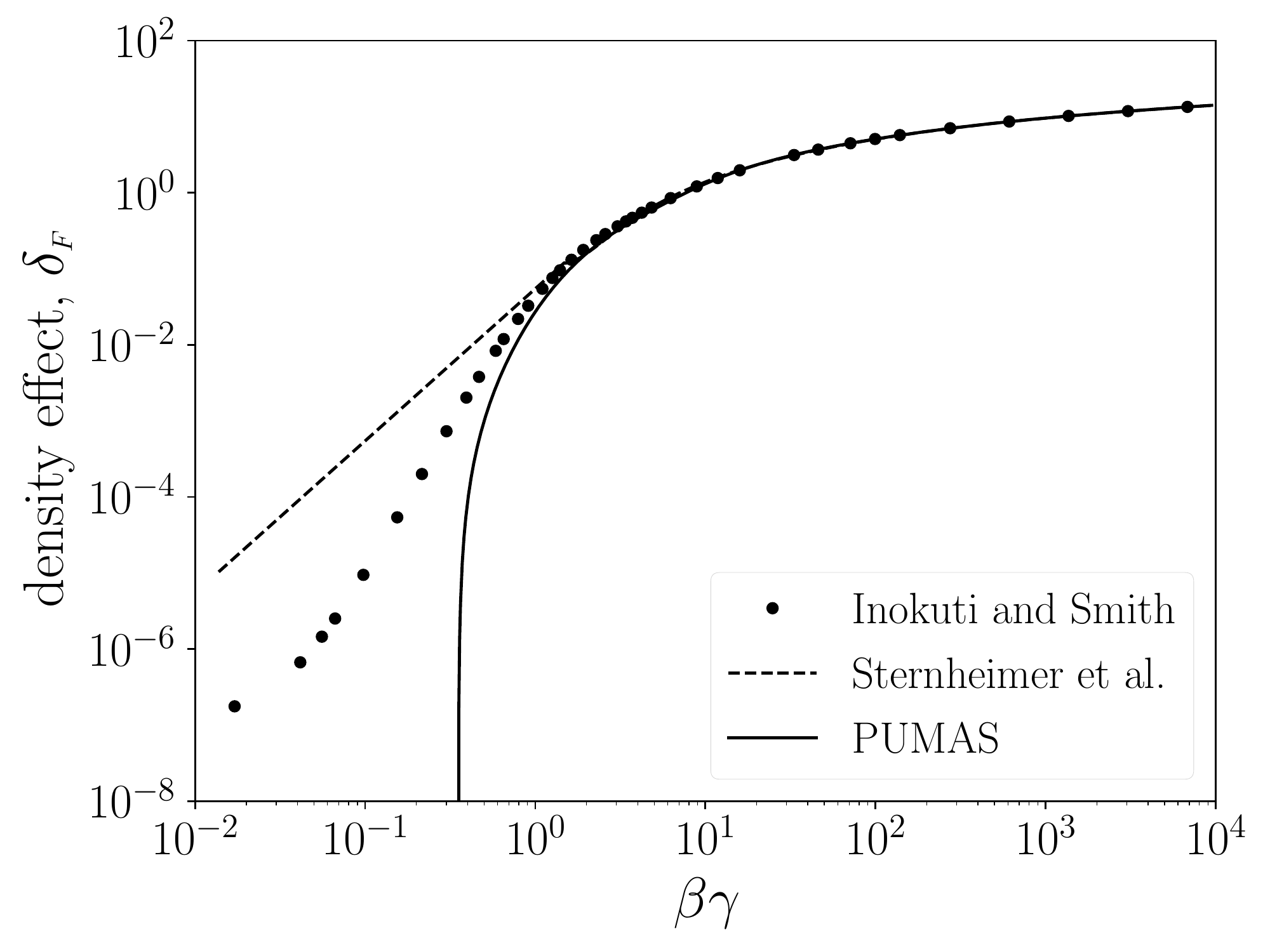}}
    \subfloat{\includegraphics[width=0.5\textwidth]{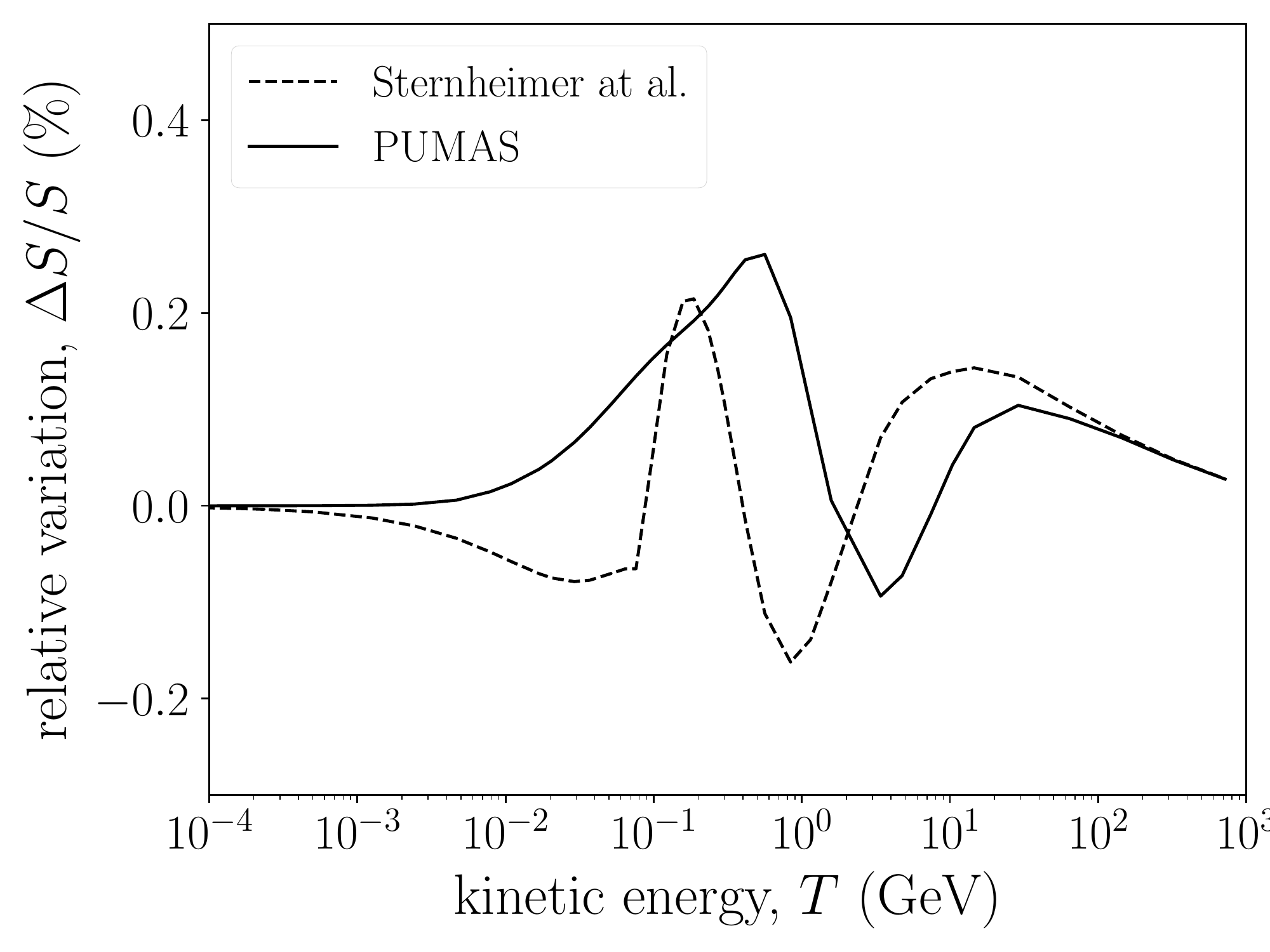}}
    \caption{Density effect in aluminium and variation of the electronic
    stopping power for a muon using various computations. Left: density effect
    parameter, $\delta_{\scriptscriptstyle{F}}$, according to
    \citet{Inokuti1982}, \citet{Sternheimer1984} and this work (PUMAS). Right:
    variation of the electronic stopping power w.r.t. \citet{Inokuti1982} by
    using this work (PUMAS) or \citet{Sternheimer1984} computation of the
    density effect.
    \label{fig:density-effect}}
\end{figure}

A detailed computation of the density effect in aluminium was done by
\citet{Inokuti1982}. This computation proceeds directly from the aluminium
dielectric response, which was accurately modelled from experimental data. The
detailed computation of \citea{Inokuti1982} allows a cross-check of the accuracy
of other models.  Figure\,\ref{fig:density-effect} shows a comparison of the
density effect computed by \citea{Inokuti1982} to \citet{Sternheimer1984} or
using equation~\eqref{eq:resonance-energies}. In the high energy limit where the
density effect matters most, $\delta_{\scriptscriptstyle{F}}$ is insensitive to
the details of the electronic structure. Then, electrons can be considered as
free and all models give the same result. In this case, the density effect
depends only on the total electron density of the target, through the plasma
energy $E_p$. At low energies the three results disagree. However, the
differences on the electronic stopping power are small, as can be seen on the
right of Figure\,\ref{fig:density-effect}.  Overall, for a muon we find a
maximal deviation on the electronic stopping power of $0.26\,\%$ ($0.21\,\%$) by
using \eqref{eq:resonance-energies} (\citea{Sternheimer1984}) instead of
\citea{Inokuti1982} density effect. The standard deviation of the differences is
of $0.09\,\%$ ($0.10\,\%$).

Given the previous results, it would be tempting to consider an even simpler
electronic structure with a single shell of resonance energy $I_1 = I$. This
would yield the right behaviour for the stopping power at high energies.
However, at intermediate energies ($\sim$GeV) this model result in significantly
larger errors than \citea{Sternheimer1984} and our model. In the case of
aluminium, we find a $3.5\,\%$ maximal deviation on the electronic energy loss
when using this single shell model instead of \citea{Inokuti1982}.

The parametrized results of \citet{Sternheimer1984} are commonly used in
Monte~Carlo codes in order to compute the density effect. In PUMAS, the density
effect is estimated directly from the atomic binding energies, as described
above.  In the case of standard rock, it results in a slightly larger stopping
power than e.g. \citet{Groom2001}. The discrepancy reaches a maximum value of
$0.4\,\%$ at GeV energy. Considering the lack of data in order to discriminate
the various models, this can be considered as the uncertainty on the electronic
stopping power related to the density effect.

\subsubsection{Angular deflections}
In the general case, describing both the energy loss and the projectile
deflection requires considering the \ac{ddcs} for electronic collisions w.r.t.
to $\nu$ and to the electron recoil energy $Q$.  Detailed expressions are
provided by~\citet{Salvat2013}.

Close collisions are modelled as if the scattering occurs with a free electron
at rest. In this case the recoil energy is $Q = \nu$. The projectile
scattering angle $\theta$ is given by the kinematics, as
\begin{equation} \label{eq:close-angular}
    \cos\theta = \frac{p^2 - \nu \left( E + m_e \right)}{p
        \sqrt{p^2 + \nu^2 - 2 E \nu}}.
\end{equation}
Thus, in order to simulate close electronic collisions it is enough to
randomise only the energy loss $\nu$. The corresponding angular deflection is
given by equation\,\eqref{eq:close-angular}.

In order to compute the transport mean free path $\lambda_1$ a numeric
integration of equation~\eqref{eq:close-angular} would be needed. This
integration is delicate due to rounding errors at high energies, since both $p$
and $\nu$ converge towards $E$. Let us instead follow \citet{Salvat2013} by
assuming $\nu \leq \min(\nu_\text{max}, \nu_{\scriptscriptstyle{C}}) \ll E$.
With this assumption the angular parameter $\mu$ is
\begin{equation}
    \mu = \frac{m_e}{2 p^2} \nu + \mathcal{O}\left(\nu^2\right).
\end{equation}
Thus, for $\nu_{\scriptscriptstyle{C}} \ll E$ or $\nu_\text{max} \ll E$ (i.e. $E
\ll m^2 / 2 m_e$), the transport mean free path restricted to soft close
collisions is
\begin{equation} \label{eq:close-transport}
    \lambda_{1,\scriptscriptstyle{C}} = \frac{2 p^2}{m_e
        S_{\scriptscriptstyle{C}}} .
\end{equation}

In the case of distant collisions, the energy loss $\nu$ and the recoil $Q$ are
limited to small values of $\mathcal{O}(E_k)$. Therefore, one can safely assume
$Q \ll m_e$ as \citea{Salvat2013}. In addition, let us also assume $p \gg I_k$
and $\nu_{\scriptscriptstyle{C}} \geq I_{max}$. Thus, we obtain the following
approximation for the angular \ac{dcs} in distant collisions:
\begin{equation} \label{eq:distant-angular-dcs}
    \frac{d\sigma_{\scriptscriptstyle{D}, k}}{d \mu} = \begin{cases}
        \frac{2 \pi r_e^2 m_e}{\beta^2 I_k} \frac{1}{\mu_k + \mu} &
            \text{if } \mu_k + \mu \leq \frac{m_e E_k}{2 p^2} \\
            0 & \text{otherwise}
    \end{cases},
\end{equation}
where the screening parameter is
\begin{equation}
    \mu_k = \frac{I_k^2}{4 \beta^2 p^2}.
\end{equation}

Summing up close and distant collisions, the soft transport mean free path in
electronic collisions is approximated by
\begin{equation} \label{eq:electronic-transport}
    \frac{1}{\lambda_{1,s}} =
        \frac{2 \pi r_e^2 m_e^2 Z}{\beta^2 p^2} \frac{\mathcal{N}_A}{M} \left[
        \ln\left(\frac{a_S \nu_{\scriptscriptstyle{C}}}{I}\right) -
        \beta^2 \frac{\nu_{\scriptscriptstyle{C}}}{\nu_{max}}
        + \frac{\nu_{\scriptscriptstyle{C}^2}}{4 E^2} + \frac{1}{a_S}
        + \delta_{e\gamma}(\nu_{\scriptscriptstyle{C}})\right]
        ,
\end{equation}
where the contribution of distant collisions has been reduced to its leading
term, $1 / a_S$, with $a_S$ the scaling parameter given by
equation~\eqref{eq:sternheimer-scaling}.

\begin{figure}[th]
    \centering
    \subfloat{\includegraphics[width=0.5\textwidth]{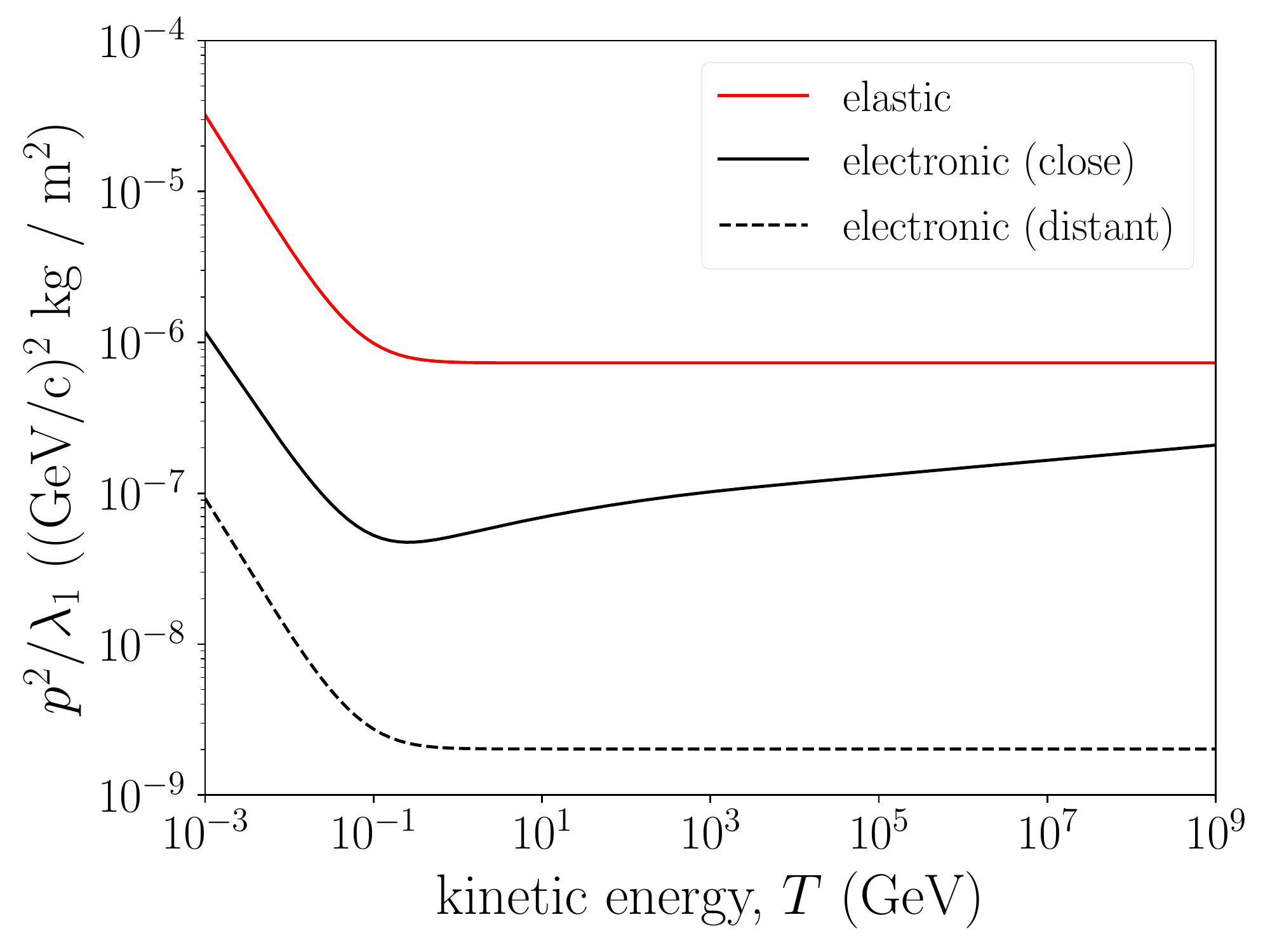}}
    \subfloat{\includegraphics[width=0.5\textwidth]{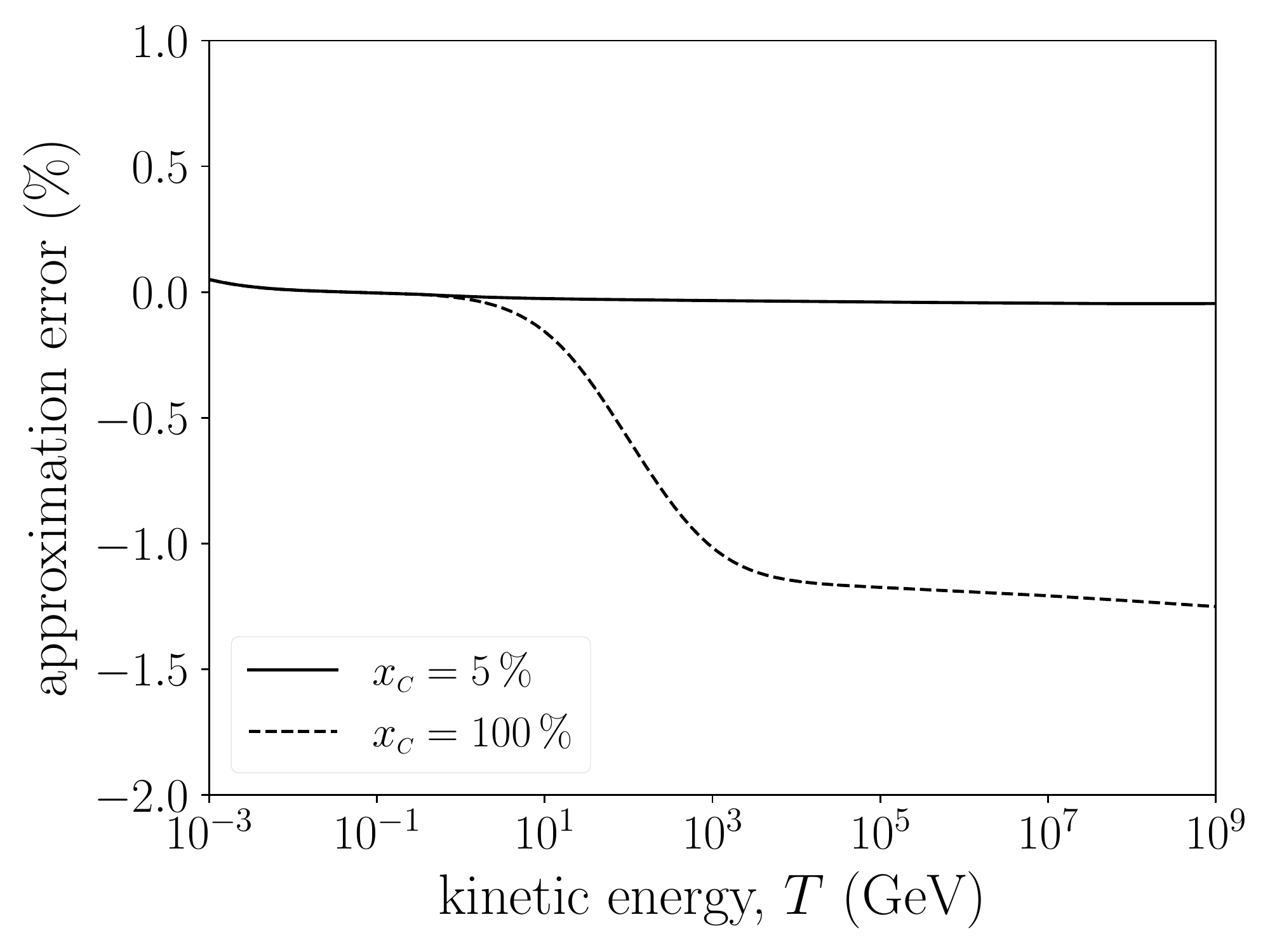}}
    \caption{Transport mean free path for electronic collisions in standard
    rock. A muon projectile is considered. Left: normalised transport, $p^2 /
    \lambda_1$, for close and distant electronic collisions
    ($x_{\scriptscriptstyle{C}} = 100\,\%$). For purpose of comparison, the
    corresponding value for elastic collisions is also indicated. Right:
    relative error on the total transport mean free path, summing up elastic and
    electronic contributions, when using the approximation of
    equation~\eqref{eq:electronic-transport}. The default cutoff value used in
    PUMAS is $x_{\scriptscriptstyle{C}} = 5\,\%$.
    \label{fig:electronic-transport}}
\end{figure}

Numerical investigations show that equation~\eqref{eq:electronic-transport}
holds well even for large cutoff values, as can be seen on
figure~\ref{fig:electronic-transport}. In standard rock, with PUMAS default
relative cutoff value of $x_{\scriptscriptstyle{C}} = 5\,\%$ the approximation
error is less than $0.1\,\%$ of the total transport mean free path, summing up
elastic and electronic contributions. At high energy, $\nu_\text{max} \to E$
(see equation~\eqref{eq:electronic-close-cutoff}). Thus, for
$\nu_{\scriptscriptstyle{C}} = T$ the assumption $\nu \leq \min(\nu_\text{max},
\nu_{\scriptscriptstyle{C}}) \ll E$ breaks.
Nevertheless, it can bee seen that even in this extreme case the approximation
error is small, $1.3\,\%$ in standard rock, above $1\,$TeV. These errors are
considered as acceptable for the present purpose.
Equation~\eqref{eq:electronic-transport} is thus used in PUMAS over all values
of $\nu_{\scriptscriptstyle{C}}$.

\subsection{Radiative collisions \label{sec:radiatives}}

When the projectile energy is high enough, the collision with target atoms might
lead to the creation of secondary particles or to the fragmentation of the
target.  Those collisions are conventionally designated as ``radiative''. At low
energy, radiative collisions contribute only marginally to the total stopping
power. However, as the projectile energy increases, radiative processes becomes
important. Their contribution to the stopping power increases approximately
linearly with the projectile energy, contrary to electronic collisions. Thus,
radiative processes are the dominant source of energy loss at high energies. The
energy $E_c$ at which half of the stopping power is due to radiative processes
is called the ``critical energy''.  The critical energy of muons in standard
rock is $E_c = 693\ $GeV\,\citep{Groom2001}.

Contrary to electronic collisions, high energy radiative collisions are likely
to result in ``catastrophic events'', where the muon looses a significant
fraction of its energy. The muon range thus fluctuates significantly above the
critical energy. On the contrary, muons have an almost deterministic
energy loss below $E_c$, with a well defined range given by \ac{csda}.

\begin{figure}[th]
    \center
    \label{fig:beta_standardrock}
    \includegraphics[width=\textwidth]{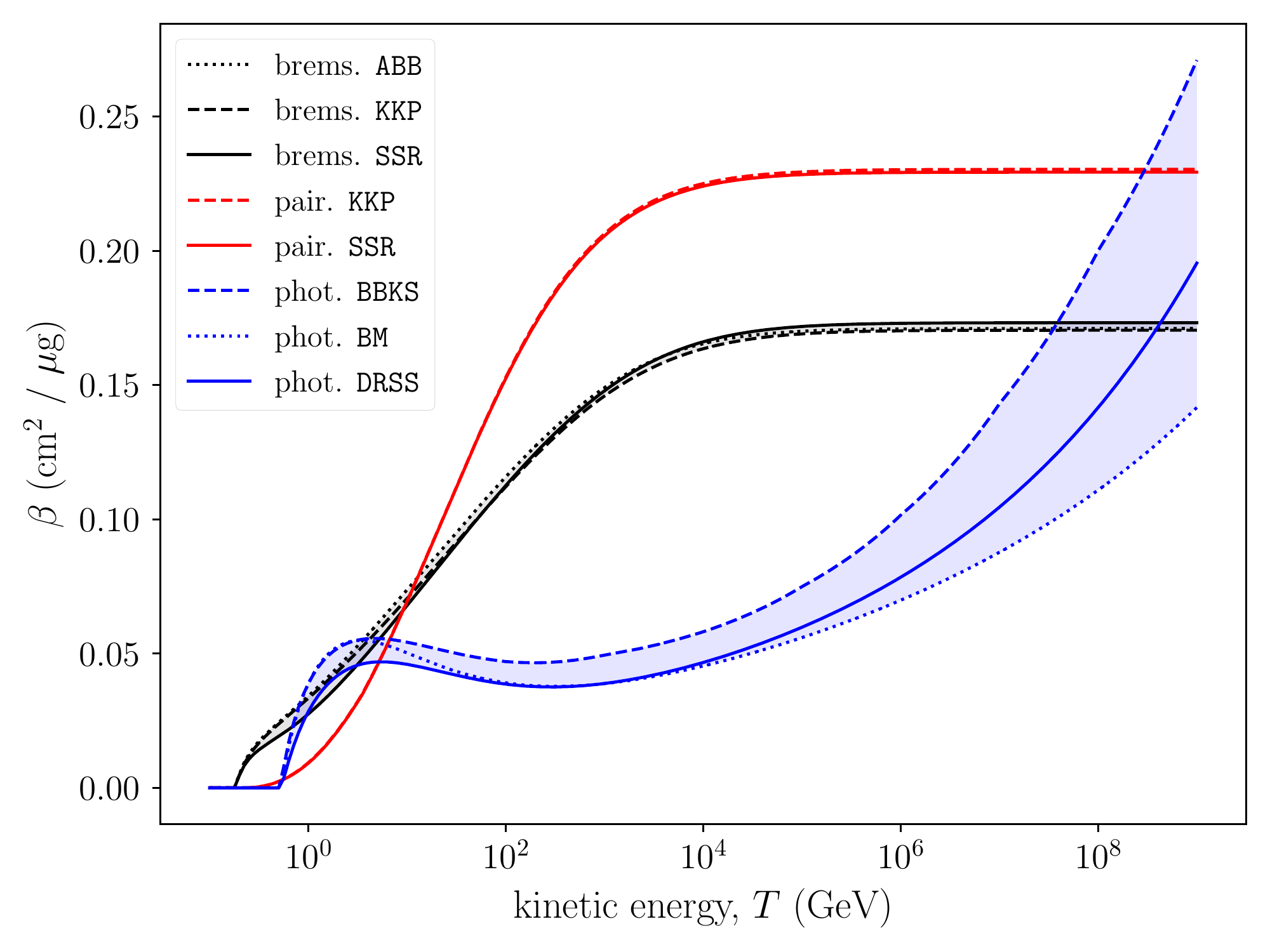}
    \caption{Normalised stopping power, $\beta = S / E$, for bremsstrahlung
    (black), $e^+e^-$~production (red) and photonuclear interactions (blue).  A
    muon in standard rock is considered. The different lines correspond to the
    models models indicated in table~\ref{tab:radiative-processes}. The default
    models used in PUMAS~v$1.2$ (\texttt{SSR} and \texttt{DRSS}) are indicated
    with a solid line. The shaded area represent the range of variation between
    models for a given process.
    \label{fig:radiative-beta}}
\end{figure}

Three radiative processes are of importance for the transport of muons and taus:
bremsstrahlung, direct $e^+e^-$ production and photonuclear interactions. These
processes have been extensively studied in the past.  Different \ac{dcs}
parametrizations are implemented in PUMAS.  The initial implementation was done
following \citet{Groom2001} and Geant4~\cite{Geant4PRF}.  More recently updated
models have been developed, especially for the PROPOSAL
Monte~Carlo~\citep{Koehne2013,Dunsch2019}. These updated models have been added
in PUMAS~v$1.1$. A summary of the \ac{dcs} parametrizations available in PUMAS
is given in table~\ref{tab:radiative-processes}. A comparison of their stopping
powers is shown on figure~\ref{fig:radiative-beta}.  A specific
parametrization can be selected by the user during the physics initialisation.
Otherwise, a default model is used as indicated in the table. 

Before discussing the details of the \ac{dcs} parametrizations, let us point out
that the Ter-Mikaelian~\citep{Ter-Mikaelian1972} and \ac{LPM} effects are not
included in PUMAS. Those lead to a suppression of radiative \acp{dcs} due to
interferences with other processes: elastic collisions of the projectile or
Compton scattering of the secondary photon. However, these effects are
important only at very high energies, above $\sim$$10^{11}\ $GeV for muons in
rocks according to \citet{Koehne2013}.  This is thus not considered relevant for
muography applications.

A priori, at high energy one must also consider the direct $\mu^+\mu^-$
production, not only the $e^+e^-$ one. The contribution to the stopping power of
the former is negligible according to \citet{Koehne2013}. However, $\mu^+\mu^-$
production increases the total transmitted flux since secondary muons are
produced. This was studied by \citet{Kelner2000}. Their results suggest that
secondary muons contribute at most at $\sim$$0.1\,\%$ to the transmitted flux of
muons at large depths.  Thus, $\mu^+\mu^-$ production is not considered in
PUMAS.

\begin{table}
    \caption{\ac{dcs} parametrizations implemented in PUMAS for radiative
    processes. The default parametrizations used in PUMAS~v$1.2$ are indicated
    with a star~($^*$) symbol.
    \label{tab:radiative-processes}}
\center
\resizebox{\textwidth}{!}{%
\begin{tabular}{llll}
    \toprule
    Process & Label & Parametrization name & References \\
    \midrule
    \multirow{3}{*}{bremsstrahlung}
    & \texttt{ABB} &  Andreev, Bezrukov and Bugaev &
        \citep{Andreev1994,Sokalski2001} \\
    & \texttt{KKP} & Kelner, Kokoulin and Petruhkhin &
        \citep{Kelner1995,Kelner1999,Groom2001,Koehne2013} \\
    & \texttt{SSR}$^*$ & Sandrock, Soedingrekso and Rhode &
        \citep{Sandrock2018,Dunsch2019,Soedingrekso2019} \\
    \midrule
    \multirow{2}{*}{$e^+e^-$ production}
    & \texttt{KKP} & Kelner, Kokoulin and Petruhkhin &
        \citep{Kokoulin1971,Kelner1998,Geant4PRF,Kelner1999} \\
    & \texttt{SSR}$^*$ & Sandrock, Soedingrekso and Rhode &
        \citep{Koehne2013,Soedingrekso2019,Sandrock2020} \\
    \midrule
    \multirow{3}{*}{photonuclear}
    & \texttt{BBKS} &  Bezrukov, Bugaev, Kokoulin and Shlepin &
        \citep{Bezrukov1981,Kokoulin1999,Bugaev2003,Bugaev2004,Koehne2013} \\
    & \texttt{BM} & Butkevich and Mikheyev &
        \citep{Butkevich2002,Koehne2013} \\
    & \texttt{DRSS}$^*$ & Dutta, Reno, Sarcevic and Seckel &
        \citep{Abramowicz1997,IyerDutta2001,Koehne2013} \\
    \bottomrule
\end{tabular}}
\end{table}

\subsubsection{Bremsstrahlung \label{sec:bremsstrahlung}}

The bremsstrahlung process models the interaction of a charged projectile
in the electromagnetic field of a nucleus resulting in the creation of a photon,
as $\ell + Z \rightarrow \ell + Z + \gamma$. Three parametrisations for the
\ac{dcs} of this process are available in PUMAS.

The Kelner, Kokoulin and Petrukhin
parametrisation\,\citep{Kelner1995}~(\texttt{KKP}) of the bremsstrahlung
\ac{dcs} was initially implemented in PUMAS according to~\citet{Groom2001}. In
version~1.1 of PUMAS this cross-section has been updated following
\citet{Koehne2013}.  The nuclear excitation term is taken into account (see e.g.
\citet{Kelner1995}), as well as a more accurate radiation logarithm computation
from \citet{Kelner1999}.

An improved parametrisation was proposed by Sandrock, Soedingrekso and
Rhode\,\citep{Soedingrekso2019} (\texttt{SSR}). This parametrisation refines the
modelling of atomic screening functions and takes into account radiative
corrections to the bremsstrahlung cross-section from \citet{Sandrock2018}. As a
result, the energy loss due to bremsstrahlung increases by $2\,\%$ at high
energies.  The PROPOSAL implementation of the \texttt{SSR} bremsstrahlung
\ac{dcs} is used in PUMAS (see e.g.~\citet{Dunsch2019}). The \texttt{SSR}
parametrization is the default bremsstrahlung model for PUMAS~v$1.2$.

The Andreev, Bezrukov and Bugaev
parametrisation\,\citep{Andreev1994}~(\texttt{ABB}) of the bremsstrahlung
\ac{dcs} has been added to PUMAS as well. MUM's implementation is used (see e.g.
\citet{Sokalski2001}).

In order to simulate angular deflections in bremsstrahlung processes the
\ac{ddcs} would be required. However, the parametrizations discussed previously
only provide the \ac{dcs}. As an alternative, we rely on the \ac{ddcs} from
\citet{Tsai1974}, but scaled to the \ac{dcs} selected in PUMAS. Thus, the
\ac{ddcs} used in PUMAS is
\begin{equation} \label{eq:angular-ddcs-scaling}
    \frac{d^2\sigma}{d\nu d\mu} =
        \frac{1}{\frac{d\sigma_{\scriptscriptstyle{T}}}{d\nu}}
        \left(\frac{d^2\sigma_{\scriptscriptstyle{T}}}{d\nu d\mu} \right)
        \frac{d\sigma}{d\nu},
\end{equation}
where $\sigma_{\scriptscriptstyle{T}}$ is given by \citet{Tsai1974} and where
$\sigma$ is the bremsstrahlung \ac{dcs} actually selected.

\citea{Tsai1974} obtains the \ac{ddcs} w.r.t. the bremsstrahlung photon
emission angle $\theta_k$. In order to express the \ac{ddcs} as function of the
muon scattering angle $\theta$, let us neglect the target recoil in the
collision.  Then, from momentum conservation one has
\begin{equation}
    p' \sin(\theta) = \nu \sin(\theta_k),
\end{equation}
where $p'$ is the muon momentum after the collision and $\nu$ the energy
transferred to the photon. Let us further consider the ultra-relativistic limit,
where $p' \simeq E - m - \nu$ and where both $\theta$ and $\theta_k$ are small.
Neglecting Coulomb corrections, one obtains
\begin{equation} \label{eq:tsai-ddcs}
    \frac{d^2\sigma_{\scriptscriptstyle{T}}}{d\nu d\mu} =
        \frac{2 \alpha r_e^2}{\nu} \left[\left(2 - 2 y + y^2\right)
        \frac{\mu_0}{(\mu_0 + \mu)^2}
        -4 (1 - y) \frac{\mu_0^2 \mu}{(\mu_0 + \mu)^4} \right] X(\mu),
\end{equation}
where $y = \nu / E$ and where the screening parameter is
\begin{equation}
    \mu_0 = \frac{m^2 \nu^2}{E^2 \left(E - \nu\right)^2} .
\end{equation}

The form factor $X$ for an exponential distribution of the nucleus charge was
computed by \citet{Tsai1974} (see also the erratum\,\citep{Tsai1977}).  Let us
recall its expression:
\begin{equation}
    X(\mu) = \begin{cases}
        Z^2 \left[(1 + 2 q r) \ln\left(\frac{1 + q r}{r + q r}\right)
            -\left(1 - r\right)\frac{1 + 2 q}{1 + q}\right] & \text{if } \mu
            \leq \sqrt{\mu_0} - \mu_0 \\
        0 & \text{otherwise}
    \end{cases},
\end{equation}
where $q = (1 + \mu / \mu_0)^2 / \mu_N$ and $r = \mu_0 (1 + \mu / \mu_0)^2$.
The nuclear cutoff parameter is
\begin{equation}
    \mu_N = \frac{6 \hbar^2}{m^2 \left<r^2\right>}
\end{equation}
where $\left<r^2\right>$ is the nuclear charge radius squared given in
\ref{sec:nuclear-radius}.

\subsubsection{$e^+e^-$ production}

The electron pair production process is similar to bremsstrahlung but with the
direct creation of an $e^+e^-$ pair, as $\ell + Z \rightarrow \ell + Z + e^+ +
e^-$. Two parametrisations for the \ac{dcs} of this process are available in
PUMAS.

The Kelner, Kokoulin and Petruhkin
parametrisation\,\citep{Kokoulin1971,Kelner1998}~(\texttt{KKP}) was the first
implemented in PUMAS. The \texttt{KKP} parametrisation provides the \ac{ddcs}
w.r.t. to the total energy loss $\nu$ and the $e^+e^-$ relative energy
difference $\rho$. The Geant4 implementation was initially used in PUMAS (see
e.g.~\citep{Geant4PRF}).  The $e^+e^-$ production \ac{ddcs} is numerically
integrated over $\rho$ using a Legendre-Gauss quadrature in $\ln(1 - \rho)$.
This cross-section has been updated in version~1.1 of PUMAS. The improved
radiation logarithm computation from \citet{Kelner1999} is used. In addition,
the order of the Legendre-Gauss numeric integration was increased from $8$ to
$12$. With these two modifications, the PUMAS pair production \ac{dcs} agrees
with \citet{Koehne2013} at better than $0.1\,\%$. Note that the resulting
\ac{dcs} is now $2\,\%$ lower than the Geant4 (10.7) one in standard rock, due
to these two modifications.

A refined parametrisation of the $e^+e^-$ pair production cross-section has been
proposed by Sandrock, Soedingrekso and
Rhode\,\citep{Soedingrekso2019}~(\texttt{SSR}), as for the bremsstrahlung
cross-section.  Improvements in the modelling of atomic screening functions
yield a $0.5\,\%$ decrease on the energy loss w.r.t. the \texttt{KKP}
parametrisation.  Higher order radiative corrections are not currently modelled.
They are estimated to be of the order of $1\,\%$ (see e.g.  \citet{Sandrock2020}
for a more detailed discussion). The PROPOSAL implementation of the \texttt{SSR}
doubly differential cross-section is used in PUMAS. It is integrated over $\rho$
using a Legendre-Gauss quadrature.

Note that the collision kinematics is not fully determined given only
$\nu$ and $\rho$, even when the target recoil is neglected. Properly simulating
scattering angles in $e^+e^-$ pair production would require a triply
differential cross-section. However, since in PUMAS the distribution of
secondary particles is not a concern, the following approximation is used. The
$e^+e^-$ pair is considered as a virtual photon of energy $\nu$. Then, the pair
production \ac{ddcs} w.r.t. the muon angular parameter $\mu$ is approximated
using equation\,\eqref{eq:angular-ddcs-scaling}, as for the bremsstrahlung
process, but using the pair production \ac{dcs} for $d\sigma/d\nu$.

\subsubsection{Photonuclear interaction}

The photonuclear process models an inelastic interaction of a lepton with a
nuclei. Three parametrisations for the  \ac{dcs} of this process are available
in PUMAS.

The Dutta, Reno, Sarcevic and Seckel\,\citep{IyerDutta2001}~(\texttt{DRSS})
parametrisation has been initially implemented in PUMAS. It relies on the
ALLM97\,\citep{Abramowicz1997} parametrisation of the $F_2$ proton structure
function.  In the initial \texttt{DRSS} paper, $Z = A / 2$ is assumed when
modelling the atomic screening function. In PUMAS v1.1, this has been refined,
e.g. as in section 2.3.2 of \citet{Koehne2013}. The DRSS parametrisation
requires integrating the doubly differential photonuclear cross-section over
the square of the four momentum transfer $Q^2$, given by
\begin{equation} \label{eq:theta-photonuclear}
    Q^2 = 2 \left(E E' - p p' \cos\theta - m^2 \right),
\end{equation}
where $p$ ($p'$) is the initial (final) momentum of the muon and $\theta$ its
scattering angle. The integration of the \ac{ddcs} is done numerically using a
Gaussian quadrature in $\ln Q^2$.

An alternative parametrisation of the $F_2$ neutron, proton and atomic structure
functions was proposed by Butkevich and Mikhailov\,\citep{Butkevich2002},
resulting in the \texttt{BM} \ac{ddcs} for photonuclear interactions.  The
PROPOSAL implementation for the \texttt{BM} photonuclear \ac{ddcs} is used in
PUMAS.  As for the \texttt{DRSS} parametrisation, the \texttt{BM} \ac{dcs} is
integrated numerically over $Q^2$ using a Gaussian quadrature in $\ln Q^2$.

The initial parametrisation of Bezrukov and Bugaev\,\citep{Bezrukov1981}~(BB)
has been widely used in the past, e.g. by \citet{Groom2001}. In this model, the
photonuclear \ac{dcs} is normalised to the photon-nucleon cross-section
$\sigma_{\gamma N}$ for the absorption of a real photon. A refined
parametrisation for the latter was proposed by \citet{Kokoulin1999}.  The BB
cross-section of 1981 misses a hard QCD component which makes it inaccurate at
high energies, e.g. as compared to the \texttt{DRSS} or \texttt{BM}
cross-sections (see \citet{Sokalski2002} for a more detailed discussion). An
improved model taking into account the hard QCD component was given more
recently by Bugaev and Shlepin~\citep{Bugaev2003} (parametrised in
\citet{Bugaev2004}). This leads to the Bezrukov, Bugaev, Kokoulin and Shlepin
(\texttt{BBKS}) cross-section for photonuclear interactions. The PROPOSAL
implementation of this parametrisation is used in PUMAS.

The \ac{ddcs} w.r.t. the angular parameter $\mu$ is related to the one in
$Q^2$ by
\begin{equation}
    \frac{d^2\sigma}{d\nu d\mu} = 4 p p' \frac{d^2\sigma}{d\nu dQ^2} .
\end{equation}
For the \texttt{DRSS} and \texttt{BM} parametrizations the scattering angle of
the muon is thus derived from the \ac{ddcs} in $Q^2$. In the case of the
\texttt{BBKS} parametrization, the \texttt{DRSS} \ac{ddcs} is used, but rescaled
to the \texttt{BBKS} \ac{dcs}, as detailed previously for bremsstrahlung.

\section{Transport algorithms \label{sec:condensed}}

From the expressions given in section\,\ref{sec:physics}, it can be observed
that \acp{dcs} are diverging functions for $\nu\to0$, down to some lower bound
cutoff.  This implies that most collisions are soft, i.e. resulting in a small
individual energy loss and deflection. However, the sum of these soft events
constitute the bulk of the energy loss and scattering over macroscopic
distances. Simulating in detail every collision would be highly inefficient
CPU-wise. This is seldom done in practice.  Instead, one relies on condensed
simulation schemes by replacing a group of collisions with an approximate
condensed model, allowing to directly render the behaviour of multiple
collisions. This procedure was outlined in detail by \citet{Berger1963}.

PUMAS implements three algorithms for the simulation of the energy loss,
designated in the following as ``\ac{csda}'', ``mixed'' and ``straggled''. In
\ac{csda} mode, the energy loss of the projectile is deterministic given by its
average value. The mixed and straggled modes are class II algorithms according
to Berger's terminology. A cutoff value $\nu_{\scriptscriptstyle{C}}$ is
selected on the projectile energy loss in individual collisions. Events are
separated into soft collisions ($\nu \leq \nu_{\scriptscriptstyle{C}}$) and hard
ones ($\nu > \nu_{\scriptscriptstyle{C}}$) accordingly. Soft collisions are
rendered collectively while catastrophic ones are simulated explicitly. In mixed
mode the soft part is rendered by CSDA while in straggled mode soft electronic
collisions are fluctuated.

The projectile deflections are rendered by a mixed algorithm as well. A cutoff
is applied on the scattering angle in individual elastic collisions, following
\citet{Fernandez-Varea1993}. This procedure reproduces the exact multiple
scattering distribution and the corresponding spatial displacement when the
number of elastic collisions is $n \gtrsim 20$.

The energy loss and the scattering can also be disabled, independently.
Disabling all physics processes can be useful in order to cross-check the
geometry implementation (see e.g. section~\ref{sec:geometry}). Furthermore, the
simulation scheme can be changed on the fly during the tracking of a particle.
For example, if the energy of a backward transported particle exceeds
$0.1$-$1$~TeV, then scattering can be turned off as a CPU optimisation.

Let us also recall that a peculiarity of PUMAS is that it can operate in both
forward and backward Monte~Carlo mode. In particular, care was taken into
implementing the mixed and straggled simulation schemes in a symmetric way. In
the following, we provide specific details on the simulation algorithms
available in PUMAS, as well as on their implementation.

\subsection{CSDA mode \label{sec:csda-mode}}

\ac{csda} is frequently used in muography applications together with the
assumption that muons follow straight paths. \ac{csda} provides accurate
estimates of the transmitted flux of muons for moderate target thickness, $d
\lesssim 300\ $m of standard rock, as illustrated in
section~\ref{sec:validation-transmission}.  For thicker targets, \ac{csda}
underestimates the transmitted flux because of strong fluctuations in the energy
loss, due to radiative processes. However, when \ac{csda} is applicable it is
particularly efficient, since it provides direct semi-analytical solutions to
the transport problem.

\subsubsection{Forward CSDA transport}

Within \ac{csda}, the projectile energy loss per unit path length is
deterministic.  It is equal to the total stopping power $S$  given by
equation~\eqref{eq:csda-energy-loss}, setting $\nu_{\scriptscriptstyle{C}} = T$
and considering all processes discussed in section\,\ref{sec:physics}.  In a
uniform medium, $S$ does not depend on the projectile position but only on its
energy. Thus, within \ac{csda} the curvilinear distance $s$ along the projectile
path and its kinetic energy $T$ are equivalent representations of the particle
state. They are related by
\begin{align}
    \rho \left(s_1 - s_0 \right) &= \int_{T_1}^{T_0}{\frac{dT}{S}} \nonumber \\
        &= R(T_0) - R(T_1),
        \label{eq:csda-distance}
\end{align}
where $\rho$ is the medium density and 
\begin{equation} \label{eq:csda-range}
    R(T) = \int_{0}^{T}{\frac{dT'}{S(T')}},
\end{equation}
is the mass \ac{csda} range. The CSDA range is a strictly increasing function of
the energy of the projectile. Let $R^{\scriptscriptstyle{(\shortminus 1)}}$
denote its inverse, i.e. the minimum kinetic energy required in order to travel
over a path length $\rho \Delta s$. Let us consider a particle with initial
energy $T_0$ travelling over a distance $\Delta s = s_1 - s_0$. The particle
final energy, $T_1$, is obtained by inverting equation~\eqref{eq:csda-distance}:
\begin{equation} \label{eq:csda-transport}
    T_1 = \begin{cases}
        R^{\scriptscriptstyle{(\shortminus 1)}}\left(R(T_0) -
            \rho \Delta s\right) &
            \text{if } \rho \Delta s \leq R(T_0) \\
        0 & \text{otherwise}
    \end{cases}.
\end{equation}
In the case where $\rho \Delta s > R(T_0)$, the particle stops after a path
length $R(T_0) / \rho < \Delta s$ with a null kinetic energy.

Note that the previous equations are not valid in a non uniform medium due to
the density effect. While it is conventional to express the stopping power and
\ac{csda} range of muons per mass, this is actually miss-leading because these
quantities depend on the medium density, due to
$\delta_{\scriptscriptstyle{F}}$.  This is especially the case at energies where
\ac{csda} is a good approximation, i.e. when the stopping power is dominated by
electronic collisions. However, there are some exceptions to this that are
discussed further in section~\ref{sec:local-density}.

Equation~\eqref{eq:csda-transport} is used in PUMAS for transporting particles
in \ac{csda} mode. The \ac{csda} range is pre-computed by PUMAS and tabulated as
function of the projectile energy using a logarithmic sampling. Then, at runtime
a lookup algorithm is used, detailed in \ref{sec:lookup}. The converse \ac{csda}
energy $R^{\scriptscriptstyle{(\shortminus 1)}}$ is obtained from the same data
and lookup algorithm, but swapping columns.

\subsubsection{Backward CSDA transport}

In order to introduce the case of the backward transport, let us perform a toy
muography experiment. Let us consider a uniform medium as discussed previously
with density $\rho$. Let there be a muon source located at $s_0$ with
differential flux $\phi_0$, and let there be a counting detector located at
$s_1$. The expected rate of muons in the detector is
\begin{equation} \label{eq:muon-rate}
    \tau = \int_0^{\infty}{\int_0^{\infty}{A(T) p(T; T_0) dT}\,
        \phi_0(T_0) dT_0} ,
\end{equation}
where $p(T; T_0)$ is the \ac{pdf} for a muon to exit the target with kinetic
energy $T$ given its initial energy $T_0$. $A$ denotes the detector acceptance,
i.e.  in this simple model the probability to detect and select a muon of energy
$T$.

Within \ac{csda}, $p$ is a Dirac $\delta$-distribution:
\begin{equation} \label{eq:csda-transmission}
    p(T; T_0) = \delta(T - T_1(T_0)),
\end{equation}
where $T_1$ is given by equation~\eqref{eq:csda-transport}. Inserting
equation~\eqref{eq:csda-transmission} into \eqref{eq:muon-rate} and integrating
out the final energy yields the \ac{csda} rate
\begin{equation} \label{eq:csda-rate}
    \tau = \int_0^{\infty}{A(T_1) H(T_0 -
        R^{\scriptscriptstyle{(\shortminus 1)}}\left(\rho \Delta s\right))
        \phi_0(T_0) dT_0},
\end{equation}
where $H$ is the Heaviside step function. When it is further assumed that the
detector acceptance $A$ is also a step function, then
equation~\eqref{eq:csda-rate} leads to an approximation frequently used in
muography applications (see e.g.~\citet{Nagamine1995}). The accuracy of
\ac{csda} is discussed more in depth in
section~\ref{sec:validation-transmission}.

The backward formulation of the previous problem consists in a change of the
integration variable from $T_0$ to $T_1$. Inverting the \ac{csda} transport
equation~\eqref{eq:csda-transport} for $T_0$ yields
\begin{equation}
    T_0 = R^{\scriptscriptstyle{(\shortminus 1)}}\left(
        R(T_1) + \rho \Delta s\right) .
\end{equation}
The Jacobian of this change of variable is
\begin{equation} \label{eq:csda-jacobian}
    \left| \frac{dT_0}{dT_1} \right| = \frac{S(T_0)}{S(T_1)} .
\end{equation}
Thus, one obtains
\begin{equation} \label{eq:csda-backward-rate}
    \tau = \int_0^{\infty}{A(T_1) \phi_0(T_0) \frac{S(T_0)}{S(T_1)} dT_1} .
\end{equation}

In PUMAS, the Jacobian factor given by equation~\eqref{eq:csda-jacobian} is
applied to the particle weight when a backward \ac{csda} transport is done. Let
us point out that this is consistent with the backward Monte~Carlo method
described in \citet{Niess2018}, summarised hereafter.

\subsection{Mixed mode \label{sec:mixed-mode}}

In mixed mode, a class II Monte~Carlo algorithm is used following
\citet{Berger1963}.  The cutoff value between soft and hard collisions is set to
a fraction $x_{\scriptscriptstyle{C}}$ of the projectile kinetic energy.  By
default $x_{\scriptscriptstyle{C}} = 5\,\%$. This can be modified during the
physics initialisation. A $5\,\%$ relative cutoff might seem rather high.
However, it was shown by \citet{Sokalski2001} to give accurate results for the
transport of a continuous flux, e.g. for atmospheric muons. It is confirmed in
section~\ref{sec:validation-transmission} that this cutoff value is appropriate
for most muography applications, i.e. for targets thinner than $\sim$3\,km of
standard rock. However, note that in order to simulate the impulse response for
a mono-energetic muon beam, a smaller cutoff would be needed as discussed by
\citet{Koehne2013}.

\subsubsection{Forward mixed transport \label{sec:forward-mixed}}

The energy loss due to soft collisions is rendered with \ac{csda}. It is
computed using equation~\eqref{eq:csda-energy-loss}, as previously, but
restricted to soft collisions with $\nu \leq \nu_{\scriptscriptstyle{C}}$. All
processes are considered except elastic collisions. However, for electronic
losses an effective model is used, described below. This model reproduces the
exact soft stopping power and \ac{dcs} for $\nu_{\scriptscriptstyle{C}} \gtrsim
I$, where $I$ is the mean excitation energy.

Let us consider the following effective \ac{dcs} for electronic collisions with
the $i^\text{th}$ atom of the target material:
\begin{equation} \label{eq:effective-inelastic-dcs}
    \frac{d\sigma_{\scriptscriptstyle{C},i}}{d\nu} = \begin{cases}
        \frac{2 \pi r_e^2 m_e Z_i}{\beta^2} \left[\frac{1}{\nu^2} -
            \frac{\beta^2}{\nu_\text{max}} \frac{1}{\nu} +
            \frac{1}{2 E^2} \right]
        & \text{if } \alpha I_i \leq \nu \leq \nu_\text{max} \\
        0 & \text{otherwise}
    \end{cases},
\end{equation}
where $\alpha = 0.62$ corresponds to the ratio of the ionisation energy of
liquid hydrogen to its mean excitation energy (see e.g.~\citet{Zyla2020}).

Summing over atomic elements yields the correct electronic \ac{dcs} for $\nu
\geq I_{max}$, apart from the radiative term, as can be seen from
section~\ref{sec:electronic-process}.  I.e., for $\nu$ large w.r.t. the atomic
binding energies, the electronic \ac{dcs} does not depend on the details of
the electronic structure, but only on the total electron number $Z = \sum{Z_i}$.
However, the stopping power does depend on the electronic structure. Therefore,
let us use the exact stopping power given by
equation~\eqref{eq:csda-energy-loss} with $\nu_{\scriptscriptstyle{C}} =
\nu_{max}$, but let us subtract from the latter the contributions of hard
electronic collisions as given by our effective \acp{dcs}, i.e.
equation~\eqref{eq:effective-inelastic-dcs}. This gives the correct soft
stopping power for $\nu \geq I_{max}$, apart from the radiative term $\Delta_{e
\gamma}$.  For the latter, all collisions are treated as soft, i.e.
$\nu_{\scriptscriptstyle{C}} = \nu_{max}$.

We could have included the radiative term to the effective \ac{dcs}.  However,
this would have significantly complicated the simulation of hard electronic
collisions while this correction is overall negligible.  Therefore, we only take
it into account for the stopping power. Note also that the ``atomic'' mean
excitation energy is used in equation~\eqref{eq:effective-inelastic-dcs} instead
of the material one. As a result, the material can be considered as a pure
atomic mixture for \emph{all} hard collisions.

The use of an effective model simplifies the implementation of electronic
processes, but at the cost of a loss of accuracy at $\sim$MeV energies and large
$Z$. However, as discussed in \ref{sec:wallace}, the range of muons is
negligible in this case, $\lesssim 10\ \mu$m. Hence, this is not relevant for
muography applications considering targets larger than $\sim$1\,m. Note also
that contrary to \citet{Salvat2013}, with this effective model we do not
explicitly simulate distant hard collisions with atomic electrons.  This is not
a problem since we are not concerned by the spectrum of secondary knock-on
electrons, but only by the projectile energy loss.

The mixed Monte~Carlo algorithm requires specifying the interaction length
restricted to hard collisions:
\begin{equation} \label{eq:hard-interaction-length}
    \frac{1}{\Lambda_h} = \mathcal{N}_A \sum_{i,j}{\frac{f_i}{A_i}
        \int_{\nu_{\scriptscriptstyle{C}}}^{\infty}{\frac{d\sigma_{ij}}{d\nu}
        d\nu}},
\end{equation}
where $f_i$ and $A_i$ are the mass fraction and the atomic weight of the
$i^\text{th}$ constituent atom of the medium. The sum over $j$ runs over all
physics processes except elastic collisions. Note that for electronic collisions
the effective \ac{dcs} described previously is used.

Let us further consider a uniform medium such that the stopping power per unit
mass does not depend on the target density. Then, as discussed previously in the
\ac{csda} case, between two hard collisions the projectile kinetic energy $T$
and its path length $s$ are analogous variables. Let $T_0$ denote the projectile
kinetic energy at $s_0$.  The probability for a hard collision to occur with $T
\leq T_1$ can be expressed semi-analytically, e.g. as \citet{Berger1963}. It is
given by
\begin{equation} \label{eq:mixed-probability}
    P(T \leq T_1; T_0) = e^{N_h(T_1) - N_h(T_0)},
\end{equation}
where
\begin{equation} \label{eq:average-hard-collisions}
    N_h(T) =
        \int_{0}^{T}{\frac{dT'}{\Lambda_h S_s}},
\end{equation}
is the average number of hard collisions over the path from $T$ to $0$ and where
$S_s$ is the stopping power restricted to soft collisions as discussed
previously.

Equation~\eqref{eq:mixed-probability} provides an easy way to sample the
energy $T_1$ at which the next hard collision would occur using the inverse
\ac{cdf} method. Let $N_h^{\scriptscriptstyle{(\shortminus 1)}}$ denote the
inverse of $N_h$. Then, $T_1$ is given by
\begin{equation} \label{eq:mixed-transport}
    T_1 = \begin{cases}
        N_h^{\scriptscriptstyle{(\shortminus 1)}}
        \left(N_h(T_0) + \ln\xi \right) & \text{if } \ln\xi > -N_h(T_0) \\
        0 & \text{otherwise}
    \end{cases},
\end{equation}
where $\xi$ is a random number uniformly distributed in $[0,1]$.  Note that if
$T_1 = 0$, then the projectile actually lost all its energy before any hard
collision occurred. The path length $s_1$ at which the hard collision occurs is
given by equation~\eqref{eq:csda-distance}, but using the range $R_s$ restricted
to soft collisions.

In practice, the quantities $N_h$ and $R_s$ are tabulated at PUMAS
initialisation, and then read back during the simulation using the lookup
algorithm detailed in \ref{sec:lookup}. Between two hard collisions, the
projectile behaves as in \ac{csda} mode, but using a soft stopping power $S_s$
and range $R_s$ instead of $S$ and $R$.

\subsubsection{Backward mixed transport}

In backward mode, the energy of the previous hard collision is obtained by
inverting equation~\eqref{eq:mixed-transport} for $T_0$. Thus
\begin{equation} \label{eq:mixed-backward}
    T_0 = N_h^{\scriptscriptstyle{{\shortminus 1}}}
        \left(N_h(T_1) - \ln\xi \right) .
\end{equation}
It is not enough to invert the transport equation in order to preserve the flux
of particles in a backward process . As demonstrated in \citet{Niess2018} a
backward Monte~Carlo weight should also be applied as well. This backward weight
is given by the Jacobian of the change of variable from $T_0$ to $T_1$.  The
differentiation of the previous equation~\eqref{eq:mixed-backward} yields
\begin{equation} \label{eq:backward-transport-weight}
    \left| \frac{dT_0}{dT_1} \right| =
        \frac{\Lambda_h(T_0) S_s(T_0)}{\Lambda_h(T_1) S_s(T_1)} .
\end{equation}
This result is similar to the \ac{csda} Jacobian factor given by
equation~\eqref{eq:csda-jacobian} but with an extra vertex weight,
$\Lambda_h(T_0) / \Lambda_h(T_1)$.  It is thus convenient to use the following
scheme for backward Monte~Carlo weights:
\begin{itemize}
    {\item When the projectile is backward transported from $s_1$ to $s_0$, its
        Monte~Carlo weight is multiplied by $S_s(T_0) / S_s(T_1)$.}
    {\item At a backward collision vertex, the particle Monte~Carlo weight is
        multiplied by $\Lambda_h(T_0) / \Lambda_h(T'_1)$, where $T_0$ ($T'_1$)
        is the energy of the particle before (after) the backward collision,
        i.e. $T'_1 > T_0$.}
\end{itemize}
This scheme is still valid in the case that the backward transport stops before
a hard collisions occurs, because the particle reached a medium boundary or an
external constraint, as demonstrated in \citet{Niess2018} (corollary 3).  Note
also that in the case where there is no continuous energy loss, the previous
vertex weighting is still correct. One just needs to set the backward transport
weight to 1.

The weighting scheme described previously only accounts for the projectile
backward transport between two hard collisions vertices. One needs to apply an
additional collision backward weight at vertices, as discussed hereafter in
section~\ref{sec:hard-collisions}.

\subsection{Straggled mode}

In straggled mode, a class II Monte~Carlo simulation is performed following
PENELOPE~\citep{Salvat2015}, with some modifications for convenient usage in
backward mode. Physics processes are split as in mixed mode, but in addition the
soft electronic energy loss is fluctuated around its mean value, $S_s$. This
implies that one cannot directly draw the position of the next hard collision as
in mixed mode. Instead, ones relies on the procedure described hereafter.

\subsubsection{Forward straggled transport}

Let $s_0$ and $T_0$ denote the initial path length and kinetic energy of the
projectile. A tentative Monte~Carlo step length $\Delta
s_{\scriptscriptstyle{E}}$ is set as
\begin{equation} \label{eq:step-straggling}
    \Delta s_{\scriptscriptstyle{E}} = \frac{\epsilon_s}{\rho} R_s(T_0),
\end{equation}
where $\epsilon_s$ is a configurable parameter allowing to tune the simulation
accuracy. By default $\epsilon_s = 1\,\%$. This tentative step length is
compared to other processes as detailed in~\ref{sec:stepping}. Thus, the actual
step length $\Delta s = s_1 - s_0$ might be smaller than $\Delta
s_{\scriptscriptstyle{E}}$ in practice.

Using the notations of PENELOPE manual, the mean energy loss over the step and
the variance are given by
\begin{align}
    \label{eq:momentum-forward}
    \left<\omega\right> =& {} T_0 - \overline{T}_1  +
        \mathcal{O}\left(\Delta s^3 \right), \\
    \text{var}(\omega) =& {} \frac{\rho \Delta s}{2} \left(
        \Omega^2_s(T_0) +
        \Omega^2_s\left(\overline{T}_1\right) \right) \times \nonumber \\
        & {} \left[1 +
        \frac{\rho \Delta s}{\left<\omega\right>} \left(
        S_s\left(\overline{T}_1\right) - S_s(E_0)
        \right) \right] + \mathcal{O}\left(\Delta s^3\right),
        \label{eq:momentum-2}
\end{align}
where $\overline{T}_1 = R_s^{\scriptscriptstyle{(\shortminus 1)}}\left(R_s(T_0)
- \rho \Delta s\right)$ is the energy at $s_1$ assuming \ac{csda} for soft
collisions. The soft energy straggling is
\begin{equation} \label{eq:soft-straggling}
    \Omega^2_s(T) = \frac{\mathcal{N}_A}{M}
        \int_{0}^{\nu_{\scriptscriptstyle{C}}}{\frac{d\sigma}{d\nu}(T)\nu^2
        d\nu},
\end{equation}
where only soft electronic collisions are considered, as discussed hereafter.

Equations~\eqref{eq:momentum-forward} and \eqref{eq:momentum-2} differ from the
original energy corrected expressions found in PENELOPE~\citep{Salvat2015}.  It
is shown in \ref{sec:detailed-momenta} that those are equivalent to PENELOPE at
order $2$ of Taylor expansion in the step length $\Delta s$. The present
expressions are symmetric by exchange of $T_0$ and $\overline{T}_1$, which is
convenient for the backward formulation of the transport.

The energy loss $\omega$ is drawn from equations~(4.59) to (4.63) of the
PENELOPE manual\,\citep{Salvat2015}, using the procedure described there, and
the energy $T_1$ at $s_1$ is set to $T_1 = T_0 - \omega$. Note that this
procedure reproduces the correct mean and variance for the energy loss $\omega$,
i.e. equations~\eqref{eq:momentum-forward} and \eqref{eq:momentum-2}. For large
energy losses ($\left<\omega\right>^2 > 9\,\text{var}(\omega)$), a truncated
Gaussian distribution is used. For intermediate energy losses
($3\,\text{var}(\omega) < \left<\omega\right>^2 \leq 9\,\text{var}(\omega)$), a
uniform distribution is used. For small energy losses ($\left<\omega\right>^2
\leq 3\,\text{var}(\omega)$), an admixture of a delta and a uniform distribution
is used. The delta distribution allows for null losses with a non null
probability.

Once $T_0$ and $T_1$ are known, we are left with randomising a possible hard
collision over $\Delta s$. Since the soft energy loss is randomised, the mixed
procedure described in section~\ref{sec:forward-mixed} (see e.g.
equation~\eqref{eq:mixed-transport}) is rigorously no more valid. Therefore, a
rejection sampling method is used instead, following section 4.3 of the PENELOPE
manual\,\citep{Salvat2015}. The total cross-section is regularised with a
virtual ``do nothing'' ($\delta$) process. Let $\Lambda_{h,0}$ and
$\Lambda_{h,1}$ denote the interaction lengths for $T_0$ and $T_1$. Let the
regularised interaction length be $\Lambda_{min} = \min\left(\Lambda_{h,0},
\Lambda_{h,1}\right)$. The distance to the next event is
\begin{equation} \label{eq:straggled-next-event}
    \Delta s_h = - \frac{\Lambda_{min}}{\rho} \ln \xi,
\end{equation}
where $\xi$ is uniformly distributed over $[0,1]$. If $\Delta s_h \leq \Delta
s$, then a hard collision or a $\delta$ event occurred over the step at $s_h =
s_0 + \Delta s_h \leq s_1$.

The kinetic energy $T_h$ at $s_h$ is determined by linear interpolation, by
assuming that the ratio of the actual energy loss to the \ac{csda} expectation,
$(T_0 - T) / (T_0 - \overline{T})$, is constant over the step. Thus,
\begin{equation} \label{eq:detailed-interpolation}
    T_h = T_0 - \frac{T_0 - T_1}{T_0 -
        \overline{T}_1} \left(T_0 - \overline{T}_h\right),
\end{equation}
where $\overline{T}_h = R_s^{\scriptscriptstyle{(\shortminus 1)}}\left(R_s(T_0)
- \rho \Delta s_h\right)$.

Let $\Lambda_h(T_h)$ denote the interaction length for $T_h$. Then, a hard
collision occurs with a probability $p_{\scriptscriptstyle{H}} = \Lambda_{min} /
\Lambda_h(T_h)$.  Otherwise, a $\delta$ event occurred, in which case the event
can be ignored, i.e.  the projectile kinetic energy and its path length are set
to $T_1$ and $s_1$ respectively.

Let us recall that only electronic collisions are considered in PUMAS, when
computing the soft energy straggling $\Omega^2_s$. The soft energy loss due to
radiative processes is thus not fluctuated. It is deterministic even in
straggled mode. This is valid since at high energies where radiative processes
represent the bulk of the energy loss, fluctuations are dominated by
catastrophic events.  These catastrophic fluctuations are efficiently rendered
by the mixed algorithm, as shown in section~\ref{sec:validation-transmission}.

However, if the energy straggling $\Omega^2_s$ includes radiative processes,
then it becomes very large at high energies, in the radiative regime. This would
complicate the backward simulation as discussed below. Thus, we decided not to
fluctuate soft radiative processes in PUMAS. In addition, PENELOPE's straggling
model would not be appropriate to us in this case. Indeed, in the radiative
regime, since $\Omega^2_s$ would grow very large, PENELOPE's model would result
in a significant number of Monte~Carlo steps without any energy loss at all
($\left<\omega\right>^2 \leq 3\,\text{var}(\omega)$ case). Although the correct
energy loss distribution would be recovered over a large enough number of steps,
this behaviour does not seem natural for a high energy projectile.

\subsubsection{Backward straggled transport
\label{sec:backward-straggling}}

The backward transport procedure in straggled mode would be obtained by
inverting the algorithm described previously. However, this is not possible
directly because the projectile initial energy is not known a priori in a
backward Monte~Carlo step. This can be circumvented by using a \ac{BIS} method,
detailed previously in \citet{Niess2018} (see corollary~1 of the latter).
That is, the energy loss straggling is backward sampled from an alternative
(biased) process, whose inverse is known. Then, Monte Carlo events are weighted
in order to correct for the sampling bias, as discussed below.

As biased process, in backward mode let us randomise the energy loss $\omega$
with the same procedure as in forward mode, but swapping $T_0$ and $T_1$ in
the previous equations. I.e. $\left<\omega\right> = T_1 - \overline{T}_0 \leq 0$
in the backward case, where $\overline{T}_0 =
R_s^{\scriptscriptstyle{(\shortminus 1)}}\left(R_s(T_1) + \rho \Delta s\right)$.
Note that in the backward case
\begin{equation}
    \Delta s_{\scriptscriptstyle{E}} = \frac{\epsilon_s}{\rho} R_s(T_1)  > 0.
\end{equation}
Since $R_s(T_1) < R_s(T_0)$, in backward mode one tends do do smaller steps than
in forward mode.

The biased backward process described previously is a good approximation of the
true reverse process as long as $T_1$ fluctuates closely around its \ac{csda}
expectation, $\overline{T}_1$, i.e. for large enough steps or summing up several
small steps.  In the limit $T_1 \to \overline{T}_1$ ($T_0 \to \overline{T}_0$)
equations~\eqref{eq:momentum-forward} and \eqref{eq:momentum-2} yield the same
result for the forward and approximate backward procedure.

In order to correct for the biased process, in principle the projectile
Monte~Carlo weight must be multiplied by the ratio of the true \ac{pdf} to the
biased one.  In addition, a Jacobian backward weight must be applied
corresponding to the change of variable from $T_0$ to $T_1$, given by the biased
transform. The complete backward weighting procedure would be rather complicated
considering the straggling procedure that is used in PUMAS. Thus, an approximate
backward weighting is used instead.

When considering multiple steps yielding $T_1$, $T_2$, \ldots $T_n$, the
Jacobian of the total transform is equal to the product of the Jacobians of each
step. For a large enough number of steps, we might expect $T_n$ to be close to
its \ac{csda} expectation. Hence, for an elementary step, let us use the
Jacobian obtained previously in \ac{csda} mode, i.e.
equation~\eqref{eq:csda-jacobian}, but substituting the \ac{csda} expectation
$\overline{T}_0$ with the outcome of the backward procedure, $T_0$. This
procedure yields a backward weight satisfying to the Jacobian composition law
and having the correct asymptotic behaviour for $T_0\to\overline{T}_0$, as
\begin{align}
    \left|\frac{dT_0}{dT_n}\right| =&
        \frac{S_s(T_0)}{S_s(T_1)} \frac{S_s(T_1)}{S_s(T_2)} \ldots
        \frac{S_s(T_{n-1})}{S_s(T_n)} \nonumber \\
        =&  \frac{S_s(T_0)}{S_s(T_n)} .
\end{align}
Note that if we use $\overline{T}_0$ for the backward weight instead of $T_0$,
the intermediary terms do not simplify out.

Similarly, when a hard collision occurs, let us use $\Lambda_h(T_h) /
\Lambda_h(T'_1)$ as backward weight at the vertex, where $T_h$ is the energy
interpolated using equation~\eqref{eq:detailed-interpolation}, swapping $T_0$
and $T_1$, and $T'_1$ the projectile energy after the backward collision.

When fluctuating only soft electronic collisions, the approximate backward
procedure described previously is accurate at $0.2\,\%$ for transmission
muography applications, as illustrated in
section~\ref{sec:validation-transmission}. For large values of the straggling,
when soft radiative losses are also fluctuated, the approximate procedure is
still surprisingly accurate, to a few percent.

\subsection{Multiple scattering \label{sec:multiple-scattering}}

As for the energy loss discussed previously, most collisions are soft, resulting
in a small angular deflection.  However, the collective effect of multiple soft
collisions can lead to a sizeable total deflection over the projectile path
length.  This is efficiently taken into account with a mixed Monte~Carlo
algorithm, where the soft scattering is rendered by a ``multiple scattering''
process.

\subsubsection{Transport mean free path}

For small values, the multiple scattering deflection angle $\theta$ after a
path length $\Delta s$ is approximately Gaussian, with standard deviation
\begin{equation} \label{eq:sigma-theta}
    \sigma_\theta = \sqrt{\rho \Delta s / \lambda_{1,s}},
\end{equation}
where $\rho$ is the density of the target material and $\lambda_{1,s}$ the
transport mean free path path restricted to soft collisions.

As can be seen from equation~\eqref{eq:sigma-theta}, the soft transport path
$\lambda_{1,s}$ directly quantifies the magnitude of the multiple scattering.
It is computed by summing up the contributions of all physics processes as
\begin{equation}
    \frac{1}{\lambda_{1,s}} = \frac{1}{\lambda_{1,e}} +
        \sum_j{\frac{1}{\lambda_{1,j}}},
\end{equation}
where $\lambda_{1,e}$ is the contribution from elastic collisions and the
$\lambda_{1,j}$ the contributions from all other processes restricted to soft
collisions with $\nu \leq \nu_{\scriptscriptstyle{C}}$. The elastic contribution
is given by
\begin{equation} \label{eq:elastic-mfp}
    \frac{1}{\lambda_{1,e}} = 2 \frac{\mathcal{N}_A}{M}
        \int_0^{\mu_{\scriptscriptstyle{C}}}{\frac{d\sigma_e}{d\mu}\mu d\mu}.
\end{equation}

Let us recall that, for elastic collisions, the angular parameter $\mu$ and the
energy loss $\nu$ are directly related by the kinematics. However, since in
PUMAS we neglect the energy loss in elastic collision, the soft cutoff is
applied on the angular parameter $\mu$. Following \citet{Fernandez-Varea1993},
the value of the angular cutoff parameter, $\mu_{\scriptscriptstyle{C}}$, is set
such that a large enough number of hard elastic collisions occur on the
projectile path. Thus, $\mu_{\scriptscriptstyle{C}}$ is obtained by solving
\begin{equation} \label{eq:elastic-ratio}
    \Lambda_e(\mu_{\scriptscriptstyle{C}}) = C_e
        \min\left(\lambda_{1,e}(1), R) \right),
\end{equation}
where $R$ is the CSDA range given by equation~\eqref{eq:csda-range} and
$\Lambda_e$ the interaction length for hard elastic collisions, defined as
\begin{equation}
    \frac{1}{\Lambda_e} =
        \frac{\mathcal{N}_A}{M} \int_{\mu_{\scriptscriptstyle{C}}}^{1}
        {\sigma_e d\mu}.
\end{equation}
In practice, $R \leq \lambda_{1,e}$ for muon and tau particles. Thus, the
cross-section for hard elastic collisions is driven by the projectile range.

The elastic ratio $C_e$ appearing in equation~\eqref{eq:elastic-ratio} is a
configurable parameter. By default, it is set to $C_e = 5\,\%$, following
\citet{Fernandez-Varea1993}. This results in $20$ hard elastic collisions on
average on the projectile full range. We did not observe any gain in accuracy by
using smaller values for muography applications. However, it might be required
to decrease $C_e$ for specific applications with thin targets.

Equation~\eqref{eq:elastic-ratio} is solved numerically
when tabulating material properties. Note that it is possible that no solution
exist for $\mu_{\scriptscriptstyle{C}} \in [0,1]$, e.g. for low projectile
energies and/or small values of $C_e$. In this case,
$\mu_{\scriptscriptstyle{C}}$ is set to $0$, i.e.  all elastic collisions are
simulated individually.

The non-elastic contributions to multiple scattering are given by
\begin{equation}
    \frac{1}{\lambda_{1,j}} = 2 \frac{\mathcal{N}_A}{M}
        \int_0^{\nu_{\scriptscriptstyle{C}}}\int{
            \frac{d^2\sigma_j}{d\nu d\mu}\mu d\mu d\nu} .
\end{equation}
For electronic collisions, the approximate result of
equation~\eqref{eq:electronic-transport} is used. For radiative processes, the
\acp{ddcs} discussed in section~\ref{sec:radiatives} are integrated numerically
with a Gaussian quadrature.

\begin{figure}[th] \center
    \includegraphics[width=\textwidth]{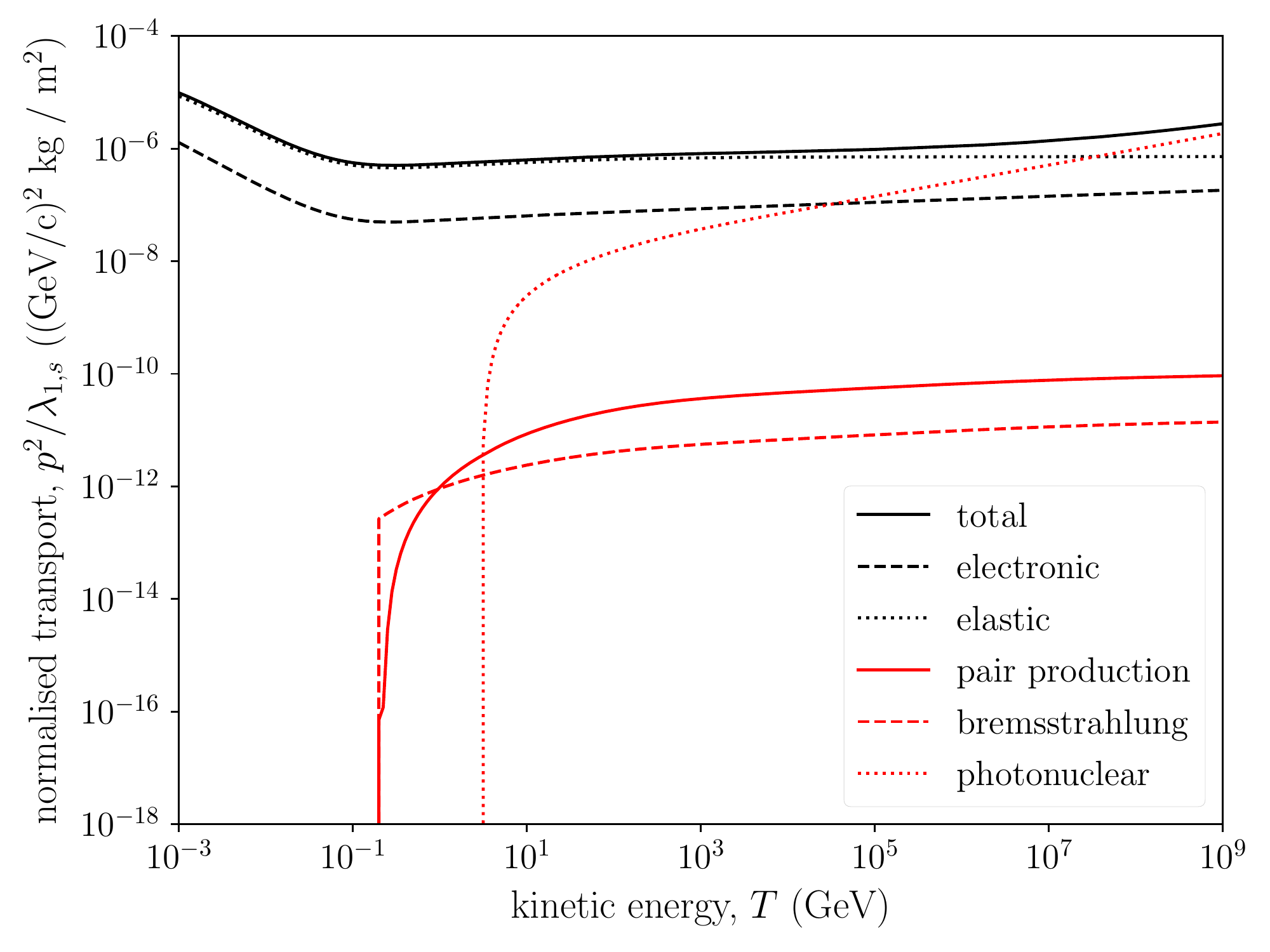}
    \caption{normalised transport cross-section, $p^2 / \lambda_{1,s}$, for soft
    interactions. A muon in standard rock is considered. The total transport
    cross-section is indicated as well as the individual contributions of the
    different processes discussed in section~\ref{sec:physics}. PUMAS default
    cutoff values are used, i.e. an elastic ratio of $C_e = 5\,\%$ and a
    relative cutoff of $x_{\scriptscriptstyle{C}} = 5\,\%$ on the projectile
    energy loss.
    \label{fig:transport-path}}
\end{figure}

The results obtained for the soft transport path, using PUMAS default settings,
are shown on figure~\ref{fig:transport-path}, for a muon in standard rock. It
can be seen that the soft part of the multiple scattering is dominated by
elastic collisions up to very high energies. Above $\sim$$10\ $PeV, photonuclear
interactions are the dominant source of soft scattering. It would be tempting
from this figure to conclude that angular deflections in bremsstrahlung and pair
production events are negligible. However, this is only true for soft
collisions.  At PeV energies and above, the total transport path length is
actually dominated by the latter processes with catastrophic collisions, in
which the projectile looses most of its initial energy.

\subsubsection{Forward multiple scattering}

The multiple scattering and energy loss procedures implemented in PUMAS are
similar. First, the energy at which the next hard elastic collision would occur
is randomised using equation~\eqref{eq:mixed-transport}, but substituting
$\Lambda_h$ with $\Lambda_e$. Then, similarly to straggled mode, Monte~Carlo
steps are limited to a fraction of the soft transport path, as
\begin{equation} \label{eq:step-msc}
    \Delta s_e = \frac{\epsilon_s}{\rho} \lambda_{1,s} .
\end{equation}
If both energy loss and scattering are enabled, then the smaller of
equation~\eqref{eq:step-straggling} or \eqref{eq:step-msc} is used as the
Monte~Carlo step length $\Delta s$. Other conditions might further limit the
step length as detailed in~\ref{sec:stepping}. E.g. if a hard elastic collision
occurs over the step, then the step length is shortened accordingly.  In
practice, for muons and taus the step length is driven by
equation~\eqref{eq:step-straggling}, i.e. by the energy loss.

A soft multiple-scattering process is applied at the end of each Monte~Carlo
step. The soft angular parameter $\mu_s$ is randomised using
equation~\eqref{eq:sigma-theta} and a small angle approximation:
\begin{equation} \label{eq:soft-msc-sampling}
    \mu_s = -\frac{\rho \Delta s}{4} \left(
        \frac{1}{\lambda_{1,s}(T_0)} +
        \frac{1}{\lambda_{1,s}(T_1)} \right) \ln \xi,
\end{equation}
where $T_0$ ($T_1$) is the initial (final) kinetic energy over the step and
$\xi$ a random number distributed uniformly over $[0, 1]$. Note that
$\lambda_{1,s}$ might vary significantly over the Monte~Carlo step due to energy
loss. Thus, an average estimate of the transport cross-section $1/\lambda_{1,s}$
is used in equation~\eqref{eq:soft-msc-sampling}.

Once the angular parameter $\mu_s$ has been sampled the projectile momentum
direction is rotated accordingly. This is done at the end step location as if a
discrete collision had occurred. Note that \citet{Fernandez-Varea1993} instead
apply the soft scattering at a random position over the step. Both procedures
are asymptotically valid, i.e. they both yield correct multiple scattering
distributions when considering several ($\sim$20 or more) Monte Carlo steps. The
method of \citea{Fernandez-Varea1993} is however more accurate when considering
only a few steps. It reproduces the correct spatial distribution over the step
at leading order in the step length (see e.g. section~3 of
\cite{Fernandez-Varea1993}). But, as a result, soft scattering vertices are
interleaved with other processes (energy loss, hard collisions, etc.) occurring
at step ends. This is not convenient. Therefore, in PUMAS, the soft scattering
is instead applied at end step, i.e. synchronously with other processes.

\subsubsection{Backward multiple scattering}

The backward simulation of the multiple scattering is almost identical to the
forward one. The energy at which the next hard elastic collision would occur is
sampled using equation~\eqref{eq:mixed-backward} instead of
\eqref{eq:mixed-transport}. Otherwise, the exact same procedure as in forward
mode is applied. Thus, as discussed previously for the straggled mode, backward
Monte~Carlo steps tend to be smaller than forward ones.

Let us recall that the energy loss is neglected in elastic collisions.
Therefore, there is no transport backward weight to apply when such a collision
occurs, since $\Lambda_e$ is the same before and after the collision. In
addition, the soft multiple scattering process applied at the end of each
Monte~Carlo step is a pure rotation. The Jacobian of this transformation is
unity.  As a result, the backward simulation of the multiple scattering induces
no additional backward weight.  Only the already discussed transport weight
needs to be applied for the used energy loss algorithm, i.e.
equation~\eqref{eq:csda-jacobian} for \ac{csda} or mixed mode or its equivalent
approximation in straggled mode.

\subsection{Hard collisions \label{sec:hard-collisions}}

When a hard collision occurs, an interaction with an individual atom of the
target medium is explicitly simulated. This is done in two steps. First the
energy loss $\nu$ of the projectile is randomised. For this step a different
algorithm is used in forward and in backward mode.  Second, if scattering is
enabled, then the angular parameter $\mu$ is randomised according to the
\ac{ddcs}, given the randomised energy loss $\nu$.

In the case of an elastic collision, the energy loss is approximated by zero.
The first step is thus skipped, and one directly randomises the angular
parameter $\mu$ of the collision from the elastic \ac{dcs}.

\subsubsection{Forward collision}

In forward mode, first the target atom and the interaction process are
randomised using the classical procedure. The probability $p_{ij}$ that the hard
collision occurs on the $i^\text{th}$ atom with the $j^\text{th}$ physics
process is given by the ratio of interactions lengths, as
\begin{equation} \label{eq:target-probability}
    p_{ij} = \frac{\Lambda_{h}(T)}{\Lambda_{ij}(T)},
\end{equation}
where $\Lambda_{h}$ is given by equation~\eqref{eq:hard-interaction-length} and
where $\Lambda_{ij}$ is the summand in the latter, i.e. the interaction length
for $i$ and $j$.

Once the target atom and the physics process have been determined, the energy
loss $\nu$  is randomised by rejection sampling. The efficiency of this method
depends critically on the \ac{dcs} bounding envelope that is used. In the case
of electronic collisions, a simple yet efficient envelope is provided by keeping
only positive terms in the \ac{dcs} given by
equation~\eqref{eq:effective-inelastic-dcs}. The latter envelope can be
randomised with the inverse \ac{cdf} method.

For radiative processes, the \ac{dcs} expressions are more complex and numeric
methods are used. In previous versions of PUMAS, a Zigurrat like envelope was
used for projectile energies above $10\ $GeV. This method requires the
restricted \ac{dcs} for hard collisions to be monotone.  It was satisfied for
projectile energies larger than $10\,$GeV with a fixed (hard-coded) relative
cutoff of $x_{\scriptscriptstyle{C}} = 5\,\%$. Below $10\ $GeV, a weighted
procedure was used as described in \citet{Niess2018}. Since PUMAS v$1.1$ the
relative cutoff between soft and hard collisions can be varied.  Consequently
the restricted \ac{dcs} is not guaranteed to be monotone. Thus, the Ziggurat
method is no longer valid. It has been replaced by a power law envelope
described in \ref{sec:analogue-sampling}.  The new procedure typically requires
$1$ or $2$ evaluations of the \ac{dcs} per hard collision, and it is valid over
all projectile energies. Thus, since PUMAS v$1.1$ the Monte~Carlo particle
states are no longer weighted in forward mode, except possibly from the decay
probability as discussed in section~\ref{sec:decays}.

At high energies w.r.t. the critical energy, hard collisions are frequent even
with the high relative cutoff $x_{\scriptscriptstyle{C}} = 5\,\%$ used by
default in PUMAS. Then, a single computation of the \ac{dcs} for a radiative
process is already expensive CPU-wise. Therefore, a cubic spline interpolation
is used instead for radiative \acp{dcs}, as described in \ref{sec:lookup}. The
spline parameters are tabulated at PUMAS initialisation. The spline grid is set
in order to achieve a better than $0.1\,\%$ relative accuracy on the \ac{dcs}.
However, close to kinematic boundaries the \acp{dcs} can vary sharply, which
would require a very dense grid for the spline. These boundary regions are thus
removed from the spline interpolation and an exact computation is used instead.
Note that those cases are rare. Thus, it is not expensive CPU-wise to perform a
full computation when they occur.

Let us point out that the procedure used in PUMAS for the sampling of the energy
loss in hard collisions differs from what is traditionally done. Usually, the
inverse \ac{cdf} method is used instead of rejection sampling. We do not use the
former for two reasons. First, the backward sampling described hereafter also
requires the \ac{dcs} not the \ac{cdf}. Thus, an extra tabulation would be
required if using the inverse \ac{cdf} method. Secondly, accurately tabulating
the \ac{cdf} close to kinematic boundaries could be numerically difficult. With
the rejection sampling method this is avoided by using the exact \ac{dcs} in
those cases.

\subsubsection{Backward collision \label{sec:backward-collisions}}

In backward mode, the projectile energy before the collision is not known a
priori. Thus, the rejection sampling method described previously cannot be used.
Instead, a \ac{BIS} procedure is used as previously for the energy loss
straggling (see section~\ref{sec:backward-straggling}). First, the initial
kinetic energy $T$ is randomised from a biased distribution. Let us consider a
biased process whose \ac{dcs} is given by a power law function of exponent
$\alpha > 1$. Let its normalized \ac{dcs} be given by
\begin{equation} \label{eq:backward-collision-bias}
    p_b(\nu; T) = \begin{cases}
        \frac{1}{T}
        \frac{(\alpha - 1) x_{\scriptscriptstyle{C}}^{\alpha - 1}}
        {1-x_{\scriptscriptstyle{C}}^{\alpha - 1}}
        \frac{1}{x^{\alpha}} &
        \text{if } x \in [x_{\scriptscriptstyle{C}}, 1] \\
        0             & \text{otherwise}
    \end{cases},
\end{equation}
where $x = \nu / T$ and where $x_{\scriptscriptstyle{C}} =
\nu_{\scriptscriptstyle{C}} / T$ is the relative cutoff for hard collisions. In
order to properly apply the \ac{BIS} procedure, one needs to consider the
mathematical support of the true and biased \acp{dcs}, as real-valued functions
of $\nu$. In this case, the support of the biased \ac{dcs} is bounded from below
by the cutoff for hard collisions and from above by the kinetic energy.  In
practice, the true \ac{dcs} has a smaller support, included into the biased
\ac{dcs} one. This is correct for the biasing procedure, the only downside being
that the \ac{BIS} procedure might result in null Monte Carlo weights. The
converse would be wrong however, i.e. if the biased \ac{dcs} would take null
values on the support of the true \ac{dcs}. Then, the \ac{BIS} procedure would
be biased, since it would miss the corresponding cases.

The value set for $\alpha$ significantly impacts backward Monte~Carlo results,
especially when backward transporting muons over large distances ($\gtrsim
3\,$km of standard rock). In PUMAS~v$1.2$, $\alpha = 2$ is used, which fits the
total radiative \ac{dcs} within $5\,\%$ for $x \in [0.6$, $7]\,\%$. Previously,
a value of $\alpha = 1.4$ was preferred (see \citep{Niess2018}). However, the
validation procedure described in section~\ref{sec:validation-transmission},
indicates that the latter value is less efficient. The updated value ($\alpha =
2$) results in lower variance over all distances, and it is numerically more
robust. Thus, it must be concluded that the previous estimate of $\alpha = 1.4$
was inaccurate.

For very small cutoff values ($x_{\scriptscriptstyle{C}} \lesssim 0.1\,\%$) and
large distances, the backward collision method described below is numerically
unstable, whatever the selected value of $\alpha$. Therefore, in mixed or
straglled backward modes, PUMAS prevents using a cutoff value lower than
$1\,\%$. This is not expected to be a limitation for most muography
applications, as shown in section~\ref{sec:validation-transmission}.

The fractional energy loss $x$ for the biased process can be sampled
semi-analytically with the inverse \ac{cdf} method:
\begin{equation} \label{eq:bias-forward-sampling}
    x = x_{\scriptscriptstyle{C}}
        \left[ 1 - \xi \left(1 - x_{\scriptscriptstyle{C}}^{\alpha-1}\right)
        \right]^{\frac{1}{1 - \alpha}},
\end{equation}
where $\xi$ is a random number uniformly distributed over $[0,1]$. Conversely,
equation~\eqref{eq:bias-forward-sampling} can be solved for the initial kinetic
energy $T$ by substituting $x = 1 - T' / T$. Thus, the initial kinetic energy is
backward sampled as
\begin{equation} \label{eq:bias-backward-sampling}
    T = \frac{T'}{1 - x_{\scriptscriptstyle{C}}
        \left[ 1 - \xi \left(1 - x_{\scriptscriptstyle{C}}^{\alpha-1}\right)
        \right]^{\frac{1}{1 - \alpha}}}.
\end{equation}

Let us point out that the backward weight for the biased \ac{dcs} is
\begin{equation}
    \frac{dT}{dT'} = \frac{T}{T'}.
\end{equation}

Once the initial energy has been backward sampled, the target atom and the
physics process can be randomised. This is done with a biased distribution as
well. For this purpose, it is convenient to consider the macroscopic
cross-section, $\Sigma_h = 1 / \Lambda_h$, instead of the interaction length.
The probability to select the $i^\text{th}$ atom and the $j^\text{th}$ process
is set to
\begin{equation}
    p_{ij} = \frac{d\Sigma_{ij}}{d\nu}(T, T - T') \bigg/
        \frac{d\Sigma_h}{d\nu}(T, T - T').
\end{equation}
Note that for the true distribution, $p_{ij}$ is proportional to the macroscopic
cross-section $\Sigma_{ij}$ instead of the \ac{dcs}, see previous
equation~\eqref{eq:target-probability}. However, since both the initial and the
final energies are known at this stage, we prefer to use the \ac{dcs}. This
choice is justified hereafter.

The backward sampling method requires applying a collision backward weight,
$\omega_c$, in addition to the transport backward weight discussed previously,
e.g.  given by equation~\eqref{eq:backward-transport-weight}. A detailed proof
is provided in~\citet{Niess2018} (see e.g. corollary~3 and equation~(15) of the
latter). For the procedure described herein, the collision backward weight is
given by
\begin{equation}
    \omega_c = \frac{\Lambda_h(T) T}{p_b(T - T', T) T'}
        \frac{d\Sigma_h}{d\nu}(T, T - T').
\end{equation}

Note that the collision backward weight does not depend on the selected target
atom and physics process. This is due to the choice of randomising those
according to the \acp{dcs} instead of the cross-sections. Consequently, the
dispersion of backward weights is reduced with the biased target selection. In
addition, since we do not simulate secondary particles, the target selection
step might be skipped when scattering is disabled.

In principle, the collision backward weight can be null if the kinematics does
not match the \ac{dcs} support of any physics process.  This almost never occurs
in practice. Still, whenever a backward collision results in a null weight, the
backward transport stops and the corresponding event can be dropped.
Nevertheless, one should not silently discard these events. They must be counted
when computing the Monte Carlo estimate (average), otherwise the result would be
biased. In order to make this point clear, let us denote $N_\text{gen}$ the
total number of generated Monte Carlo events, among which $n_0 \leq
N_\text{gen}$ have a null weight. Then, one must normalise the Monte Carlo
estimate by $N_\text{gen}$, not by $N_\text{gen} - n_0$.

An additional technical difficulty arises from the fact that the total
cross-section for hard collisions is null at low energies, below a value $E_0$.
This breaks the symmetry between forward and backward collisions, since a
backward collision could occur at any energy, but a forward one only for $E >
E_0$. This problem is solved by adding a $\delta$-process (``do nothing'') in
order to regularize the total cross-section. More details on this can be found
in \citet{Niess2018}, e.g. in section~3.3 of the latter.

\subsubsection{Scattering angle}

The method used for the randomisation of the scattering angle in a hard
collision depends on the physics process. For electronic collisions, the
effective model discussed in section~\ref{sec:mixed-mode} only considers close
interactions. The corresponding scattering angle is directly computed from
equation~\eqref{eq:close-angular}. For other processes, rejection sampling is
used.

For elastic collisions, the scattering angle is randomised in the CM frame, and
then it is transformed back to the laboratory frame. The exact \ac{dcs}, as well
as its envelope, are computed as explained in \ref{sec:elastic-implementation}.

For bremsstrahlung and $e^+e^-$ pair production, the \citet{Tsai1974} \ac{ddcs}
is used as discussed previously in section~\ref{sec:radiatives}. An upper bound
is given by considering only the first (positive) term in
equation~\eqref{eq:tsai-ddcs} and substituting $X(\mu)$ with $X(0)$ since
$X(\mu) \leq X(0)$ for all $\mu \in [0,1]$. Thus,
\begin{equation}
    \frac{d^2\sigma_{\scriptscriptstyle{T}}}{d\nu d\mu} \leq
        \frac{2 \alpha r_e^2}{\nu} X(0) \left(2 - 2 y + y^2\right)
        \frac{\mu_0}{(\mu_0 + \mu)^2} .
\end{equation}
The latter function is used as envelope for the rejection sampling of the
angular parameter $\mu$.

For photonuclear interactions, the exact \ac{ddcs} is used when available.
Otherwise, the \texttt{DRSS} one is used, rescaled as discussed in
section~\ref{sec:radiatives}. The squared four momentum transfer $Q^2$ is
sampled, and then the scattering angle is computed from
equation~\eqref{eq:theta-photonuclear}. As envelope for the rejection sampling,
a $1 / Q^2$ law is used:
\begin{equation}
    \frac{d^2\sigma}{d\nu dQ^2} \leq \sup_{Q^2}{
        \left[Q^2 \frac{d^2\sigma}{d\nu dQ^2} \right]} \frac{1}{Q^2}.
\end{equation}
The maximum is determined numerically using Brent's algorithm. Note that the
latter is done on the fly during the course of the simulation. This might look
very expensive CPU-wise. However, when scattering is enabled the number of
Monte~Carlo steps per track is large and so is the associated CPU cost. As a
result, with PUMAS default relative cutoff of $x_{\scriptscriptstyle{C}} =
5\,\%$ we observe no significant slow down from the simulation of the scattering
in hard photonuclear collisions, even at PeV energies.

In forward and backward modes, the same algorithm is used for simulating the
scattering of hard collisions. Since this transform is a rotation, its Jacobian
is $1$. Thus, no extra backward weight arises from this step.

\subsection{Decays \label{sec:decays}}

Muon and tau decays are not explicitly simulated by PUMAS. However, the proper
time $t$ of the projectile is computed over the course of the simulation.  Thus,
the reduction of the flux due to decays can be accounted for. PUMAS proposes two
methods for this, ``weighted'' or `randomised`, described hereafter. In
addition, decays can also be totally disabled.

\subsubsection{Proper time}

In PUMAS, the trajectory of a transported particle is approximated by a
succession of line segments connected by vertices. The vertices correspond to
hard collisions or to Monte Carlo steps end points. Between two vertices, the
energy is assumed to vary continuously, even in straggled mode. The proper time
variation between two successive vertices of index $i$ and $i+1$ is given by
\begin{equation}
    t_{i+1} - t_ {i} = \int_{s_i}^{s_{i+1}}{\frac{m}{p_{i,i+1}(s)}ds},
\end{equation}
where $p_{i,i+1}$ is the momentum of the projectile when going from $s_i$ to
$s_{i+1}$, varying continuously. At hard collision vertices, the particle energy
changes discontinuously. Since hard collisions are considered as point-like, no
proper time is needed for those. Thus, the total proper time variation is
obtained by summing up the contributions of all ``soft'' segments, over which
the projectile momentum varies continuously.

In \ac{csda} or in mixed mode, the proper time elapsed between two hard
collisions depends only on the kinetic energy at end points. It is given by
\begin{equation}
    \rho \left( t_{i+1} - t_{i} \right) = R_t(T_{i}) - R_t(T_{i+1}),
\end{equation}
where $R_t$ is the proper time range defined as
\begin{equation} \label{eq:time-range}
    R_t(T) = \int_0^{T}{\frac{dT'}{\beta' \gamma' S(T')}}.
\end{equation}
Note that a uniform medium is assumed as discussed previously, e.g. in
section~\ref{sec:csda-mode}. Note also that, in mixed mode, the total stopping
power $S$ must be replaced with the soft one, $S_s$, in
equation~\eqref{eq:time-range}.

In straggled mode, the proper time is integrated over Monte-Carlo steps using
trapezoidal rule. Thus,
\begin{equation} \label{eq:time-trapezoidal}
    \rho \left(t_{i+1} - t_{i} \right) = \left(\frac{1}{p_{i+1}} +
        \frac{1}{p_{i}} \right) \frac{m (s_{i+1} - s_{i})}{2},
\end{equation}
where $p_i$ ($p_{i+1}$) is the projectile initial (final) momentum for the Monte
Carlo step.

The proper time is identical in backward and in forward modes. But, in backward
mode, the proper time integration proceeds in reverse order, i.e.  from the
target to the source.

\subsubsection{Weighted mode}

The weighted mode accounts for decays globally by weighting Monte Carlo states
by the survival probability for decays, $P_s$. This is possible since the
survival probability $P_s$ depends only on the total elapsed proper time, $t_1 -
t_0 \geq 0$, as
\begin{equation} \label{eq:decay-probability}
    P_s(t_0, t_1) = e^{-(t1 - t_0) / \tau_0},
\end{equation}
where $t_0$ ($t_1$) is the initial (final) proper time and $\tau_0$ the
projectile proper lifetime.

The weighting method does not depend on the direction of the Monte Carlo flow,
i.e. forward or backward sampling. However, a time reference must be defined in
both cases. In forward mode, it seems natural for the user to specify $t_0$ at
input of the transport procedure, and to get $t_1 \geq t_0$ at output.  In
backward mode, it would be tempting to instead specify $t_1$, and to decrease
the proper time in order to deliver $t_0$ at output of the transport.  This
would be valid as well, but considering the final state as time reference.
Nevertheless, this could be error prone. Let us elaborate more on this point.

A common confusion is to identify backward transport with time reversal, in the
sense that it would ``rewind'' the Monte~Carlo as a movie. This would be the
case if collision physics was deterministic. But, except when using
deterministic approximations, the like \ac{csda}, it is not. Let us consider the
example of a point source in order to illustrate this case. Backward
transporting particles originating from a point source does not focus them back
onto the source, despite collisions are reverse sampled. Instead, the backward
transported particles would be spread around the point source. This is contrary
to the time rewinding picture described above, showing that this picture is
not correct. Indeed, forward and backward Monte~Carlo both generate stochastic
trajectories according to the collision physics. However, while in forward
Monte~Carlo one specifies the initial state of a trajectory, backward
Monte~Carlo let us specify its final state. This is achieved by biasing and
weighting. Therefore, backward Monte~Carlo should actually be recognised as a
particular \ac{IS} method.

Fixing $t_1$ and decreasing the proper time, in backward mode, would contribute
to propagate the erroneous belief that backward transport would time rewind the
Monte Carlo. Therefore, it was decided to use the same time reference in forward
and backward modes in PUMAS. Thus, in a backward transport, one actually
specifies the initial time $t_0$, and then one gets the final time $t_1$ from
the transport routine. In practice, this implies that the reported proper time
can only increase, in both forward and backward modes.

For muons, the weighting method is the default algorithm in order to account for
decays. Muons have a large lifetime compared to their interaction length.
Therefore, it is usually more instructive to weight Monte Carlo events rather
than dropping some of them.  On the contrary, taus have a short lifetime.  As a
result, the weighting method becomes highly inefficient as the tau energy
decreases.  Thus, PUMAS actually forbids weighting for forward taus. By default,
taus decays are randomised as explained hereafter.

\subsubsection{Randomised mode}

The second way to account for decays in PUMAS is by Monte Carlo. At the
beginning of the transport, the proper time of the decay is drawn from
equation~\eqref{eq:decay-probability} using the inverse \ac{cdf} method, as
\begin{equation} \label{eq:decay-vertex}
    t_d = t_0 - \tau_0 \ln \xi,
\end{equation}
where $\xi$ is a random number distributed uniformly over $[0,1]$. If the
projectile proper time reaches $t_d$, then the transport stops and the
Monte~Carlo state is flagged as ``decayed''.

In backward mode, ``randomising decays'' can be understood in different ways.
The most comprehensive one would be to generate decay vertices of muons or taus
over a volume of interest, and to do a backward transport from there on. This is
what is done in the DANTON Monte-Carlo~\citep{Niess2018a}. However, this is not
what is relevant for muography. In muography, one is interested in surviving
muons not in their decay products. In this case, it is enough to randomise over
the survival probability $P_s$ of a backward transported particle.  Considering
the proper time reference used in PUMAS, this is done with the exact same
algorithm as in forward mode. It is correct because when the proper time $t$
of the backward projectile exceeds $t_d$ on its course from $s_1$ to $s_0$, then
$t_1 \geq t \geq t_d$ as well. However, with this backward procedure the
stopping point does not represent the decay vertex.  When a backward state is
flagged as decayed, it must be understood that it actually did not reach the
final position $s_1$, but instead decayed at an unknown location before.

\section{Miscellaneous \label{sec:miscellaneous}}

In the present section we discuss miscellaneous features of PUMAS not covered by
the previous sections.

\subsection{Composite materials \label{sec:composites}}

A type of material frequently encountered in muography applications is
``rocks''. A rock is an aggregate of minerals, e.g quartz, calcite, dolomite,
etc. The typical grain size of these minerals is of order $0.1$-$1\ $mm.
Though, large variations can be observed in the grain size. In addition rocks
are porous. Pores also have millimetric size. They can be filled with a gas
e.g.  air and/or a liquid e.g. water.

A mineral has a well defined atomic and electronic structure. Minerals
correspond to the definition that has been used so far for materials in PUMAS.
On the contrary rocks do not. As a particle traverses a rock, it crosses
different mineral grains each of which has its specific dielectric response,
hence density effect.  Minerals are large enough such that most individual
electronic collisions can be approximated as occurring inside a material of
infinite extension.  However, at the boundary between two minerals transition
radiations occur. The latter are beyond the scope of this work, and they are
neglected in PUMAS.

On the other hand, the minerals size is small w.r.t. the typical range of
atmospheric muons in rocks. E.g. a $1\ $GeV muon has a range of $2\ $m in
standard rock. When the projectile range is much larger than the minerals grains
size, the stopping power of a rock, $S_c$, can be approximated by the weighted
sum of its minerals ones, $S_k$, as:
\begin{equation} \label{eq:composite-stopping-power}
    S_c = \sum_k{w_k S_k},
\end{equation}
where $w_k$ is the mass fraction of the $k^\text{th}$ mineral.

In PUMAS, materials such as rocks are modelled as ``composite'' materials. A
composite material is defined as a macroscopic mixture of ``base'' material. It
is specified by the mass fractions, $w_k$, of its base materials components. A
base material has been discussed in previous sections. It is an atomic mixture
of elements together with an electronic structure. In practice, the electronic
structure is defined from the atomic elements mixture as discussed in
section~\ref{sec:electronic-process}.

The density, $\rho_c$, of a composite material depends on the densities,
$\rho_k$, of its parts as:
\begin{equation}
    \frac{1}{\rho_c} = \sum_k{\frac{w_k}{\rho_k}} .
\end{equation}
The stopping power of a composite is given by
equation~\eqref{eq:composite-stopping-power}. Similar rules are derived for
other macroscopic properties, e.g. the transport cross-section, $1/\lambda_1$,
or the energy straggling, $\Omega^2$. For hard collisions, the composite behaves
as a mixture of atomic elements, i.e. identically to a base material.

The impact of rock composition on the accuracy of muography measurements has
been studied by \citet{Lechmann2018}. Variations of several percent are reported
depending on the mineral content and on the rock depth. However, let us point
out that while \citea{Lechmann2018} consider the same definition of rocks
stopping power as us (i.e. equation~\eqref{eq:composite-stopping-power}), they
use the approximate parametrisation of \citet{Sternheimer1971} for the density
effect. As a result, their electronic stopping power might differ by
$\sim$1\,\% with PUMAS (see e.g. section~\ref{sec:density-effect} and
\cite{Sternheimer1971}).

\subsection{Bulk density \label{sec:local-density}}

In section~\ref{sec:condensed}, we considered a target material with a constant
composition and density. This was required in order that the stopping power does
not depend on the particle location. However, there are cases were the previous
assumption(s) can be relaxed while preserving an (approximately) constant mass
stopping power. Then, the transport algorithms discussed in
section~\ref{sec:condensed} have simple generalizations. Thus, PUMAS allows
users to override the default material density, $\rho_0$, with a bulk
(effective) value, $\rho(s)$, depending on the projectile location, $s$.  In
order to motivate the present discussion, let us first present two particular
use cases for this. Then, we discuss the general case and its implementation in
PUMAS.

\subsubsection{Porous material}

Let us consider a composite material made of $n-1$ solids and of a gas. This
could be for example a porous rock made of several minerals and filled with air.
The gas density, $\rho_n$, is much lower than solids ones, $\rho_k$, typically
by 3 orders of magnitude. Thus its mass fraction, $w_n$, is negligible w.r.t.
others. However, the gas might occupy a large fraction of the composite volume,
such that the bulk density of the composite, $\rho_c$, is significantly lower
than the ones of its solids components, $\rho_k$.

Let us further assume that the relative composition of solids components is
constant.  Then, neglecting the gas contribution, the mass stopping power of the
composite, $S_c$, is approximately constant as can be seen from
equation~\eqref{eq:composite-stopping-power}. However, the bulk density
(porosity) of the composite material can vary spatially. Thus, instead of
considering a composite of solids and gas, it is equivalent to consider a
material made only of solids, but with a bulk density reflecting porosity. Note
that the material can be a base material as well, e.g. standard rock, not
necessarily a composite. Let us also point out that due to the density effect
correction, $\delta_F$, this bulk density approach differs from a material whose
mineral density would be set lower.

\subsubsection{Gas material}

A second use case is a pure gas material. The density of a gas is variable, e.g.
with temperature and pressure conditions. Due to the density effect correction,
$\delta_F$, an accurate computation of the gas mass stopping power would require
to takes its density variations into account. However, the variations of
$\delta_F$ are usually a second order correction to the projectile energy loss
in comparison to the density variation itself. Let us consider Earth's
atmosphere as a practical illustration. The air density decreases by a factor
of $3$ between sea level and an altitude of $10\ $km. In comparison, the mass
stopping power varies at most by $2\,\%$ over the same range.

In addition, for similar path lengths the projectile energy loss in gas is
negligible in comparison to the one in liquids and solids, due to the large
density difference. Thus, depending on the geometry of the simulation one might
simply neglect the density effect corrections for gases and use an average
density value. However, density variations are mostly taken into account by
using a local density value that might differ from the one used for computing
the gas mass stopping power.

For projectiles with energy much larger than the critical energy, the
contribution of electronic collisions to the stopping power is negligible. Thus,
the density effect can be neglected at high energy. However, in the radiative
regime the stopping power depends strongly on the target atomic content as
$\sim$$Z^2 / A$. Provided that the material composition is constant, at high
energy the stopping power per unit mass would be approximately independent of
the material density.  In practice, having a constant composition but variable
density seldom happens, apart from the two cases discussed previously, i.e. for
a gas material like Earth's atmosphere or, approximately, for a porous rock
with a uniform mineral content.

\subsubsection{General case}

Let us assume that the mass stopping power is constant, but that the local
target density depends on the location $s$ along the projectile path. Let $X$
denote the column density along the path $s$, as:
\begin{equation} \label{eq:column-density}
    X = \int_{0}^s{\rho(s') ds'} .
\end{equation}

For a constant mass stopping power, the \ac{csda} range defined by
equation~\eqref{eq:csda-range} only depends on the projectile energy not on its
location. Thus, equation~\eqref{eq:csda-distance} can be generalized as:
\begin{equation}
    X_1 - X_0 = R(T_0) - R(T_1).
\end{equation}
For a constant mass stopping power, assuming \ac{csda}, the column density and
the particle kinetic energy are related independently of the details of the
density distribution of the target medium. Therefore, the \ac{csda} transport
equations of section~\ref{sec:csda-mode} can be generalized as well by
substituting $X$ for $\rho s$. The relation between the particle location, $s$,
and $X$ is given by the previous equation~\eqref{eq:column-density}. It is a
priori complex for an arbitrary density distribution.

The cross-sections for hard collisions do not depend on the target density.
Thus, the mixed algorithm discussed in sections~\ref{sec:mixed-mode}
and~\ref{sec:multiple-scattering} is unchanged. The average number of hard
collisions, $N_h$, depends only on the projectile initial energy or equivalently
on the column density, $X$.  The expressions for the soft multiple scattering
and for the straggled energy loss can be generalized as well by substituting $X$
for $\rho s$, as for the \ac{csda} mode. E.g. in
equation~\eqref{eq:straggled-next-event} the column density to the next hard
collision, $\Delta X_h$, is randomised instead of $\rho \Delta s_h$.

In addition, the Monte~Carlo stepping algorithm requires some modifications when
letting the density vary. The local density value is provided by the user with a
callback function. This function must also indicate the typical length,
$L_\rho$, over which the density varies. For example, for a density gradient,
$\vec{\nabla}\rho$, it would return:
\begin{equation}
    L_{\rho} = \frac{\rho}{\left|\vec{\nabla}\rho \cdot \vec{u}\right|},
\end{equation}
where $\vec{u}$ is the momentum direction of the projectile. Then, the
Monte~Carlo local step length is defined as
\begin{equation}
    \Delta s_l = \epsilon_s L_{\rho},
\end{equation}
where $\epsilon_s$ a configurable parameter allowing to tune the accuracy of the
simulation.  This local step length is compared to other conditions, as detailed
in~\ref{sec:stepping}, leading to an actual Monte~Carlo step $\Delta s \leq
\Delta s_l$.

The column density over a Monte~Carlo step, $\Delta X$, is computed with the
trapezoidal rule from the local density values, $\rho_0$ and $\rho_1$, at the
beginning and end of the step. Thus,
\begin{equation} \label{eq:step-grammage}
    \Delta X = \frac{\rho_0 + \rho_1}{2} \Delta s .
\end{equation}
If a hard collision occurs over the step at a column density $\Delta X_h <
\Delta X$, then its location is interpolated by assuming that the density varies
linearly over the step. Note that this assumption is consistent with the
trapezoidal rule used in equation~\eqref{eq:step-grammage}.

The stepping algorithm used in PUMAS relies on the assumption that the local
density variations of the material are continuous. If this is not the case, then
different propagation media should be used instead, as described in
section~\ref{sec:geometry}.

\subsection{Magnetic field \label{sec:magnetic-field}}

PUMAS allows an external magnetic field to be supplied. However, electric fields
are not supported, as in PUMAS v$1.2$.

Magnetic fields are defined, by the user, with the same callback as for local
density models, discussed in previous section~\ref{sec:local-density}. If the
magnetic field is not uniform, then the user must indicate the typical length
over which it varies. As for the density model, the variations are expected to
be continuous. If this is not the case, then different media should be used. If
both a variable local density and a variable magnetic field are used, then the
smallest of both variation lengths should be returned.

The interaction with the external magnetic field $\vec{B}$ is treated using a
classical formalism. Magnetic fields are assumed to be weak enough such that
synchrotron radiation can be neglected. This is for example valid for a muon
evolving in Earth's geomagnetic field. For a point-like particle of electric
charge $q = z e$, the relativistic equation of motion writes
\begin{equation} \label{eq:magnetic-motion}
    \frac{d\vec{u}}{ds} = \frac{z e}{p} \vec{u} \times \vec{B}
\end{equation}
where $\vec{u}$ is the momentum direction. Note that $e \simeq
0.2998\ $GeV/$c\,$T$^{-1}\,$m$^{-1}$, which is convenient when $p$ is expressed
in GeV/$c$ and $B$ in Tesla.

The classical magnetic deflection conserves the particle energy. However, the
projectile looses energy over its path due to collisions. In addition, the
magnetic field might vary over the path. In order to take these effects into
account, the equation of motion is discretized over Monte~Carlo steps. In
practice, the method used in PUMAS is equivalent to modelling the interaction
with the external magnetic field as an additional, and independent, scattering
process. It is similar to the soft multiple scattering, i.e. it is applied at
end-step (see section \ref{sec:multiple-scattering}), but a deterministic
deflection angle is used, derived from the equation of
motion~\eqref{eq:magnetic-motion}.

The update rule is derived from finite differences of the equation of motion.
Let $\vec{u}_0$ denote the initial momentum direction of the projectile and
$p_0$ its absolute value. Let $\vec{B}_0$ denote the magnetic field at the step
starting location. Then, at the end of the Monte Carlo step, the projectile
momentum direction would be updated as
\begin{equation} \label{eq:magnetic-finite-diff}
    \vec{u}_1 = \vec{u}_0 + \frac{z e \Delta s}{2} \vec{u}_0 \times \left(
        \frac{\vec{B_0}}{p_0} + \frac{\vec{B_1}}{p_1} \right)
\end{equation}
where $p_1$ and $B_1$ are the projectile momentum and the magnetic field at the
end of the Monte~Carlo step. Let us recall that the projectile follows a
straight line trajectory over the step, along $\vec{u}_0$. Its direction is only
modified at the end of the Monte~Carlo step.

The previous update rule does not properly conserve the norm of the momentum
direction, $\vec{u}$. This could be solved by re-normalising the direction after
each update. An alternative solution is to instead use the following update
rule, equivalent at leading order in $\Delta s$ to
equation~\eqref{eq:magnetic-finite-diff}.  In order to simplify the notations
let us first define:
\begin{equation}
    \vec{\ell}_{0,1} =
    \vec{u}_0 \times \left(
        \frac{\vec{B_0}}{p_0} + \frac{\vec{B_1}}{p_1} \right).
\end{equation}
Then, the update rule, used in PUMAS, for magnetic deflection is
\begin{equation} \label{eq:magnetic-update}
    \vec{u}_1 = \cos\left(\frac{\Delta s}{r_B}\right) \vec{u}_0 +
        \sin\left(\frac{\Delta s}{r_B}\right) \vec{u}_{\bot}
\end{equation}
with
\begin{equation}
    \frac{1}{r_B} = \frac{z e}{2} \left| \vec{\ell}_{0,1} \right|
\end{equation}
and
\begin{equation}
    \vec{u}_{\bot} = \vec{\ell}_{0,1}
        \bigg/ \left| \vec{\ell}_{0,1} \right| .
\end{equation}
Equation~\eqref{eq:magnetic-update} conserves the norm of $\vec{u}$ since
$\{\vec{u}_0, \vec{u}_{\bot}\}$ forms an orthonormal basis.

In order for the finite difference method to be accurate, the Monte~Carlo step
length $\Delta s$ must be small w.r.t. the magnetic bending radius. Thus,
when a magnetic field is defined, the Monte~Carlo step length is limited by
$\Delta s \leq \Delta s_{\scriptscriptstyle{B}}$ with:
\begin{equation} \label{eq:step-magnetic}
    \Delta s_{\scriptscriptstyle{B}} = \frac{\epsilon_s p_0}
        {z e \left| \vec{u}_0 \times \vec{B}_0 \right|}
\end{equation}
where $\epsilon_s$ is a configurable parameter allowing to tune the accuracy of
the simulation.

In the case of a uniform magnetic field, a muon would follow a circular (helix)
trajectory in the void. Instead, PUMAS approximate this trajectory with line
segments. The relative difference on the path length, between both trajectories,
is of $\mathcal{O}\left(\Delta s^2 / r^2_B\right)$. It is assumed that
$\epsilon_S \lesssim 1\,\%$. Thus, the path length difference would be
negligible, and no additional correction is applied in PUMAS.

In backward Monte~Carlo mode, the same algorithm is used for magnetic
deflections as in forward mode. The backward update rule is obtained by
swapping $\vec{u}_0$ and $\vec{u}_1$, as well as other quantities in
equation~\eqref{eq:magnetic-update}. In addition, one must take care that the
magnetic deflection is reversed since the projectile propagates along
$\shortminus \vec{u}_1$. This can be accounted for by changing the sign of the
deflection angle $\Delta s / r_B$, in equation~\eqref{eq:magnetic-update}. Note
also that, since the magnetic deflection is a pure rotation, the corresponding
backward weight is unity.

A typical use case of magnetic field in PUMAS is the deflection of low energy
muons by Earth's geomagnetic field, as they travel through the atmosphere. Let
us point out that Earth's geomagnetic field is too weak to significantly deflect
muon or tau leptons in dense materials, like rocks or water.  In the latter
case, the projectile range is much lower than the magnetic bending radius, over
all energies. Thus, as an optimisation, the geomagnetic field can be deactivated
in dense materials without loss of accuracy.

\subsection{Geometry \label{sec:geometry}}

So far, we considered a single target material potentially with continuous bulk
density variations. However, in a practical muography use case the projectile
would traverse different materials, e.g. air and rocks. Modelling this requires
a geometry description. The PUMAS library does not include a geometry engine
with primitive shapes and related functions, such as ray tracers. Instead, the
geometry must be supplied by the user with a callback. Nevertheless, this
generic mechanism allows one interfacing PUMAS with external geometry engines.
The PUMAS website~\citep{PUMAS:GitHub} provides examples of interfaces with
Geant4~\citep{Agostinelli2003,Allison2006,Allison2016}, using a G4Navigator, or
with the TURTLE library~\citep{Niess2019}.  In principle, PUMAS could also be
interfaced with ROOT~\citep{Brun1997} using a TGeo object. However, at the time
of this writing we are not aware of any publicly available example of this.

Let us briefly describe the generic geometry representation used in PUMAS. Let
us define a propagation medium as a material with possibly a local density
model. If no local density is provided, then the material density is assumed for
the medium. The geometry of the simulation is supplied by the user as a
``medium'' callback function. This function gets as input the projectile
location, $\vec{r}$, and its momentum direction, $\vec{u}$. In return, it is
expected to indicate the corresponding medium at $\vec{r}$ and the distance
$\Delta s_m$ to the medium boundary, assuming a straight line propagation. Let
us recall that in backward mode the projectile propagates along
$\shortminus{\vec{u}}$ instead of $\vec{u}$. Note also that on some calls both
$\Delta s_m$ and the medium are required, while in other calls only one of those
is expected. This is indicated to the user by passing a null pointer when the
corresponding information is not needed by PUMAS. Additional technical details
are given in \ref{sec:stepping}.

\subsection{Software}

The present document focuses on the physics used in PUMAS and on its
implementation. However, some particular software aspects of PUMAS deserve
attention as well. Those are discussed below.

The PUMAS library has been implemented in C99 with the C standard library as
sole external dependency. It is LGPL-$3.0$ licensed. The PUMAS source is $12\,$k
\ac{LOC} in total. It is contained in two files \texttt{pumas.c} and
\texttt{pumas.h}. Thus, PUMAS can easily be embedded in any software project
compatible with the C \ac{ABI}, e.g. in C++ or Python.

Examples of installation of PUMAS are given on the website~\citep{PUMAS:GitHub}.
On Linux and OSX, an example of \texttt{Makefile} is available for compiling
PUMAS with \texttt{Make}. However, since PUMAS is standard C we do not expect any
specific difficulty for compiling it. PUMAS can also be installed with
\texttt{CMake} using the provided \texttt{CMakeLists.txt} file. This has been
tested on Linux, OSX and Windows.

The PUMAS library uses a single namespace. All publicly exported enums,
functions, macros and structs are prefixed with \texttt{pumas\_} or
\texttt{PUMAS\_}.

The library design follows an object oriented pattern with three main C struct
objects: \texttt{pumas\_context}, \texttt{pumas\_physics} and
\texttt{pumas\_state}. These objects correspond to specific memory usages,
detailed hereafter.

The \texttt{pumas\_physics} object is the first that needs to be instantiated.
A physics instance is specific to one type of particle, i.e. muon or tau.  It is
an opaque C struct that contains the tabulations needed during the Monte Carlo
transport. These table are computed when instantiating the physics. Then, they
are readonly at runtime. The initial computation of the physics tables can take
tens of seconds to a few minutes, depending on the number of defined materials.
This can be problematic in some use cases. For faster initialisation, the physic
tables can be dumped to disk and loaded back as well. Note that the raw content
of the C struct is written to disk. Therefore, this format is not portable.

The materials to tabulate are defined in a specific XML file called \ac{MDF}.
The user must supply this file. It specifies the properties of atomic elements,
of base materials and of composite materials, used during the simulation.
Examples of \ac{MDF} are available from the PUMAS website\,\citep{PUMAS:GitHub}.
The tabulation generates stopping power text files in the \ac{pdg} format. These
files can be browsed as a cross-check. They can also be overridden in order to
rescale PUMAS stopping power to a different model.

The \texttt{pumas\_context} object manages a simulation stream for a given
physics object. It contains configuration parameters for the simulation, which
the user can set directly, e.g. the energy loss mode, the transport direction,
etc. Limits can be set on the projectile properties as well, e.g. on its
travelled path length or on its kinetic energy.  In addition, this object
manages a set of opaque data needed during the Monte~Carlo transport. These data
are thread specific. For multithreaded applications one \texttt{pumas\_context}
must be created per thread. The simulation context is forwarded to user
callbacks e.g. for local models or for the geometry. Extra user memory can be
reserved when instantiating the context object. As for the context data, this
user memory is intended to be thread specific.

The \texttt{pumas\_state} object contains a minimal set of data needed by PUMAS
for describing the Monte~Carlo state of a transported particle. Those data are
the particle charge, its kinetic energy, its position, etc. A PUMAS state is
both read and written by the user. For example it is initialised by the user
before the transport. Then, the transport routine updates the state and the
result is read back by the user. A state structure can be extended by the user,
as long as the extension can be cast back to a \texttt{pumas\_state}.  Depending
on the use case, this can be more convenient than reserving extra user memory
for the simulation context.

In addition to these three structures, a geometry must be defined by the user.
This is done by providing a medium callback to the simulation context, as
described in section~\ref{sec:geometry}. This callback points to the current
propagation medium. Propagation media are defined as \texttt{pumas\_medium}
structures. A propagation medium is specified by a material of constant
composition and an optional \texttt{pumas\_locals\_cb} callback. The latter
allows the user to specify local values of the bulk density, or of the magnetic
field, over the medium. If not provided, then the filling material density is
used. As for a PUMAS state struct, the media can be extended by the user, as
long as the extension can be cast back to a \texttt{pumas\_medium}.

\section{Validation \label{sec:validation}}

The PUMAS library has been continuously validated since its initial
implementation. The source code is unit tested with a coverage of $91\,\%$ of
\ac{LOC} for v$1.2$. The library results are validated on each update with
various tests. Still, PUMAS is software. If malfunctioning is observed, these
issues can be reported on the PUMAS website~\citep{PUMAS:GitHub}.  Let us also
point out that, although PUMAS can run on Windows and OSX systems, it is
developed, tested and mainly used on Linux.

PUMAS physics results have been compared to other Monte~Carlo codes showing
consistent results. For example PUMAS was compared to
Geant4\,\citep{Agostinelli2003,Allison2006,Allison2016} and
MUM\,\citep{Sokalski2002} in \citet{Niess2018}.  In the following, we provide an
updated validation test for a transmission muography problem. In particular, we
performed a comparison to PROPOSAL\,\citep{Koehne2013,Dunsch2019}.  In addition,
we also  consider scattering applications with two new cases: a comparison to
experimental muon scattering data, and a background estimate with a toy geometry
of volcano.

The validation results presented in this section were obtained with v$1.1$ of
PUMAS for default settings ($x_{\scriptscriptstyle{C}} = 5\,\%$), and with
v$1.2$ in other cases. The main difference between both version is an
improvement in the sampling of backward collisions (see
section~\ref{sec:backward-collisions}). It has been checked that v$1.1$ and
v$1.2$ yield consistent results when using PUMAS default settings.

\subsection{Transmission muography \label{sec:validation-transmission}}

Let us consider a toy transmission muography problem similar to
\citet{Niess2018}, but using updated inputs and software. Let us compute the
flux of atmospheric muons $\Phi_1$ transmitted through a given thickness $d$ of
standard rock. Let us recall that current muographs cannot measure the energy of
detected particles. Thus, only the total flux is observed, not the differential
one, $\phi_1$, w.r.t. the energy. As initial differential spectrum $\phi_0$, let
us input the sea level parametrisation of \citet{Guan2015} at an observation
angle of $20\ $deg w.r.t.  the horizon.  The latter parametrization was fitted
to a compilation of experimental data. Contrary to Gaisser's
parametrization~\citep{Gaisser2016,Zyla2020} used previously
in~\citep{Niess2018}, \citet{Guan2015} extends to low energies by correcting for
the Earth's curvature.  The observation angle of $20\ $deg, used herein, is a
typical pointing direction for surface muography experiments.

For the present purpose of validating the implemented algorithms, let us
symmetrise the problem. Let us clip the initial spectrum $\phi_0$ to kinetic
energies $T_0 \in [T_{min}, T_{max}]$, and let us assume that transmitted muons
are detected provided that their final kinetic energy $T_1$ is also in
$[T_{min}, T_{max}]$.  Let us set $T_{min} = 10^{-3}\ $GeV and $T_{max} =
10^{9}\ $GeV. This corresponds to the default energy range of PUMAS.

Given the previous toy problem, the transmitted flux $\Phi_1$ is computed
using the different transport modes of PUMAS. For the \ac{csda} mode
equation~\eqref{eq:csda-rate} is used in forward mode and
equation~\eqref{eq:csda-backward-rate} in backward mode. For other energy loss
modes a Monte~Carlo simulation is done.

In previous work~\citep{Niess2018}, we compared the transmitted flux values in
order to asses the accuracy of PUMAS.  However, transmission muography is an
inverse problem in which one actually determines the column density, $X = \rho
d$, from the observed transmitted flux, $\Phi_1$. Thus, let us instead interpret
flux differences as errors on the bulk density. For this purpose, let us
consider as reference the flux $\Phi_{1,FS}$ obtained in forward straggled with
PUMAS default cutoff (i.e. $x_{\scriptscriptstyle{C}} = 5\,\%$), and let us
invert it using other computations.  This yields an estimate for the column
density differing from the ``true'' value obtained in forward straggled mode by
$\Delta X$. Let us further assume that the rock thickness $d$ is known. Then the
error $\Delta X$ translates to an error, $\Delta \rho = \Delta X / d$, on the
reconstructed bulk density given by
\begin{equation} \label{eq:delta-rho}
    \frac{\Delta \rho}{\rho} =
        \frac{\Phi_1^{(\shortminus 1)} \circ \Phi_{1,FS} (X) - X}{X} .
\end{equation}

Let us point out that equation~\eqref{eq:delta-rho} provides values that differ
significantly from $\Delta \Phi_1 / \Phi_1$ for large $X$. For thick targets,
the transmitted flux drops sharply with $X$, as can be seen on figure~4 of
\citet{Niess2018}. Consequently, large $\Delta \Phi_1$ lead to small $\Delta
\rho$ in this regime. On the contrary, for very thin targets small relative
errors on the flux result in larger errors on the density.

\begin{figure}[th]
    \center
    \includegraphics[width=\textwidth]{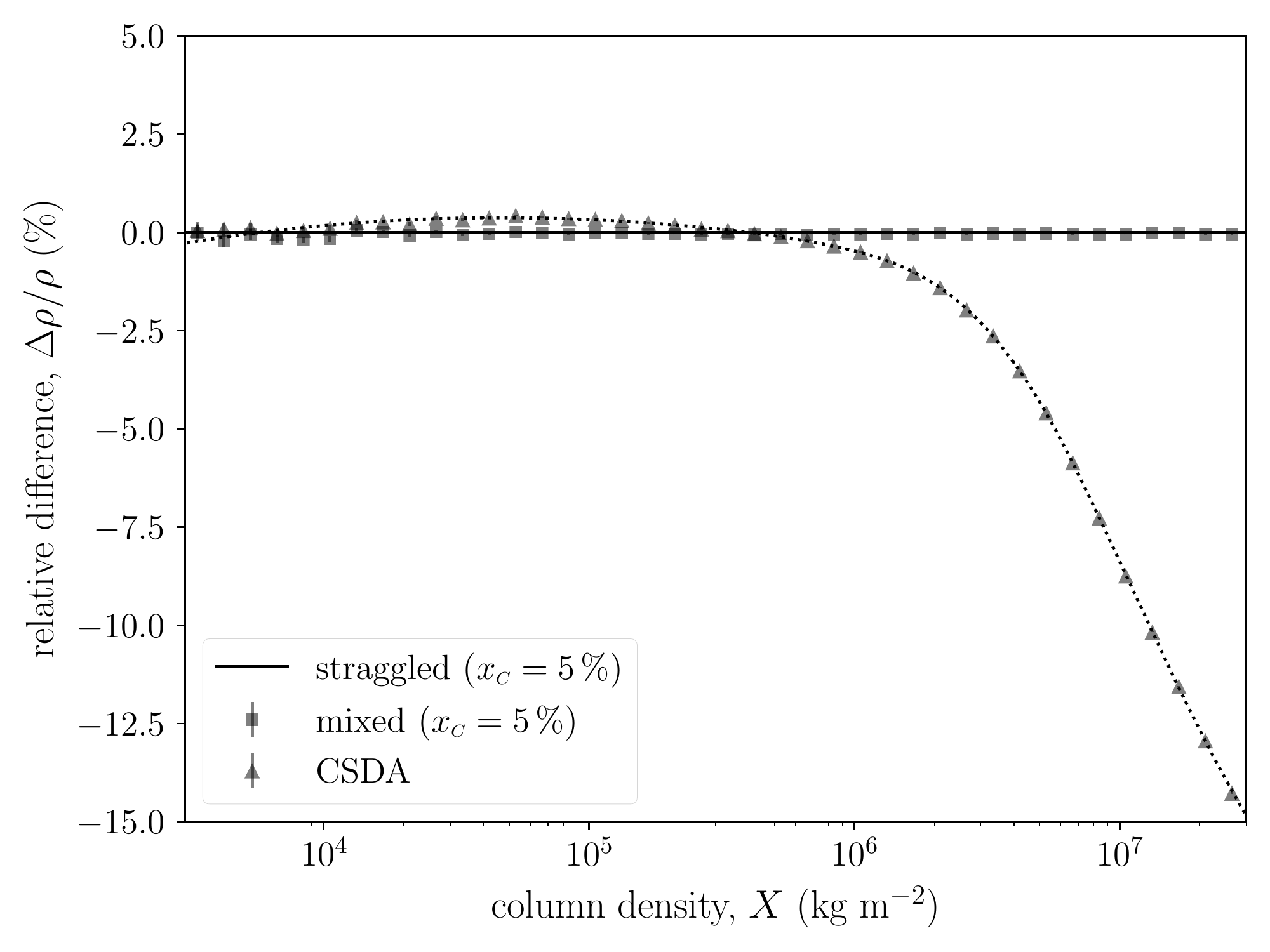}
    \caption{Relative differences on the reconstructed bulk density using
    \ac{csda} or mixed mode instead of straggled one. A muon in standard rock is
    considered. Monte Carlo simulations are performed with PUMAS default cutoff
    for radiative losses, i.e. $x_{\scriptscriptstyle{C}} = 5\,\%$.
    \label{fig:flux-csda}}
\end{figure}

The values obtained for $\Delta \rho / \rho$ in \ac{csda} and mixed mode
($x_{\scriptscriptstyle{C}} = 5\,\%$) are shown on figure~\ref{fig:flux-csda}.
It can be seen that \ac{csda} is a good approximation in this case, up to column
densities of $\sim$10$^6\ $kg$/$m$^{-2}$, i.e. $\sim$$300\ $m of standard rock.
For thicker targets, \ac{csda} underestimates the transmitted flux, as was
previously reported by \citet{Sokalski2001}. Consequently, the bulk density is
underestimated when assuming \ac{csda} for thick targets.

\begin{figure}[th]
    \center
    \includegraphics[width=\textwidth]{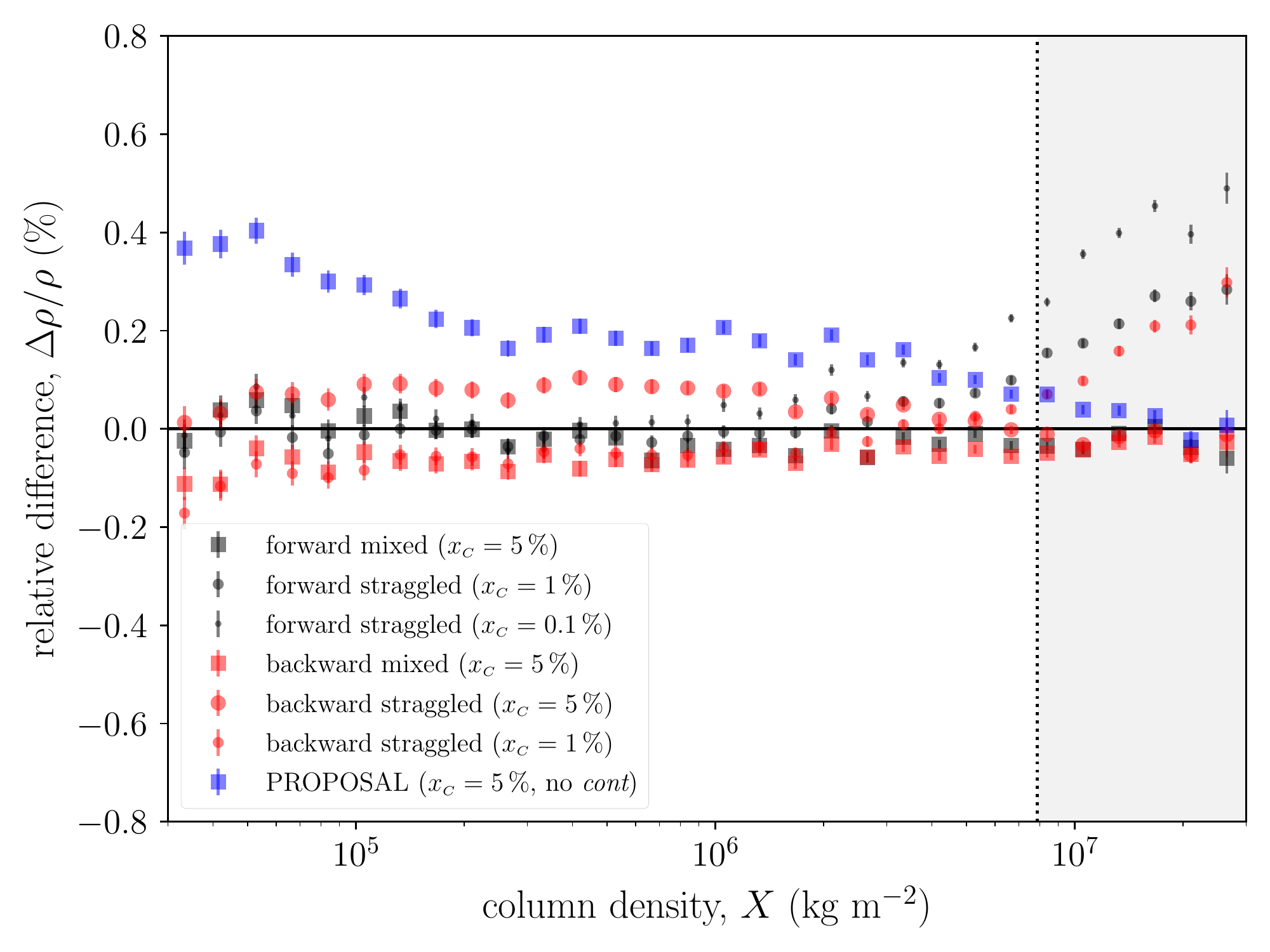}
    \caption{Relative differences on the reconstructed bulk density using mixed
    or straggled modes for the energy loss simulation, and using backward or
    forward Monte~Carlo. A muon in standard rock is considered. The results
    obtained for various cutoff values are reported, i.e.
    $x_{\scriptscriptstyle{C}} = 0.1\,\%$, $1\,\%$ or 5$\,\%$ (PUMAS default). A
    comparison to PROPOSAL~\citep{Koehne2013,Dunsch2019} is also shown. The
    shaded area indicates the region where the rate of events drops below 1 per
    year, considering an exposure of $2\pi\,\mathrm{m}^2\,\mathrm{sr}$.
    \label{fig:flux-delta}}
\end{figure}

A more detailed comparison of the mixed and straggled modes is provided on
figure~\ref{fig:flux-delta} considering various cutoff values for radiative
losses, i.e. $x_{\scriptscriptstyle{C}} = 0.1\,\%$, $1\,\%$ or $5\,\%$. For a
given cutoff, mixed and straggled results agree at better than $0.2\,\%$.
Forward and backward computations also agree within $0.2\,\%$, at least down to
$x_{\scriptscriptstyle{C}} = 1\,\%$. As of PUMAS~v$1.2$, lower cutoff values are
not allowed in backward mode (see section~\ref{sec:backward-collisions}).

Comparing results for different cutoff values, it is seen that PUMAS default
setting ($x_{\scriptscriptstyle{C}} = 5\,\%$) is accurate up to column densities
of $\sim$10$^7\ $kg$/$m$^{-2}$, i.e. $\sim$3$\ $km of standard rock. For thicker
targets, slight differences are observed, increasing with $X$ up to $0.5\,\%$
for 10$\ $km of standard rock. However, for such thick targets the transmitted
flux is practically extinguished, i.e. it is out of experimental reach. Indeed,
let us point out that current muography detectors have typical exposures smaller
than $\sim$2$\pi\,$m$^2\,$sr. With the flux model considered herein, this would
result in rates lower than 1 event per year for targets thicker than $3.0\ $km
of standard rock (indicated as the shaded region in
figure~\ref{fig:flux-delta}). Thus, in practice PUMAS default cutoff is expected
to be valid for most muography applications.

In principle, higher cutoff values than $x_{\scriptscriptstyle{C}} = 5\,\%$
could be considered as well, depending on the target thickness and on the
desired accuracy. However, in practice we observed that $5\,\%$ is usually fast
enough. For example, if straggling or elastic scattering is activated, then it
makes no difference, CPU-wise, to set $x_{\scriptscriptstyle{C}}$ to a value
higher than $5\,\%$.

In addition, we also performed a comparison to
PROPOSAL~v$7.0.6$~\,\citep{Koehne2013,Dunsch2019}  using the same  models for
radiative \acp{dcs} as PUMAS, i.e.  \texttt{SSR} for bremsstrahlung and $e^+e^-$
production and \texttt{DRSS} for photonuclear processes. Note however that
PROPOSAL uses the parametrization of \citet{Sternheimer1984} for the density
effect. Thus, in PROPOSAL the electronic stopping power is slightly lower than
in PUMAS. The discrepancy reaches $0.4\,\%$ at GeV energy in standard rock, as
discussed in section~\ref{sec:density-effect}.

PROPOSAL cuts are set as PUMAS default setting, i.e. $e_{cut} = +\infty$ (see
\citep{Koehne2013}) and $x_{\scriptscriptstyle{C}} = 5\,\%$. Continuous
randomisation is disabled since PUMAS does not include radiative processes in
its straggling. Using this setting, an excellent agreement is found between
PROPOSAL and PUMAS for thick targets. For thin targets, a discrepancy is
observed. The inverted densities using PROPOSAL are slightly larger than the
ones obtained with PUMAS. The discrepancy reaches a value of $0.4\,\%$ at small
$X$, i.e. depths of a few metres in standard rock. This is consistent with
differences in the electronic stopping power discussed previously.

\subsection{Scattering muography \label{sec:validation-scattering}}

PUMAS was primarily designed in order to solve the transmission muography
forward problem. However, it also integrates a detailed elastic scattering
model. An accurate description of the scattering of low energy muons ($E
\lesssim 10\ $GeV) is needed in order to simulate background events as discussed
in the following section~\ref{sec:validation-showa-shinzan}. Thus, a priori,
PUMAS could also be used for simulating scattering muography experiments.
However, to our knowledge Geant4~\citep{Agostinelli2003,Allison2006,Allison2016}
fulfils well this use case.

Various internal tests have been carried out in order to validate the software
implementation of single and multiple elastic scattering of muons in PUMAS. In
particular, we compared PUMAS' mixed simulation results to a ``brute force''
Monte~Carlo simulation, where every single elastic collision is simulated, i.e.
setting $\mu_c = 0$. This test was performed for low energy ($\sim$1\, GeV)
muons crossing thin foils of standard rock of tens of $\mu$m thickness. We found
a better than $\sim$$1\,\%$ agreement for the angular and spatial distributions
of exiting muons when using PUMAS or a brute force simulation of every elastic
collision.

\begin{figure}[th]
    \center
    \includegraphics[width=\textwidth]{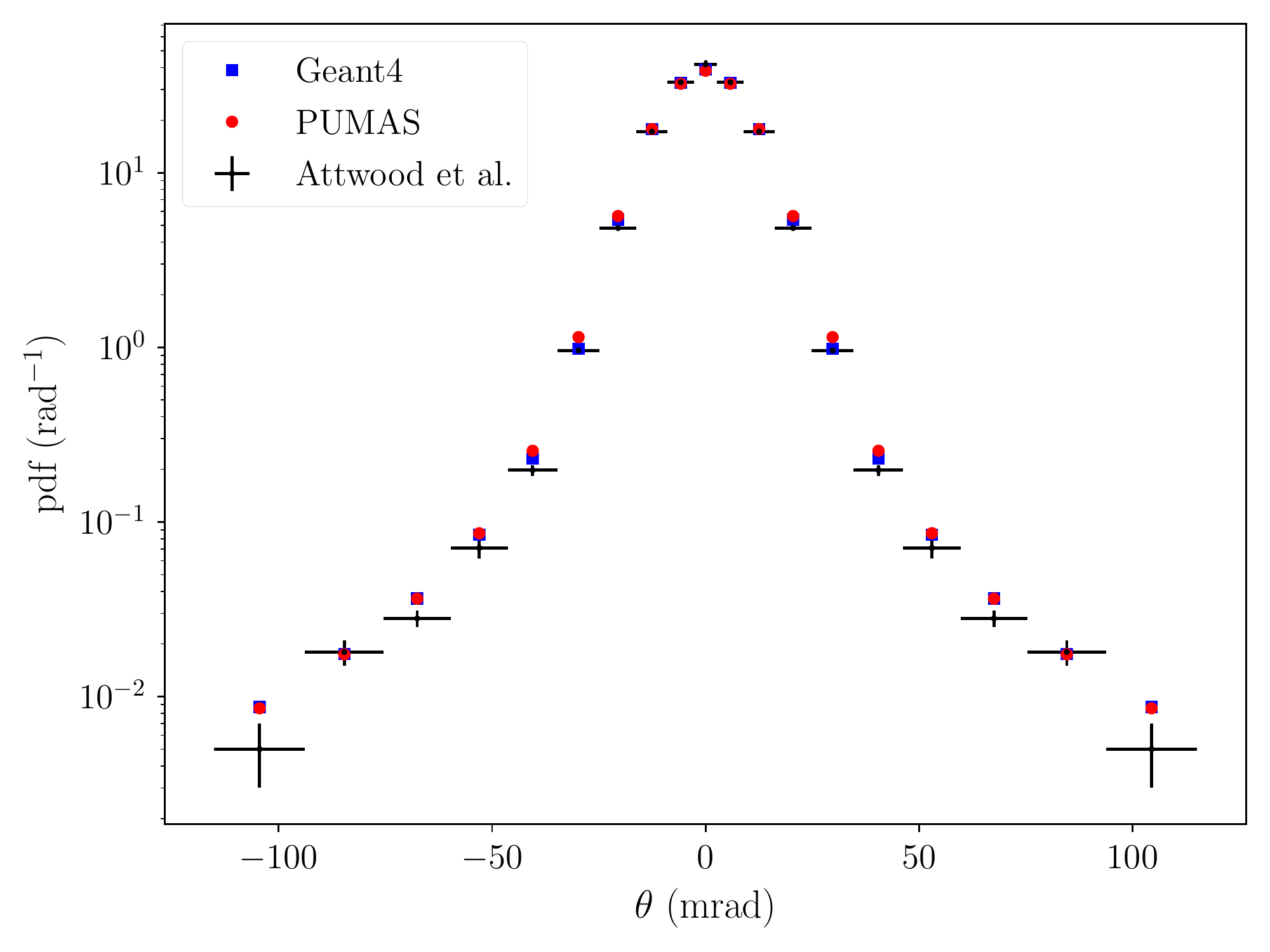}
    \caption{Angular distribution of muons exiting an aluminium foil of
    $1.5\ $mm thick. The error bars correspond to the experimental results of
    \citet{Attwood2006} reporting a $172\ $MeV/c momentum for incident muons.
    The markers represent the results of PUMAS simulations for $172$ and
    $179\ $MeV/c incident momentum.
    \label{fig:validation-scattering}}
\end{figure}

Validating the physics model used in PUMAS for elastic collisions requires a
comparison to experimental data. Systematic measurements have been performed by
\citet{Attwood2006} for $172\ $MeV/c muons crossing thin foils of various
low-$Z$ materials, ranging from H to Fe. \citea{Attwood2006} provide unfolded
angular distributions for the multiple scattering angle, which allow for direct
comparison to Monte~Carlo simulations.  Figure~\ref{fig:validation-scattering}
shows a comparison of PUMAS results to \citea{Attwood2006} in the case of an
aluminium foil of $1.5\ $mm. Satisfactory agreement is observed, over five
orders of magnitude, for the \ac{pdf}. However, closer examination shows that
the \acp{pdf} simulated by PUMAS are systematically broader than the unfolded
data, for all target materials. The standard deviation of the simulated
scattering angle is $3$ to $6\,\%$ higher than what is expected from the
unfolded data. A systematic comparison is provided by
table~\ref{tab:standard-muscat}.

\begin{table}[th]
    \caption{Comparison of PUMAS to experimental values of \citet{Attwood2006}.
    The relative difference on the standard deviation of the multiple scattering
    angle is reported for various targets.  PUMAS simulations have been done for
    two values of the beam momentum, $p$, i.e. $172$ and $179\ $MeV/c. The
    experimental value measured by \citea{Attwood2006} is $p = 172\ $MeV/c.
    \label{tab:standard-muscat}}
    \center
    \begin{tabular}{lllllll}
        \toprule
        $p$ (MeV/c) & H (\%) & Li (\%) & Be (\%) & C (\%) & Al (\%) & Fe (\%) \\
        \midrule
        $172$ &  3.2 & 6.6 & 5.1 & 4.9 & 5.2 &  2.9 \\
        $179$ & -0.6 & 2.4 & 0.7 & 0.5 & 0.9 & -1.2 \\
        \bottomrule
    \end{tabular}
\end{table}

The origin of this discrepancy is not known. It is not specific to PUMAS. The
data of \citea{Attwood2006} seem to disagree with the elastic \ac{dcs} obtained
from the first Born approximation, even though Coulomb corrections are applied.

In Geant4, this discrepancy has been reduced by assuming that the screening
parameter, $\mu_i$ in equation~\eqref{eq:screening-parameter}, is inaccurate. A
``correction'' is applied such that, for low-Z materials, the screening
parameter is multiplied by $\sim$$2$. See e.g. line 168 of
\texttt{G4WentzelOKandVIxSection.cc}.

An alternative explanation for the discrepancy could be an experimental bias.
For example, if the momentum of incident muons would actually be higher than
$172\ $MeV/c, estimated by \citea{Attwood2006}, then the experimental \acp{pdf}
would be systematically narrower than the ones from simulations with a $172\
$MeV/c beam.  This hypothesis has been tested by performing a global fit of
experimental data, but letting the beam energy free.  This fit yields a beam
momentum of $\hat{p} = 179 \pm 0.5\ $MeV/c with a $p$-value of $27\,\%$.

\citea{Attwood2006} estimated the beam momentum from time of flight differences
between different particles (see e.g. figure~10 of their paper). The reported
error is $\pm 2\ $MeV, i.e. a $1\,\%$ relative accuracy. But, this might have
been underestimated. A $4\,\%$ higher beam momentum would be preferred in order
to agree with theoretical expectations.

In addition, PUMAS results have also been compared to the experimental
scattering data collected by \citet{Akimenko1986}. A $1.44\ $cm thick Copper
target is used with incident muons of $7.3$ and $11.2\ $GeV/c.  No discrepancy
is observed. PUMAS agrees with \citea{Akimenko1986} to $0.6\,\%$ on the standard
deviation of the multiple scattering angle. Let us point out that the expansion
parameter is $\xi = 21\,\%$ in this case, as given by
equation~\eqref{eq:expansion-parameter}.  This leads to a $5\,\%$ Coulomb
correction according to \citet{Kuraev2014}. In comparison, for the targets used
in \citea{Attwood2006} $\xi$ ranges from $0.8\,\%$ for H ($Z=1$) to $22\,\%$ for
Fe ($Z=26$). Thus, the discrepancy between \citea{Attwood2006} and
\citea{Akimenko1986} would not be consistent with misestimated Coulomb
corrections.

Given the previous results, it has been decided to trust the theoretical
prediction for the elastic scattering \ac{dcs}. We consider a $4\,\%$
experimental bias plausible in comparison to a systematic error by a factor of
$\sqrt{2}$ on atomic charge radii.  As a result of this choice, PUMAS predicts
larger scattering angles than Geant4, v$10.7$ at the time of this writing.

\subsection{Background computation \label{sec:validation-showa-shinzan}}

In previous section~\ref{sec:validation-transmission}, we considered a simple
transmission muography experiment. This simplified problem allows us to
cross-check the impact on the transmitted flux of uncertainties related to the
muon energy loss. However, in a more realistic experiment, not all observed
muons propagate straight through the target. In particular, in surface
experiments, some down-going atmospheric muons scatter on the target surface,
and they are observed as horizontal muons that would have crossed the target.
These events are a background for the determination of the target inner
structure.

Forward Monte~Carlo studies of this background have been performed by
\citet{Nishiyama2016} and \citet{Gomez2017}. However, forward simulations suffer
from a large Monte~Carlo inefficiency in this case. The reason for this is well
understood from geometric considerations. Contrary to transmitted muons,
scattered ones might originate from a large area of the sky, with extents of
kilometres. But, only a tiny fraction finally reaches the muography detector, of
approximately $1\ $m extent. Thus, simplifications are needed in order to
increase the efficiency of forward Monte~Carlo simulations. Therefore,
\citea{Nishiyama2016} approximated mount Showa-Shinzan, in Japan, using a
rotation symmetry around the vertical axis. Reducing the problem dimension,
increases the forward Monte~Carlo efficiency by approximately 4 orders
of magnitude, in this case. This makes forward Monte~Carlo simulations of the
scattered background tractable on a batch system, but at the cost of a loss of
realism.

The backward Monte~Carlo procedure available in PUMAS was developed specifically
in order to address the problem of the background induced by scattered muons.
The reverse Monte~Carlo method guarantees that every simulated muon crosses the
muography detector. However, it does not guarantee that all events are relevant.
For example, not all backward muons reach the sky. Some backward muons end up
deep below the ground with energy diverging to absurdly high values. However,
this is simple to cope with, e.g. by setting an upper bound on the projectile
energy.

In order to validate PUMAS backward method for background computations, let us
consider the toy geometry of \citet{Nishiyama2016}, i.e. a symmetric dome of
$270\ $m high made of standard rock with a bulk density of $2.0\ $g/cm$^3$. Let
us point out that the background method would operate with equal performances if
a realistic geometry was used instead, e.g. provided by a \ac{DEM}. However, the
symmetry assumption is needed in the present case in order to compare forward
and backward Monte~Carlo results.

Let us assume that the differential flux of atmospheric muons, $\phi_0$, is
known on a cylindrical ``sky'' surface, of $300\ $m high and $500\ $m radius,
surrounding the target dome. Let us use the parametrisation of \citet{Guan2015}
for $\phi_0$, as previously in section~\ref{sec:validation-transmission}. Note
that this parametrization is provided for sea level. Thus, in principle a height
correction should be applied. This height correction is neglected for the
present study.

Following \citea{Nishiyama2016}, let us consider that the detector is a
cylindrical belt, of $500\ $m radius and of $10\ $m height, surrounding the
dome.  The problem consists in computing the flux of atmospheric muons,
$\Phi_1$, that crosses this detector belt as function of the direction of
observation w.r.t.  the local normal to the detector surface. Owing to the
symmetry of the problem, the detector flux $\Phi_1$ does not depend on the
radial spherical coordinate.  The direction of observation is parametrized with
horizontal angular coordinates.  An azimuth of $0\ $deg points toward the
central axis of the dome.

Muons are transported between the sky surface and the detector with PUMAS.
The transport mode varies depending on the muon energy. Straggling is enabled
below $10\ $GeV. Otherwise, the mixed mode is used. In addition, scattering is
disabled above $100\ $GeV. Both forward and backward Monte~Carlo simulations are
done using this variable scheme.

In forward mode, muons are generated on the sky surface, while in backward mode
they are generated on the detector belt. In both cases, the kinetic energy at
generation is randomised from a $1/T$ power law. This results in a log-uniform
distribution of generated values, which is efficient for a flux of atmospheric
muons spanning several decades in energy. As in
section~\ref{sec:validation-transmission}, the initial and final energy of muons
are clipped to $[T_{min}, T_{max}]$, where $T_{min} = 10^{-3}\ $GeV and $T_{max}
= 10^{6}\ $GeV. In this case, a lower value than previously is used for
$T_{max}$ since one is concerned with the background of low energy muons. In
addition, the target model is of moderate size, such that TeV muons are able to
cross its largest depth.

\begin{figure}[th]
    \center
    \includegraphics[width=\textwidth]{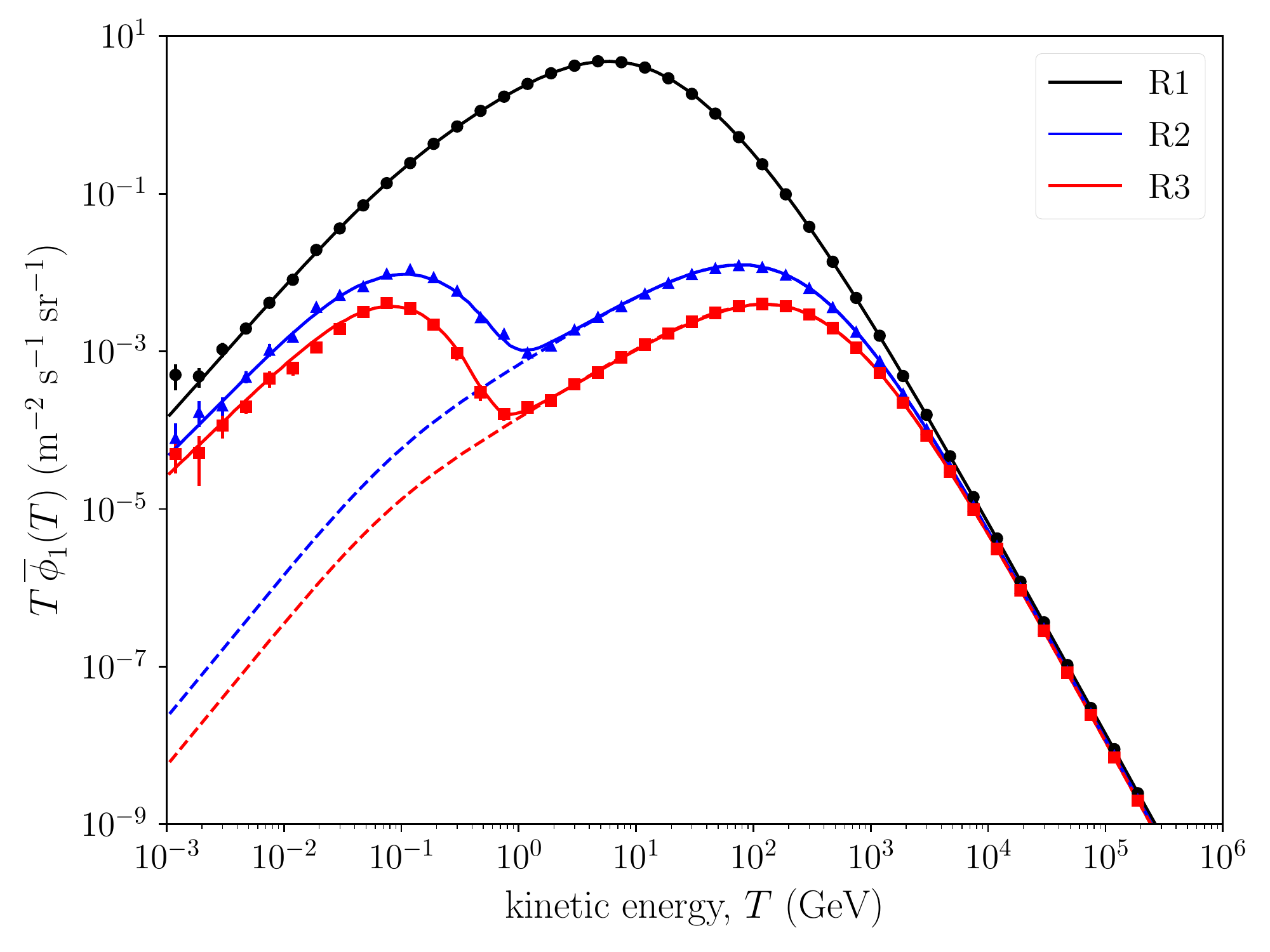}
    \caption{Mean differential flux of muons in the control regions $R1$
    (black), $R2$ (blue) and $R3$ (red) for the Showa-Shinzan toy geometry of
    \citet{Nishiyama2016}.  The flux is computed with PUMAS as indicated in the
    text. A standard rock of bulk density $2.0\ $g/cm$^3$ is considered for
    mount Showa-Shinzan. The solid lines stand for the backward Monte~Carlo
    simulation.  The markers with error bars indicate the forward Monte~Carlo
    results. The dashed lines show the flux obtained when scattering is
    disabled.
    \label{fig:showa-shinzan-flux}}
\end{figure}

\begin{figure}[th]
    \center
    \includegraphics[width=\textwidth]{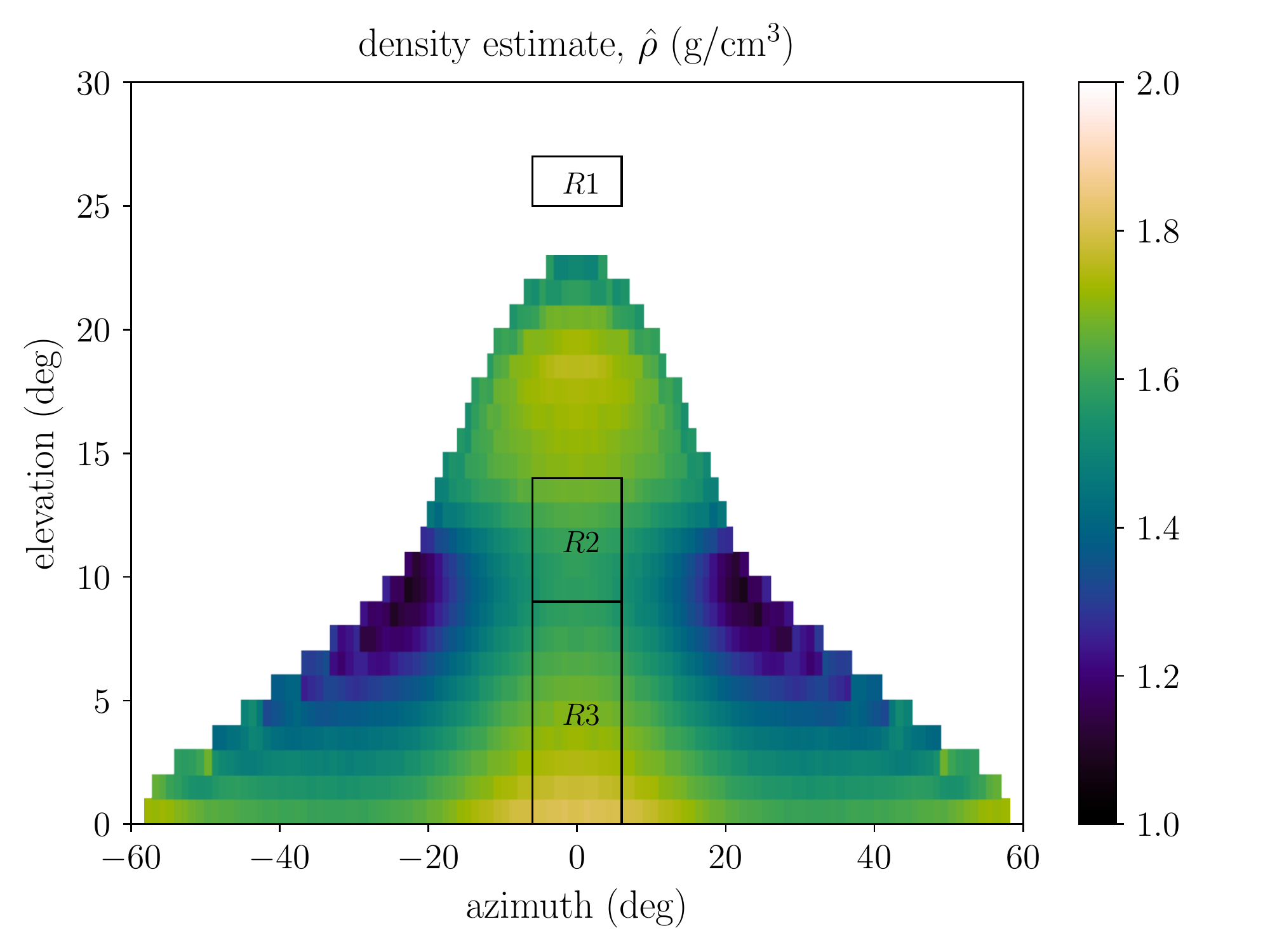}
    \caption{Impact of the scattering on the inverted bulk densities. The
    Showa-Shinzan geometry of \citet{Nishiyama2016} is considered as target.
    The reported densities, $\hat{\rho}$, are computed by inverting the flux
    values with scattering using the flux values without scattering.  The three
    control regions, $R1$, $R2$ and $R3$ are also indicated on the figure.
    \label{fig:showa-shinzan-density}}
\end{figure}

The mean values of the detector differential flux, $\overline{\phi}_1$, are
computed over three angular control regions: $R1$, $R2$ and $R3$. These regions
correspond to the ones defined by \citet{Nishiyama2016}, but we set a two times
larger width in azimuth in order to increase the Monte~Carlo statistic. The
control regions are visible on figure~\ref{fig:showa-shinzan-density}. The
corresponding mean differential fluxes are shown on
figure~\ref{fig:showa-shinzan-flux}. At high energy, perfect agreement is found
between the forward and backward Monte~Carlo results as in
section~\ref{sec:validation-transmission}. The lower the observed muon energy,
the less efficient the forward Monte~Carlo and the larger the Monte~Carlo
uncertainties. Forward and backward computations agree within those
uncertainties.

The detector mean differential flux has also been computed, in backward
Monte~Carlo mode, with scattering completely disabled.  The results are shown as
a dashed line on figure~\ref{fig:showa-shinzan-flux}.  The effect of muon
scattering is clearly visible on this figure. Down going atmospheric muons
scattering on the target appear as an additional component peaking out at low
energy, i.e.  $\sim$$100\ $MeV for $R2$ and $R3$.

The Monte~Carlo simulations have been done on the batch system of CC-IN2P3. The
used simulation queue was heterogeneous with both AMD EPYC 7302 and
Intel$^\copyright$ Xeon$^\copyright$ E5-2650~v4 CPUs. The Monte~Carlo
performances are measured by the mean time $\tau_{100}$ needed in order to
achieve a $1\,\%$ relative accuracy on the detector flux $\Phi_1$. The values
obtained for the three regions of interest are summarised in
table~\ref{tab:tau100}. When pointing toward the sky, i.e. region $R1$, the
backward Monte~Carlo is faster than the forward one by 3 orders of magnitude.
The most impressive gains are however obtained when pointing through the target,
with almost 5 orders of magnitude difference in region $R3$.

The previous gain estimate does not take into account the fact that in forward
mode one benefits from using a symmetric target. The corresponding gain can be
estimated as the ratio of the belt detector surface, $S_b = 3.1 \cdot 10^4\
$m$^2$, to the one of a realistic muography detector, i.e. $S_d \simeq 1\
$m$^2$. This would add 4 extra orders of magnitude speed up in favour of the
backward Monte~Carlo. However, the gain estimated that way is an upper limit
since additional optimizations might be used in forward mode, that have not been
considered in the present work. In addition, the reported numbers depend on the
detection threshold, $T_{min}$.  The higher the detection threshold the less
scattering matters and thus the lower the gain of the backward method.

In any case, it is observed that for transmission muography problems impressive
gains can be obtained with backward Monte~Carlo.  This is particularly true when
considering the computation of the background due to low energy scattered muons.
Accurate background estimates for realistic topographies are hardly achievable
with classical forward techniques.

\begin{table}[th]
    \caption{Mean CPU time, $\tau_{100}$, in order to achieve a $1\,\%$
    Monte~Carlo accuracy on the flux of muons, $\Phi_1$. The results for the
    three control regions, $R1$, $R2$ and $R3$ are reported for both forward and
    backward Monte~Carlo modes.
    \label{tab:tau100}}
    \center
    \begin{tabular}{lll}
        \toprule
        region & mode & $\tau_{100}$ (h) \\
        \midrule
        \multirow{2}{*}{$R_1$} & forward  & $32.9$   \\
                               & backward & 0.145    \\ \midrule
        \multirow{2}{*}{$R_2$} & forward  & $1\,317$ \\
                               & backward & 0.540    \\ \midrule
        \multirow{2}{*}{$R_3$} & forward  & $3\,082$ \\
                               & backward & 0.614    \\
        \bottomrule
    \end{tabular}
\end{table}

In order to assess the impact of the background due to scattered muons, let us
perform a density inversion. Let us first recall that current muographs cannot
measure the energy of detected particles. Thus, only the total flux can be
observed, not the differential one w.r.t. the energy.

Let us consider the detector flux computed with scattering enabled as the true
values of $\Phi_1$. Let us estimate the target bulk density using the flux
obtained without scattering. The reconstructed density values $\hat{\rho}$ are
shown on figure~\ref{fig:showa-shinzan-density}. Systematically lower values are
found than the true density, i.e. $2.0\ $g/cm$^3$. The reconstructed densities
vary from $1.1\ $g/cm$^3$ to $1.8\ $g/cm$^3$.

Interestingly, the highest bias is not observed in the central control regions,
$R1$, $R2$ or $R3$, but instead on the flanks at points where the slope of mount
Showa-Shinzan changes. This indicates that the background due to scattered muons
depends on the details of the target shape. Studies of real world targets, with
topographies given by \acp{DEM}, have shown that some detector locations can be
significantly better than others in order to reduce the background. PUMAS can
help with that respect, in the process of selecting a proper deployment site.

As stated previously, the present results depend on the detection energy
threshold $T_{min}$, thus on the detector characteristics. The higher this
threshold, the lower the bias. In addition, as observed by
\citet{Nishiyama2016}, not only muons contribute to the background but also low
energy protons and electrons. PUMAS cannot simulate those.

\section{Conclusion}

Muography has been a booming spin-off of astroparticle physics over the last
decade, with a plethora of potential applications. Some of these applications
require sophisticated computations, whereas faster but more approximate
computations are appropriate for others.  The PUMAS library is designed to cope
with these various cases. It provides a framework with a configurable accuracy,
from fast \ac{csda} estimates to detailed backward Monte~Carlo computations.

The PUMAS library is an open source project available online under the terms of
the LGPL-3.0 license. It benefited from other open source projects,
e.g.~Geant4\,\citep{Agostinelli2003,Allison2006,Allison2016} and
PROPOSAL\,\citep{Koehne2013,Dunsch2019}, as well as from a detailed related
literature, in particular for the PENELOPE
Monte~Carlo~\citep{Baro1995,Sempau1997,Salvat2015}.

In this article, we gave a comprehensive description of the physics implemented
in PUMAS. Particular care is taken in being accurate both on the energy loss and
on the scattering, over a large range of energies relevant to muography. For
transmission muography, good agreement, within $0.4\,\%$, is found between PUMAS
and PROPOSAL. The observed differences could be related to the use of two
different models for the density effect. Considering existing data, both models
are valid. Concerning multiple scattering, a $4\,\%$ discrepancy is observed
between experimental results of \citet{Attwood2006}, on one side, and
\citet{Akimenko1986} and the theory, on the other side.  Additional measurements
would be needed in order to assess if this is a theoretical or experimental
issue. In the current state, PUMAS follows theoretical predictions.  As a result
it differs from Geant4, which has been tuned according to the experimental data
of \citea{Attwood2006}

The unique feature of PUMAS lies in its backward Monte~Carlo algorithm. A
mathematical description of this technique was provided previously by
\citet{Niess2018}.  In the present article, the main results, relevant to PUMAS,
have been summarised in physicist language. Moreover, the efficiency of the
backward Monte~Carlo method for background estimates has been illustrated.

The PUMAS library is implemented in C99 with no other external dependency but
the standard library. It exposes a low level C API whose description is
available from the website~\citep{PUMAS:GitHub}. The Monte~Carlo geometry is
supplied by the user as a callback function. This design makes PUMAS very
portable, flexible and easy to interface with other codes, not necessarily
limited to muography.  However, this flexibility comes at a price. It implies
that PUMAS cannot be used out of the box without a minimum of coding. This can
be a limitation for the community. Thus, a future improvement would be the
development of a Python wrapper.

\section*{Acknowledgements}

The author thanks J.~Linnemann and an anonymous reviewer for their critical
reading which contributed to improve the present paper. In addition, we
gratefully acknowledge support from the CNRS/IN2P3 Computing Center (Lyon -
France) for providing computing and data-processing resources needed for this
work. Monte Carlo data analysis was done with numpy~\cite{Harris2020}, and
figures were produced with matplotlib~\cite{Hunter2007}.

\appendix

\section{Wallace correction to the eikonal approximation \label{sec:wallace}}

Following \citet{Wallace1971}, the eikonal phase is given by
\begin{align}
    \varphi(r) &= -2 \xi \int_{r}^{+\infty}{\frac{\omega(r')}{\sqrt{r^2 - r'^2}}
        \left( 1 + \lambda \dot{\omega}(r') \right) dr'}, \\
    \lambda & = \frac{\alpha z Z \hbar}{\beta p_0},
\end{align}
where $\dot{\omega}$ is the first derivative of $\omega$ with respect to $r$.
This expression is equivalent to \citet{Moliere1947}, i.e.
equation~\eqref{eq:eikonal_dcs} herein, but using an effective atomic charge
$Z_{eff}$ and an effective screening function $\omega_{eff}$, as
\begin{align}
    Z_{eff}         &= Z \left( 1 +
        \lambda \dot{\omega}(0) \right), \\
    \omega_{eff}(r) &= \frac{Z}{Z_{eff}} \omega(r) \left( 1 +
        \lambda \dot{\omega}(r) \right) .
\end{align}

\begin{figure}[th]
    \center
    \includegraphics[width=\textwidth]{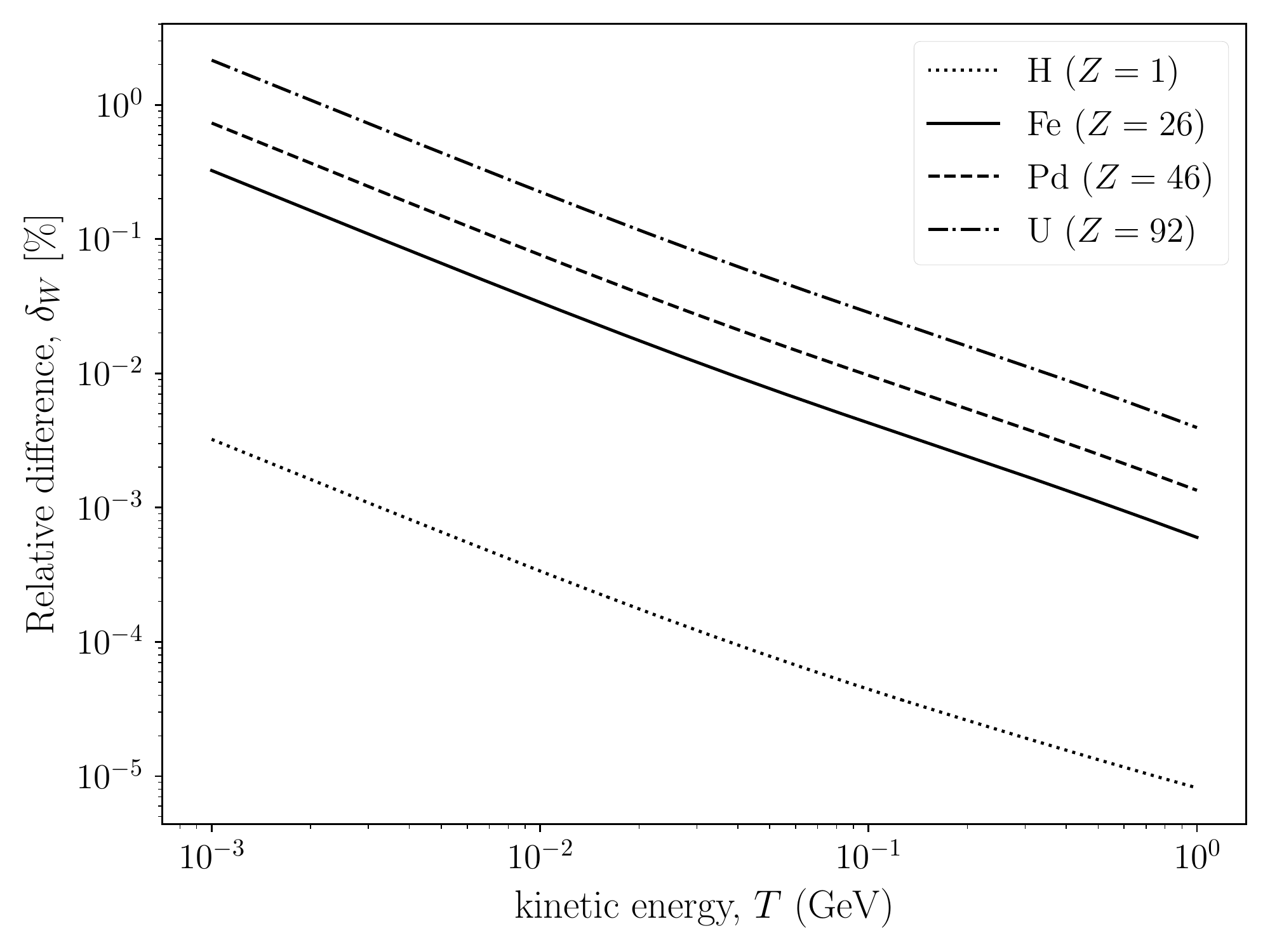}
    \caption{Relative difference on the CM elastic \ac{dcs} using the Wallace
    correction or not. A muon is considered. The different curves correspond to
    different target atoms as indicated on the figure.
    \label{fig:wallace}}
\end{figure}

Let us now consider the screening function defined by
equation~\eqref{eq:screening}.  The corresponding effective parameters are
\begin{align}
    Z_{eff} = {} & Z \left( 1 - \lambda \sum_{i=1}^n{a_i b_i} \right) \\
    \omega_{eff}(r) = {} & \sum_{i=1}^n{a_i e^{-b_i r}} -
        \lambda \sum_{i=1}^n{a_i^2 b_i e^{-2 b_i r}} \notag \\
                    & {}- \lambda \sum_{i=1}^n\sum_{j=i + 1}^n
                    {a_i a_j (b_i + b_j) e^{-(b_i + b_j)r}}
\end{align}
The latter is also a sum of exponentials but with $n (n + 1) / 2$ extra factors.
Hence, the same elastic \ac{dcs} could be used than in
equation~\eqref{eq:elastic-dcs}, but using an effective charge $Z_{eff}$ and
with $n (n + 3) / 2$ screening terms instead of $n$. Doing so would however
significantly increase the CPU cost related to the elastic cross-section, since
the number of terms to evaluate is increased quadratically. Thus, let us
first estimate the magnitude of the Wallace correction in the case of muons. Let
us denote $\delta_W$ the maximum relative difference of the CM elastic \ac{dcs}
using the Wallace correction or not.  In practice, the difference is largest at
large scattering angles, reaching a plateau value. Values of $\delta_W$ are
represented on figure~\ref{fig:wallace} as function of the muon kinetic energy,
for different target atoms. The difference increases as $\sim$$Z^{1.4}$, and it
decreases approximatively as the inverse of the muon kinetic energy. For
uranium, $\delta_W$ reaches $2\,\%$ at $1\ $MeV.  Let us point out that similar
results are obtained when considering the elastic cross-section or the transport
path instead of the maximum \ac{dcs} deviation. Hence, the Wallace correction is
significant only below $\simeq 1\ $MeV for muons colliding on heavy elements.
But, at this energy the muon range in heavy elements is very small w.r.t. the
typical extent of muography targets, i.e. metres to kilometres.  E.g. a 1 MeV
muon has a range below $10\ \mu$m in uranium.  Therefore, such corrections are
irrelevant for the present purpose.

\section{Coulomb correction with a sum of exponentials
    \label{sec:coulomb_correction}}

Using \citet{Kuraev2014} notations, for a sum of exponentials as given by
equation~\eqref{eq:screening} herein, the screening function $q(\chi)$ is
given by
\begin{equation}
    q(\chi) = \left( \sum_{i=1}^n{a_i \frac{\chi^2}{\chi^2 + \chi_i^2}}
        \right)^2,
\end{equation}
where $\chi^2_i = 4 \mu_i$. It follows that equation~(16) of \citea{Kuraev2014}
is modified as
\begin{multline} \label{eq:kuraev_16}
    \lim_{\zeta\to\infty}{\left[ \int_0^\zeta{\frac{d\chi}{\chi}q(\chi)}
        + \frac{1}{2} - \ln\zeta \right]} = \\
    -\sum_{i=1}^n{a_i^2 \ln\chi_i} - 2\sum_{i=1}^n\sum_{j=i+1}^n{a_i a_j
        \frac{\chi_i^2\ln\chi_i - \chi_j^2\ln\chi_j}{\chi_i^2 - \chi_j^2}}
        + \frac{1}{2} \left(1 - \sum_{i=1}^n{a_i^2} \right),
\end{multline}
where we have made use of $\sum{a_i} = 1$. Comparing with the original result,
one has to substitute $-\ln\chi_a$ in equation~(18) with the r.h.s. of the
latter equation~\eqref{eq:kuraev_16}. Let $L$ be the latter limit and let
us rescale all screening factors $\chi_i$ by a single factor $e^\delta$ such
that $\chi_i' = e^\delta \chi_i$. Then, the limit is modified as
\begin{equation} \label{eq:kuraev_scaling}
    \Delta L = \delta,
\end{equation}
where again we make use of $\sum{a_i} = 1$. Thus, in equation~(36) of
\citea{Kuraev2014} we can substitute $\delta = f(\xi)$ for
$\Delta_{CC}\left[\ln\left(\chi_a'\right)\right] = f(\xi)$ .  This shows that
for a sum of exponentials, rescaling all screening parameters $\chi_i$ (i.e.
$b_i$) by $e^{f(\xi)}$ allows one to recover the eikonal multiple scattering
distribution, in the limit of small angles assumed in \citea{Kuraev2014}

Let us now consider the nucleus finite extent where the screening function is
modified by a factor $\Delta \omega_N$, i.e.
equation~\eqref{eq:nuclear-screening} herein.  Let us consider the following
parametrization of the nuclear charge density:
\begin{equation} \label{eq:nuclear_charge_2}
    \rho_{\scriptscriptstyle{N}}(r) = \frac{1}{4\pi r \left(R_2^2-R_1^2\right)}
        \left( e^{-\frac{r}{R_2}} - e^{-\frac{r}{R_1}} \right),
\end{equation}
where $R_2 > R_1$. Note that this differs from the exponential density
considered by \citet{Butkevich2002a} by a factor $1/r$. The nuclear charge
density \eqref{eq:nuclear_charge_2}  yields the following modification to the
screening function:
\begin{equation}
    \Delta\omega_{\scriptscriptstyle{N}}(r) = \frac{R_2^2}{R_2^2-R_1^2} e^{-\frac{r}{R_2}} -
        \frac{R_1^2}{R_2^2-R_1^2} e^{-\frac{r}{R_1}} .
\end{equation}
The resulting screening function is a sum of exponentials but with two extra
nuclear terms. It still satisfies $\sum a_i = 1$. Therefore, with the nuclear
charge density given by equation~\eqref{eq:nuclear_charge_2} the Coulomb
correction from \citet{Kuraev2014} is still valid.

Let us further consider the case where $R_1 \simeq R_2$ and let us write $R1 =
R(1 - \epsilon)$ and $R_2 = R(1 + \epsilon)$. Substituting into
equation~\eqref{eq:nuclear_charge_2} yields
\begin{equation}
    \rho_{\scriptscriptstyle{N}}(r) = \frac{1}{8\pi R^3} e^{-\frac{r}{R}} +
    \mathcal{O}(\epsilon) .
\end{equation}
In the limit $\epsilon \to 0$, one recovers the exponential nuclear charge
distribution used e.g. by \citet{Butkevich2002a}. Considering the limit
$\epsilon \to 0$ in equation~\eqref{eq:kuraev_16} we need to scrutinize the
term
\begin{equation}
    \lim_{\epsilon \to 0}\left[
        \frac{\chi_2^2\ln\chi_2 - \chi_1^2\ln\chi_1}{\chi_2^2 -
        \chi_1^2}\right] = \ln\chi_{\scriptscriptstyle{N}} + \frac{1}{2},
\end{equation}
where $\chi_1 = \frac{\chi_{\scriptscriptstyle{N}}}{1 - \epsilon}$, $\chi_2 =
\frac{\chi_{\scriptscriptstyle{N}}}{1 + \epsilon}$ and
$\chi_{\scriptscriptstyle{N}} = \frac{\hbar}{p_0 R}$. Thus, in the limit of the
exponential nuclear charge distribution equation~\eqref{eq:kuraev_scaling} is
unchanged. This proves that the scaling $\delta = f(\xi)$ can also be applied to
the exponential nuclear density, e.g. to its inverse radius, $\frac{1}{R}$.

\section{Nuclear charge radii \label{sec:nuclear-radius}}

\citet{DeVries1987} provides an extensive compilation of data relative to
nuclear charge distributions. In particular, table~I collects estimates of
the r.m.s. radius, $\sqrt{\left<r^2\right>}$. For the nuclear charge
distribution of equation~\eqref{eq:nuclear_density} one obtains
\begin{equation}
    R_{\scriptscriptstyle{N}}^2 = \frac{5}{6} \left< r^2 \right>,
\end{equation}
which let us set the radius parameter $R_{\scriptscriptstyle{N}}$.

Note that most entries in table~I of \citet{DeVries1987} concern specific
isotopes, e.g. $^{12}$C or $^{14}$C. In PUMAS, we usually do not make this
distinction, although it could be done if desired. Instead, for a given atomic
number $Z$, a single element is used with an average atomic mass $A$,
considering isotope natural abundances. In order to determine the nuclear
charge radius of these average elements the following procedure is used. For a
given $Z$, we consider all isotope data available in \citea{DeVries1987} and
select the bracketing ones, i.e. the largest below $A$ and the lowest above $A$
in atomic mass. Let $(A_0, r_0)$ and $(A_1, r_1)$ denote the corresponding
bracketing values with $A_0 < A_1$, and let $r_0$, $r_1$ be the corresponding
nuclear charge radii. If no isotope is found with atomic mass above or below, we
repeat the search for the missing(s) bound, but considering all data available
this time, i.e. other atomic elements with different $Z$. Once a proper
bracketing is found, the mean nuclear charge radius $\sqrt{\left<r^2\right>}$ is
estimated from a linear interpolation in log-log using the bracketing values,
as
\begin{equation}
    \ln\sqrt{\left<r^2\right>} = \ln r_0 +
        \frac{\ln\left(A / A_0\right)}{\ln\left(A_1 / A_0\right)} 
        \ln\left(\frac{r_1}{r_0}\right) .
\end{equation}

The procedure described previously works for $Z \leq 92$, i.e. uranium. For
atomic elements with larger $Z$ there are no entries in the compilation of
\citet{DeVries1987}. Then, an extrapolation is used instead, as
\begin{equation}
    \sqrt{\left<r^2\right>} = 1.32 A^{0.27} \text{ fm},
\end{equation}
following \citet{Butkevich2002a} but with a higher prefactor matching high $Z$
data.

\begin{figure}[th]
    \center
    \includegraphics[width=\textwidth]{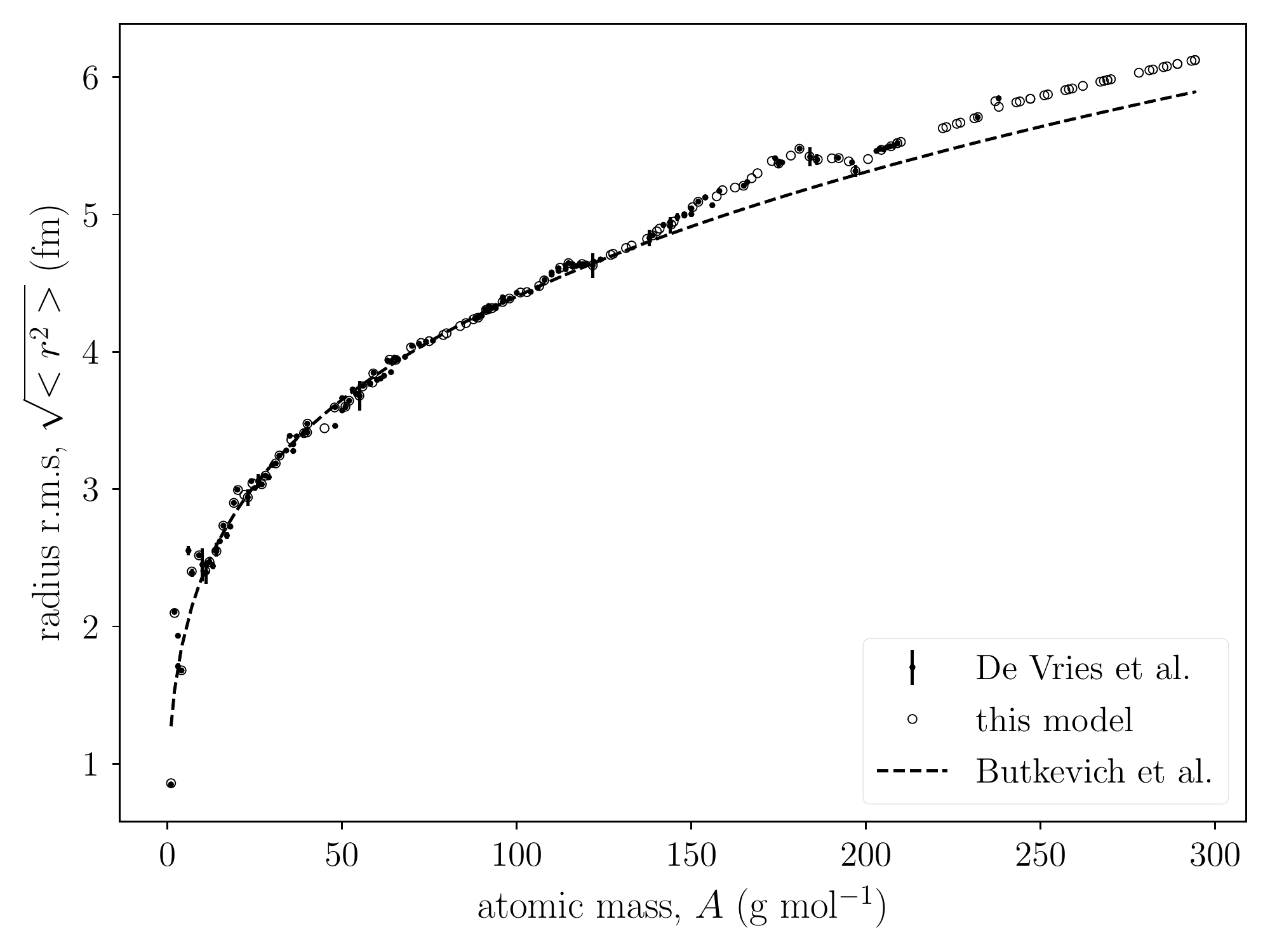}
    \caption{Nuclear charge radii, $\sqrt{\left<r^2\right>}$. The black dots
    with error bars correspond to the compilation of \citet{DeVries1987}.  The
    open dots are the fit results used in PUMAS. The dashed line indicates the
    model used by \citet{Butkevich2002a}.
    \label{fig:nuclear-radii}}
\end{figure}

The results of this procedure are compiled in table~\ref{tab:nuclear-radii} and
summarised on figure~\ref{fig:nuclear-radii}.  The corresponding mean atomic
masses $A$, taken from the \ac{pdg}, are also indicated in the table. Note that
for $Z=1$, we make an explicit distinction between hydrogen and its other
isotopes, e.g.  Deuterium, since in this case the nuclear charge radii are
significantly different.

\begin{table}[th]
    \caption{Nuclear charge radii, in fm, used in PUMAS for various atomic
    elements.  The nuclear charge radii, $\sqrt{\left<r^2\right>}$, are given
    for isotopic mixtures with atomic mass $A$ according to the \ac{pdg}. The
    radii have been fitted from \citet{DeVries1987} using the procedure
    described in \ref{sec:nuclear-radius}. ($^*$) Note that ``rockium'' (Rk) is
    a fictitious element used in standard rock.
    \label{tab:nuclear-radii}}
    \scriptsize\center
    \begin{tabular}{llll}
        \toprule
        symbol & $Z$ & $A$ & $\sqrt{\left<r^2\right>}$ \\
        \midrule
        H  &   1 &   1.008 & 0.858 \\
        D  &   1 &  2.0141 & 2.098 \\
        He &   2 &  4.0026 & 1.680 \\
        Li &   3 &    6.94 & 2.400 \\
        Be &   4 & 9.01218 & 2.518 \\
        B  &   5 &   10.81 & 2.405 \\
        C  &   6 & 12.0107 & 2.470 \\
        N  &   7 &  14.007 & 2.548 \\
        O  &   8 &  15.999 & 2.734 \\
        F  &   9 & 18.9984 & 2.900 \\
        Ne &  10 & 20.1797 & 2.993 \\
        Rk$^*$ &  11 &      22.00 & 2.958 \\
        Na &  11 & 22.9898 & 2.940 \\
        Mg &  12 &  24.305 & 3.043 \\
        Al &  13 & 26.9815 & 3.035 \\
        Si &  14 & 28.0855 & 3.098 \\
        P  &  15 & 30.9738 & 3.187 \\
        S  &  16 &  32.065 & 3.245 \\
        Cl &  17 &  35.453 & 3.360 \\
        Ar &  18 &  39.948 & 3.413 \\
        K  &  19 & 39.0983 & 3.408 \\
        Ca &  20 &  40.078 & 3.477 \\
        Sc &  21 & 44.9559 & 3.443 \\
        Ti &  22 &  47.867 & 3.595 \\
        V  &  23 & 50.9415 & 3.600 \\
        Cr &  24 & 51.9961 & 3.644 \\
        Mn &  25 &  54.938 & 3.681 \\
        Fe &  26 &  55.845 & 3.748 \\
        Co &  27 & 58.9332 & 3.843 \\
        Ni &  28 & 58.6934 & 3.776 \\
        Cu &  29 &  63.546 & 3.943 \\
        Zn &  30 &   65.38 & 3.942 \\
        Ga &  31 &  69.723 & 4.032 \\
        Ge &  32 &   72.63 & 4.065 \\
        As &  33 & 74.9216 & 4.078 \\
        Se &  34 &  78.971 & 4.123 \\
        Br &  35 &  79.904 & 4.135 \\
        Kr &  36 &  83.798 & 4.188 \\
        Rb &  37 & 85.4678 & 4.209 \\
        Sr &  38 &   87.62 & 4.237 \\
        Y  &  39 & 88.9058 & 4.249 \\
        Zr &  40 &  91.224 & 4.306 \\
        Nb &  41 & 92.9064 & 4.318 \\
        Mo &  42 &   95.95 & 4.363 \\
        Tc &  43 & 97.9072 & 4.388 \\
        Ru &  44 &  101.07 & 4.432 \\
        Rh &  45 & 102.906 & 4.435 \\
        Pd &  46 &  106.42 & 4.479 \\
        Ag &  47 & 107.868 & 4.520 \\
        Cd &  48 & 112.414 & 4.612 \\
        In &  49 & 114.818 & 4.646 \\
        Sn &  50 &  118.71 & 4.640 \\
        Sb &  51 &  121.76 & 4.630 \\
        Te &  52 &   127.6 & 4.714 \\
        I  &  53 & 126.904 & 4.706 \\
        Xe &  54 & 131.293 & 4.756 \\
        Cs &  55 & 132.905 & 4.774 \\
        Ba &  56 & 137.327 & 4.823 \\
        La &  57 & 138.905 & 4.848 \\
        Ce &  58 & 140.116 & 4.878 \\
        \bottomrule
    \end{tabular}
    \quad
    \begin{tabular}{llll}
        \toprule
        symbol & $Z$ & $A$ & $\sqrt{\left<r^2\right>}$ \\
        \midrule
        Pr &  59 & 140.908 & 4.897 \\
        Nd &  60 & 144.242 & 4.927 \\
        Pm &  61 & 144.913 & 4.948 \\
        Sm &  62 &  150.36 & 5.054 \\
        Eu &  63 & 151.964 & 5.093 \\
        Gd &  64 &  157.25 & 5.133 \\
        Tb &  65 & 158.925 & 5.177 \\
        Dy &  66 &   162.5 & 5.197 \\
        Ho &  67 &  164.93 & 5.210 \\
        Er &  68 & 167.259 & 5.264 \\
        Tm &  69 & 168.934 & 5.301 \\
        Yb &  70 & 173.054 & 5.390 \\
        Lu &  71 & 174.967 & 5.371 \\
        Hf &  72 &  178.49 & 5.429 \\
        Ta &  73 & 180.948 & 5.479 \\
        W  &  74 &  183.84 & 5.423 \\
        Re &  75 & 186.207 & 5.400 \\
        Os &  76 &  190.23 & 5.409 \\
        Ir &  77 & 192.217 & 5.411 \\
        Pt &  78 & 195.084 & 5.387 \\
        Au &  79 & 196.967 & 5.318 \\
        Hg &  80 & 200.592 & 5.404 \\
        Tl &  81 &  204.38 & 5.471 \\
        Pb &  82 &   207.2 & 5.498 \\
        Bi &  83 &  208.98 & 5.520 \\
        Po &  84 & 208.982 & 5.520 \\
        At &  85 & 209.987 & 5.529 \\
        Rn &  86 & 222.018 & 5.629 \\
        Fr &  87 &  223.02 & 5.637 \\
        Ra &  88 & 226.025 & 5.661 \\
        Ac &  89 & 227.028 & 5.669 \\
        Th &  90 & 232.038 & 5.710 \\
        Pa &  91 & 231.036 & 5.701 \\
        U  &  92 & 238.029 & 5.784 \\
        Np &  93 & 237.048 & 5.825 \\
        Pu &  94 & 244.064 & 5.824 \\
        Am &  95 & 243.061 & 5.817 \\
        Cm &  96 &  247.07 & 5.843 \\
        Bk &  97 &  247.07 & 5.843 \\
        Cf &  98 &  251.08 & 5.868 \\
        Es &  99 & 252.083 & 5.875 \\
        Fm & 100 & 257.095 & 5.906 \\
        Md & 101 & 258.098 & 5.912 \\
        No & 102 & 259.101 & 5.918 \\
        Lr & 103 &  262.11 & 5.937 \\
        Rf & 104 & 267.122 & 5.967 \\
        Db & 105 & 268.126 & 5.973 \\
        Sg & 106 & 269.129 & 5.979 \\
        Bh & 107 & 270.133 & 5.985 \\
        Hs & 108 & 269.134 & 5.980 \\
        Mt & 109 & 278.156 & 6.033 \\
        Ds & 110 & 281.164 & 6.051 \\
        Rg & 111 & 282.169 & 6.056 \\
        Cn & 112 & 285.177 & 6.074 \\
        Nh & 113 & 286.182 & 6.079 \\
        Fl & 114 &  289.19 & 6.097 \\
        Mc & 115 & 289.194 & 6.097 \\
        Lv & 116 & 293.204 & 6.119 \\
        Ts & 117 & 294.211 & 6.125 \\
        Og & 118 & 294.214 & 6.125 \\
        \bottomrule
    \end{tabular}
\end{table}

\section{Implementation of elastic collisions
    \label{sec:elastic-implementation}}

Hard elastic collisions are randomised using a rejection sampling method. The
elastic \ac{dcs} of equation~\eqref{eq:elastic-dcs} is bounded from above by
\begin{equation}
    \frac{d\sigma_0}{d\mu_0} \leq \pi \left( \frac{\alpha Z \hbar}{\beta p_0}
    \right)^2 \left(\frac{1}{\tilde{\mu}_{\text{min}} + \mu_0} \right)^2,
\end{equation}
where $\tilde{\mu}_{\text{min}} = \min(\tilde{\mu}_i)$. The bounding \ac{pdf}
can be randomised directly using the inverse \ac{cdf} method, as
\begin{equation}
    \mu_0 = \frac{\left(\tilde{\mu}_{\text{min}} +
    \mu_{\scriptscriptstyle{H}}\right)\left(\tilde{\mu}_{\text{min}} +
    1\right)}{\tilde{\mu}_{\text{min}} + 1 - \xi (1 -
    \mu_{\scriptscriptstyle{H}})} - \tilde{\mu}_{\text{min}},
\end{equation}
where $\mu_{\scriptscriptstyle{H}}$ is the cutoff value between soft and hard
elastic scattering, and where $\xi$ is a random number uniformly distributed
over $[0,1]$.

The simulation of elastic collisions also requires integrating the elastic
\ac{dcs} and its first moment, as discussed in section
\ref{sec:multiple-scattering} (see e.g. equation~\eqref{eq:elastic-mfp}).
With the U$^2$ nuclear form factor, this would require a numerical integration.
However, in this case the integral is well approximated by using a smoother
nuclear form factor, obtained from an exponential charge density, as
\begin{equation}
    F_{exp} = \left(
        \frac{\tilde{\tilde{\mu}}_{n+1}}{\tilde{\tilde{\mu}}_{n+1} +
        \mu}\right)^4,
\end{equation}
where $\tilde{\tilde{\mu}}_{n + 1} = 10 \tilde{\mu}_{n+1}$ (see
equation~\eqref{eq:elastic-nuclear-screening} for definition of
$\tilde{\mu}_{n+1}$). Consequently, the elastic \ac{dcs} is approximated by a
rational fraction in $\mu_0$ with the following shape:
\begin{equation}
    R(x) = \left( \sum_{i=1}^n{\frac{A_i}{x_i + x}} \right)^2
        \frac{x_{n+1}^4}{(x_{n+1} + x)^4} (1 - B x) .
\end{equation}
This rational fraction can be reduced to
\begin{equation} \label{eq:reduced-fraction}
    R(x) = \left(
        \sum_{i=1}^{n}{\frac{a_i}{(x_i + x)^2}} +
        \sum_{i=1}^{n}{\frac{b_i}{x_i + x}} +
        \sum_{i=1}^{4}{\frac{c_i}{(x_{n+1} + x)^i}} \right)
        (1 - B x) .
\end{equation}
The coefficients $a_i$, $b_i$ and $c_i$ of the reduction are given by
\begin{align}
    a_i &= {} r^4_{n+1,i}  A_i^2, \\
    b_i &= {} r^4_{n+1,i} \left( s_i - 4 A_i^2 d_{n+1,i} \right), \\
    c_i &= {} x_{n+1}^{i - 1} \left( \frac{5 - i}{x_{n+1}}
        \sum_{j=1}^n{A_j^2 r_{n+1,j}^{6-i} - \sum_{j=1}^n{s_j r_{n+1,j}^{5-i}}}
        \right),
\end{align}
where
\begin{equation}
    d_{ij} = \frac{1}{x_i - x_j}\text{, }
    r_{ij} = \frac{x_i}{x_i - x_j}\text{ and }
    s_i = 2 A_i \sum_{j = 1, j \neq i}^n{A_j d_{ji}} .
\end{equation}
The reduced fraction given by equation~\eqref{eq:reduced-fraction} can be
integrated analytically yielding the (restricted) elastic cross-section for
higher moments, e.g. the transport path. The $m^\text{th}$ moment is
\begin{align} \label{eq:elastic-integral}
    \int_0^{x_0}{R(x) x^m dx} =&
        \sum_{i=1}^n{a_i \left( I_{m, 2}(x_0, x_i) - B
            I_{m + 1, 2}(x_0, x_i) \right)} \, + \nonumber \\
        & \sum_{i=1}^n{b_i \left( I_{m, 1}(x_0, x_i) -
            B I_{m + 1, 1}(x_0, x_i) \right)} \, + \nonumber \\
        & \sum_{i=1}^4{c_i \left( I_{m,i}(x_0, x_{n+1}) -
            B I_{m + 1,i}(x_0, x_{n+1})\right)},
\end{align}
where
\begin{equation}
    I_{m,p}(a, b) = \int_0^a{\frac{x^m}{\left(b + x\right)^p} dx},
\end{equation}
for $m \geq 0$ and $p \geq 0$. Depending on the use case, the upper bound $x_0$
in integral~\eqref{eq:elastic-integral} can stand for the angular cutoff
parameter, $\mu_{\scriptscriptstyle{C}}$, defined in the CM frame, or $x_0 =1$
when computing the total moment. The latter family of integrals can be computed
by recurrence starting from
\begin{align}
    I_{m, 0} &= {} \frac{a^{m+1}}{m+1}, \\
    I_{0, p} &= {} \begin{cases}
        \ln\left(1 + \frac{a}{b}\right) \text{ if } p = 1 \\
        \frac{1}{p+1} \left(\frac{1}{a^{p+1}} - \frac{1}{(a + b)^{p+1}} \right)
        \text{ if } p \geq 2
    \end{cases},
\end{align}
and using the recurrence relation
\begin{equation}
    I_{m+1,p+1} = I_{m,p} - b I_{m, p+1} .
\end{equation}

While the previous expressions for the integral of elastic \ac{dcs} might seem
complicated for an analytical resolution, the implementation in a computing
language like C is rather straightforward. Nevertheless, due to rounding errors,
the integrated reduced fraction can be numerically unstable at high energies,
where the screening factor are very small. This can be solved by noting that
$c_0 + \sum b_i = 0$, e.g. by considering the limit $x R(x) / (1 - B x)$ when $x
\to \infty$ in equation~\eqref{eq:reduced-fraction}. Thus, the terms depending
only on $x_0$ can be set to zero when injecting $I_{m,1}$ and $I_{m+1,1}$ back
in equation~\eqref{eq:elastic-integral}. This was found to solve the numerical
instabilities at high energy.

Using the parametrization of \citet{Salvat1987} for the screening function, the
case of hydrogen was found numerically problematic even at low energies. This
can be understood by looking at the corresponding parameter values. The
exponents $b_i$ differ only by $0.3\,\%$ and, as a result, the prefactors
$a_i$ are very large with opposite signs. This is pathological,
suggesting that the double exponential model is not relevant in the case of
hydrogen.  Therefore, in this case we decided to use a single exponential
instead with an exponent $b_{\scriptscriptstyle{H}}$ fitted in order to
reproduce the same transport path as the original model of \citet{Salvat1987}.
This procedure yields
\begin{equation}
    b_{\scriptscriptstyle{H}} = 1.1172 \, a_0,
\end{equation}
where $a_0$ the Bohr radius. Note that the value obtained that way is close from
the one given by the Thomas-Fermi model, i.e.  $b_{\scriptscriptstyle{TF}} =
1.1299 \, a_0$.

Let us point out that equation~\eqref{eq:elastic-integral} assumes that the
integration is done in the CM frame. However, the transport path is needed in
the laboratory frame. In order to transform the integral to the laboratory,
frame we use the following approximation.  For small angles, the reduced angular
parameter in the laboratory frame, $\mu$, is proportional to the CM one,
$\mu_0$, as:
\begin{equation} \label{eq:elastic-boost}
    \mu = \frac{1}{\gamma^2_\text{\tiny CM} \left(1 + \tau\right)^2} \mu_0 +
        \mathcal{O} \left(\mu_0^2\right),
\end{equation}
where
\begin{equation}
    \gamma^2_\text{\tiny CM} = \frac{\left(E + m_A\right)^2}{m^2 + 2 m_A E +
        m_A^2},
\end{equation}
and
\begin{equation}
    \tau = \frac{m_A E + m^2}{m_A E + m_A^2} ,
\end{equation}
where $m_A$ is the target atom rest mass energy. Consequently the moments of
the elastic \ac{dcs} in the CM and laboratory frames are approximatively related
by
\begin{equation}
    \int{\mu^m\frac{d\sigma}{d\mu}d\mu} \simeq
    \frac{1}{\gamma^{2 m}_\text{\tiny CM} \left(1 + \tau\right)^{2 m}}
        \int{\mu_0^m\frac{d\sigma_0}{d\mu_0}d\mu_0} .
\end{equation}

The higher the projectile energy the more accurate the previous approximation,
due to the relativistic boost factor $\gamma_\text{\tiny CM}$ that squeezes the
scattering angle towards small values, as can be seen from
equation~\eqref{eq:elastic-boost}. On the contrary, the higher the projectile
energy the less accurate the approximation of the nuclear form factor with an
exponential distribution, because the closer to the nuclei the projectile can
reach. The combined effect of both approximations was checked by comparing the
result of the analytical integration, using equation~\eqref{eq:elastic-integral}
together with \eqref{eq:elastic-boost}, to a numeric integration of the initial
elastic \ac{dcs}, given by \eqref{eq:elastic-dcs}, but transformed to the
laboratory frame. The total approximation error was found to be at most $1\,\%$
for muon kinetic energies between $1\ $MeV to $1\ $EeV. This is considered
acceptable for the present purpose.

\section{Radiative correction to the electronic stopping power
    \label{sec:electronic-radiative-correction}}

Computing the contribution of the radiative correction to the electronic
stopping power requires integrating terms such as $\Delta_{e\gamma} \nu^p$, with
$p \in \{-1, 0, 1\}$. An analytical approximation can be obtained as following.
First, let us point out that the variation of $\Delta_{e\gamma}$ with $\nu$ is
logarithmic, hence mild. Then, inspecting the energy loss due to close
interactions without the radiative correction, it is observed that the term
$p=-1$ gives the dominant contribution. Further noting that $\Delta_{e\gamma}
\to 0$ for $\nu \ll m_e$, it is expected that at high energies the bulk of the
energy loss occurs for $m_e \ll \nu \ll E$. Thus, one has:
\begin{align}
    \delta_{e\gamma}(\nu) &\simeq
        \frac{\alpha}{2\pi}
        \int_{m_e / 2}^{\nu}{
        \ln\left(\frac{2\nu'}{m_e}\right) \left[
        2\ln\left(2 \gamma\right) - \ln\left(\frac{2\nu'}{m_e}\right) \right]
        \frac{d\nu'}{\nu'}}, \\
    &\simeq \frac{\alpha}{2\pi}
        \ln^2\left(\frac{2\nu}{m_e}\right) \left[
        \ln\left(2 \gamma\right) -
        \frac{1}{3}\ln\left(\frac{2\nu}{m_e}\right) \right],
        \label{eq:electronic-radiative-approx}
\end{align}
where the lower integration limit was arbitrarily set to $m_e /2$ in order to
simplify expressions. The approximation provided by
equation~\eqref{eq:electronic-radiative-approx} diverges for $\nu \to \infty$
instead of going to zero. This can be solved by substituting back $1 + 2 \nu /
m_e$ instead of $2\nu / m_e$ in the logarithm, yielding
\eqref{eq:electronic-radiative-energy}. The latter approximation results in an
error on the electronic loss that increases steadily with the projectile energy.
At EeV it reaches $\sim$$1\,\%$, which is considered accurate enough since at
those energies the energy loss is highly dominated by other radiative processes.

\section{Tabulations and interpolations \label{sec:lookup}}

The Monte~Carlo transport procedures discussed in section~\ref{sec:condensed}
imply several integral quantities, like the stopping power or the \ac{csda}
range. Computing those on the fly at each Monte~Carlo step would be prohibitive
CPU-wise. Instead, these quantities are tabulated, at PUMAS initialisation, as
function of the projectile kinetic energy. Then, intermediate values are
interpolated using cubic splines, as detailed below.

Let us consider a continuous property, $f$, whose values $f_i$ have been
computed at $n$ knots $x_i$. For a given $x$ in $[x_0, x_{n-1}[$ the value
$f(x)$ at $x$ is estimated in two steps, as
\begin{enumerate}[(i)]
    {\item First, the bracketing interval $[x_i, x_{i+1}[$ with $x_i \leq x <
        x_{i+1}$ is determined using a bisection.}
    {\item Then, the value $f(x)$ is estimated by interpolation on $[x_i,
        x_{i+1}]$ using an order $3$ polynomial, $H_i$, matching $f$ and its
        first derivative, $f'$, at $x_i$ and $x_{i+1}$.}
\end{enumerate}

The matching polynomial, $H_i$, can be cast as
\begin{align}
    H_i(x) &= \sum_{j=0}^{3}{a_{ij} t_i^j}, \\
    a_{i,0} &= f_i, \\
    a_{i,1} &= d_i, \\
    a_{i,2} &= -3 (f_i - f_{i+1}) - 2 d_i - d_{i+1}, \\
    a_{i,3} &= 2 (p_i - p_{i+1}) + d_i + d_{i+1},
\end{align}
where $t_i = (x - x_i) / (x_{i + 1} - x_i)$, $d_i = f'(x_i) (x_{i+1} - x_i)$ and
$d_{i+1} = f'(x_{i+1}) (x_{i+1} - x_i)$.

In some cases the derivative values $f'$ of $f$ are not known. Then,
combinations of finite differences are used instead, following
\citet{Fritsch1984}. The latter procedure ensures that the cubic spline respects
the local monoticity of the tabulated values. This is relevant in the present
case since the interpolated quantities are monotonic or have single extrema.
Therefore, even when the derivative values are known, the smoothing procedure of
\citea{Fritsch1984} is applied to knots where the derivative does not satisfy to
the local monoticity condition (see e.g.~\citet{Higham1992}).

By default, PUMAS tabulations are generated using a log like grid in kinetic
energy as in \citet{Groom2001}. The original grid extends up to $1\ $PeV. This
can be limiting for some applications. Therefore, in PUMAS~v$1.1$ the energy
grid has been extended to higher energies, $1\ $EeV for muons and $1\ $ZeV for
taus. The accuracy of the interpolation method used in PUMAS was checked to be
better than $0.1\,\%$ with those grids.

The bisection lookup method allows one to use grids with a non regular spacing,
e.g. like the grids of \citea{Groom2001} An alternative would be to enforce
grids to be regular in $\log{x}$. Then, the bracketing interval could be
inferred directly from $\log{x}$.  One might wonder what is the penalty of using
a bisection instead of enforcing a log-regular grid. For grids with
$\mathcal{O}(100)$ knots, as used in PUMAS, we did not observe any penalty. This
can be understood since computing a $\log$ is rather expensive CPU-wise, while
the bisection converges with few iterations on average. Note however that the
CPU cost of the bisection increases as $\ln{n}$, while the $\log$-regular method
has a flat cost. Therefore, for large grids, using the bisection method might be
expensive.

The bisection method was further improved by keeping a record of the two last
bracketing intervals. Then, when a new interpolation is requested the recorded
intervals are checked first before running any bisection. This is relevant since
it is frequent that different properties are requested for the same parameter
value, e.g. the kinetic energy. Hence, by keeping a record one avoids running
redundant bisections. In addition, for small Monte~Carlo steps involving many
interpolations, it is likely that the particle spends several steps within a
same bracketing interval.

\section{Moments of the soft energy loss \label{sec:detailed-momenta}}

Within \ac{csda}, the soft the energy loss between $s_0$ and $s_1$ is given by
\begin{equation} \label{eq:momentum-1}
    T_0 - \overline{T}_1 = \int_{s_0}^{s_1}{S_s ds},
\end{equation}
with $S_s$ per unit path length, following the notations of the PENELOPE
manual\,\citep{Salvat2015}.  At first order of Taylor expansion in $\Delta s$
the stopping power, $S_s$, expands as
\begin{equation} \label{eq:stopping-taylor}
    S_s = S_s(T_0) \left(1 - \dot{S}_s(T_0) \Delta s \right) +
        \mathcal{O}(\Delta s^2),
\end{equation}
where $\dot{S}_s$ is the derivative of $S_s$ w.r.t. the projectile kinetic
energy, T, and where it is made use of $dT = -S_s ds$. Substituting
equation~\eqref{eq:stopping-taylor} into \eqref{eq:momentum-1} and integrating
yields
\begin{equation}
    T_0 - \overline{T}_1 = S_s(T_0) \Delta s \left(1 - \frac{1}{2}\dot{S}_s(T_0)
        \Delta s \right) + \mathcal{O}(\Delta s^3) .
\end{equation}
The right hand side of the previous equation is identical to the result obtained
for $\left<\omega\right>$ in the PENELOPE manual, i.e. equation~(4.78). Hence,
$T_0 - \overline{T}_1$ is an approximation of $\left<\omega\right>$ at least up
to order $2$ of expansion in $\Delta s$.

Further expanding the square of the straggling parameter, $\Omega^2_s$, at first
order one obtains
\begin{equation} \label{eq:straggling-taylor}
    \Omega^2_s = \Omega^2_s(T_0) - \dot{\Omega}^2_s(T_0)
        S_s(T_0) \rho \Delta s + \mathcal{O}(\Delta s^2) .
\end{equation}
Substituting equation~\eqref{eq:straggling-taylor} into \eqref{eq:momentum-2}
and collecting terms up to $\Delta s^2$ one finds
\begin{multline}
    \text{var}(\omega) =  \Omega^2_s(T_0) \rho \Delta s \\
        - \dot{\Omega}^2_s(T_0) \rho S_s(T_0) \frac{\Delta s^2}{2}
        - \Omega^2_s(T_0) \rho \dot{S}_s(T_0) \Delta s^2
        + \mathcal{O}\left(\Delta s^3\right) .
\end{multline}
The latter is equivalent to equation~(4.79) in the PENELOPE
manual\,\citep{Salvat2015}.

\section{Forward sampling of hard energy losses \label{sec:analogue-sampling}}

Let $x = \nu / T$ denote the fractional energy loss in a radiative collision.
Let $I = [x_{min}, x_{max}]$ be the support of the \ac{dcs}, i.e. the interval
where it is strictly positive.  Then, the following function is a valid envelope
of the \ac{dcs} at the projectile energy $E$:
\begin{equation}
    f_\alpha(x) = \sup_{x \in I}{
        \left[\frac{d\sigma}{dx} x^\alpha\right]} x^{-\alpha} .
\end{equation}

The exponent $\alpha$ is selected by considering the logarithmic derivative of
the \ac{dcs}, $\beta$, defined as
\begin{equation}
    \beta(x) = \frac{d\ln\left(\frac{d\sigma}{dx}\right)}{d\ln{x}}.
\end{equation}
Values of $\beta$ are computed over a regular grid with constant spacing in
$\ln{x}$. The envelope parameter $\alpha$ is set as the median value over this
grid. The grid interval is restricted to the lower half of $I$ in logarithmic
scale. In addition, boundary values are rejected when they are close to a
kinematic threshold of the \ac{dcs}. In the latter case the \ac{dcs} has very
sharp variations that are not considered as representative.

Typical values of $\beta$ are between $1$ and $3$. Thus, when sampling $x$ from
the envelope, low values close to $x_{min}$ are more frequent than high ones.
Therefore, we select $\alpha$ only from the lower logarithmic half of $I$. This
ensures high selection efficiencies of the rejection sampling method for the
most frequent outcomes for $x$.

\section{Monte~Carlo stepping algorithm \label{sec:stepping}}

The Monte~Carlo stepping algorithm used in PUMAS proceeds as following. First, a
physical step length is determined depending on the physics
processes enabled, as
\begin{equation}
    \Delta s_p = \min\left( \Delta s_{\scriptscriptstyle{E}}, \Delta s_e, \Delta
        s_{\scriptscriptstyle{B}} \right),
\end{equation}
where $\Delta s_{\scriptscriptstyle{E}}$ is given by
equation~\eqref{eq:step-straggling}, $\Delta s_e$ by \eqref{eq:step-msc} and
$\Delta s_{\scriptscriptstyle{B}}$ by \eqref{eq:step-magnetic}. Note that
depending on the enabled processes the previous step length(s) might equal
$+\infty$. Note also that these step lengths can be tuned by the user by
modifying a shared accuracy parameter, $\epsilon_s$, configurable per simulation
context. By default $\epsilon_s$ is set to $1\,\%$.

Then, the physical step length is compared to the ones provided by the user for
the local density model, $\Delta s_l$, and for the medium, $\Delta s_m$. The
smallest value is used as step length, $\Delta s$.  If $\Delta s$ is below a
minimal resolution, $\Delta s_{min}$, then it is set to $\Delta s_{min}$. Thus,
\begin{equation}
    \Delta s = \max\left(\min\left(\Delta s_p, \Delta s_l, \Delta s_m\right),
        \Delta s_{min} \right) .
\end{equation}
Requiring that $\Delta s \geq \Delta s_{min}$ is a generic safeguard since the
user callback functions may return values close to zero. The minimal resolution
is set to $\Delta s_{min} = 0.1\ \mu$m.  Note that PUMAS is not expected
to be used for such thin targets.

Once the Monte~Carlo step length $\Delta s$ has been determined, a tentative
straight displacement is performed along $\vec{u}$ or, in backward mode, along
$\shortminus{\vec{u}}$. The medium and the corresponding step length, $\Delta
s_m$, at the would-be new position are requested. If a change of medium would
occur the projectile is moved slightly before the expected boundary at a
distance $\Delta s - \Delta s_{min} / 2$ from its initial position. The medium
at this location is requested. If it differs from the one at the initial
location, then a bisection is done in order to locate the change of medium.
Then, the projectile is moved just before the boundary. The bisection is a
safeguard in case that the users provides approximate step lengths for the
geometry.  Note that for the bisection, only the medium needs to be requested to
the user, not the distance to the next boundary.  If no change of medium is
detected, then the initial tentative displacement is confirmed.

Once the projectile location has been updated, the soft energy loss is applied,
using \ac{csda} or the straggling algorithm described in
section~\ref{sec:condensed}. If an external magnetic field was supplied, then
the projectile momentum direction $\vec{u}$ is rotated, as discussed previously
in section~\ref{sec:magnetic-field}. In addition, if scattering is enabled, then
the direction is further rotated by applying the soft multiple scattering
process discussed in section~\ref{sec:multiple-scattering}.  If a hard collision
occurred, then it is simulated at this stage. If instead a change of medium was
found, then the projectile is pushed by a distance $\Delta s_{min}$ into the new
medium.  Note that this push occurs without any energy loss. As stated
previously, PUMAS is expected to be used only for targets significantly thicker
than $\Delta s_{min} = 0.1\ \mu$m.  Finally, if the particle direction changed
in anyway, then a new medium stepping distance $\Delta s_m$ must be requested.
For this call the medium is known.  Thus, it is not requested.

PUMAS stepping algorithm ensures that for any call of the local density model, a
call to the medium function occurs before at the exact same location $\vec{r}$
and with the same momentum direction $\vec{u}$. This can be used in order to
optimise the implementation of these two callback functions.  For example, some
parameters might be recorded in the medium function for later use in the local
density model. In addition, the user provided steps lengths, $\Delta s_l$ and
$\Delta s_m$, are recorded from one Monte~Carlo step to the next one. Thus, for
a given transport no user callback should occur twice for the same projectile
state. For non uniform density models, the density values are also recorded
between steps.  These values are used for interpolating the density variations
over the Monte~Carlo step.

\bibliography{biblio}

\end{document}